\documentclass{aa}

\newcommand{\Lsun}{\hbox{L$_\odot$}}
\newcommand{\Msun}{\hbox{M$_\odot$}}

\usepackage{epsfig}  
\usepackage{setspace}
\usepackage{graphicx}
\usepackage{latexsym}
\usepackage{color}
\usepackage{savesym}
\usepackage{txfonts}
\savesymbol{bibfont}
\usepackage{longtable}
\usepackage{supertabular}
\usepackage{multirow}
\usepackage{bigdelim}
\usepackage{bigstrut}
\usepackage{natbib}
\usepackage{ulem}

\restoresymbol{TXF}{bibfont}

     \def\cut#1 {\sout{#1} }

\bibpunct{(}{)}{;}{a}{}{,}

\begin{document}

\title{A study of three southern high-mass star-forming regions}

\author{C. Dedes \inst{1,2} \and S. Leurini \inst{2,3} \and
 F. Wyrowski \inst{2} \and P. Schilke \inst{4} \and K. M. Menten 
\inst{2} \and
 S. Thorwirth \inst{4}  \and J. Ott \inst{5}} 

\authorrunning{C. Dedes et al.}

\titlerunning{three southern HMSF regions}

\institute{ETH Z\"urich, Wolfgang-Pauli-Strasse 27, 8093 Z\"urich, Schwitzerland, carolin.dedes@astro.phys.ethz.ch \and Max-Planck-Institut f\"ur Radioastronomie, Auf dem H\"ugel 69,
 53121 Bonn, Germany \and ESO,
 Karl-Scharzschild-Strasse 2, 85748 Garching-bei-M\"unchen, Germany \and I. Physikalisches Institut, Universit\"at zu
K\"oln, Z\"ulpicher Str. 77, 50937 K\"oln, Germany \and
 National Radio Astronomy Observatory, P.O. Box O, 1003 Lopezville Road,
Socorro, NM 87801-0387, USA}

\date{Received / Accepted }

\abstract{Based on color-selected IRAS point sources, we have started to
conduct a survey of 47 high-mass star-forming regions in the southern
hemisphere in 870~$\mu\rm{m}$ dust continuum and molecular line emission in
several frequency ranges between 290~GHz and 806~GHz. This paper describes the
pilot study of the three sources IRAS12326$-$6245, IRAS16060$-$5146,
and IRAS16065$-$5158.}
{To characterize the physical and chemical properties of southern massive star-forming regions, three high-luminosity
southern hemisphere hot cores are observed in continuum and molecular line
emission. Based on the
results obtained in the three sources, which served as templates for the
survey, the most promising (and feasible)
frequency setups for the remaining 44 sources were decided upon.}
{The sources were observed with the Atacama Pathfinder EXperiment (APEX) in five frequency setups aimed at groups of
 lines from the following molecules: {\hbox{${\rm CH}_3{\rm OH}$}}, {\hbox{${\rm H}_2{\rm CO}$}}, and {\hbox{${\rm CH}_3{\rm CN}$}}. Using the LTE
approximation, temperatures, source sizes, and column densities were determined
through modeling of synthetic spectra with the XCLASS program.  Dust continuum
observations were done with the Large APEX BOlometer CAmera (LABOCA) at 870~$\mu$m and the 3~mm continuum was
imaged with the Australia Telescope Compact Array (ATCA).}
{ Based on the
 detection of high-excitation {\hbox{${\rm CH}_3{\rm CN}$}} lines and lines from
 complex organic species, the three sources are classified as line rich, hot core type sources. For all three, the modeling indicates that the line emission emerges
 from a combination of an extended, cooler envelope, and a hot compact component. All three sources show an overabundance of oxygen-bearing species compared to
nitrogen-bearing species. While the chemistry in the three sources indicates
that they are already in an evolved stage, the non-detection of infrared heating sources at the dust
continuum peak of IRAS16065$-$5158 points to this source still being deeply embedded. Because this work served as a pilot study, the approach to observe the remaining 44
 massive star-forming regions was chosen based on its results.}
{The three sources are massive,
luminous hot cores. While IRAS16065$-$5158 seems to be a very young deeply
embedded object, IRAS12326$-$6245 and IRAS16060$-$5146 seem more evolved and have already developed
 {\hbox{UCH{\sc ii}}}\/ regions.}

\maketitle

\section{Introduction}
Massive stars (M $>$ 8~\Msun) are important building blocks of galaxies. Although they are rare,
they provide most of the stellar luminosity and heavily influence their surroundings
through energetic interactions throughout their short lifetime. 
A further understanding of the role of massive stars involves understanding
their formation processes, which has always posed a challenge, because the
earliest stages of massive star formation are deeply embedded in these objects' natal
dust clouds.

One approach to study high-mass star-forming regions in depth would
be to observe a large number of sources in a wide frequency range, which is not yet
feasible. Observing the sources in a number of frequency setups to cover many molecular lines is needed to
study the prevailing excitation conditions of the molecular gas. In the past, individual sources have either been studied in several
frequency setups \citep[see][to name a few]{1994ApJ...435L.137C,1996A&AS..119..333M,2003A&A...407..225O,2006A&A...454..221B},
or surveys of several sources have been performed, mostly on the northern hemisphere, with only very limited frequency coverage
\citep{1992A&A...253..541C,2002ApJ...570..758M,2002ApJS..143..469M,2002ApJ...566..931S,2003ApJS..149..375S}. On the
southern hemisphere, \citet{2006MNRAS.367..553P} and
\citet{2007A&A...474..891U} have performed line surveys in the mm regime. 
While many
samples of high-mass star-forming regions have been observed in
the dust continuum on the northern and southern hemispheres, such as the
samples of \citet{2002ApJ...566..945B,2004A&A...426...97F,2005MNRAS.363..405H} and
\citet{2006A&A...447..221B}, no project has so far observed the same sample
both in a high-frequency
line survey and in dust continuum. 

To address the lack of statistically significant sub-mm surveys of
southern hemisphere sources in both lines and continuum, we have conducted a survey of 47 high-mass star-forming regions in the southern
hemisphere in 870~$\mu\rm{m}$ dust continuum and molecular line emission in
several frequency ranges between 290~GHz and 806~GHz, based on
color-selected IRAS point source criteria. Our selection criteria follow those described in \citet{2002ApJ...566..931S}, with the exception that our sample
includes sources with and without detected radio continuum emission in medium resolution single
dish 5~GHz surveys. \\
This large sample contains sources at very different stages of
evolution and will allow a consistent and comparative analysis of these
regions. 

Out of this survey, we will present in this paper observations of three high-luminosity
southern hemisphere hot cores with the Atacama Pathfinder
Experiment\footnote{This publication is based on data acquired with the Atacama
Pathfinder Experiment (APEX). APEX is a collaboration between the
Max-Planck-Institut f\"ur Radioastronomie, the European Southern Observatory
and the Onsala Space Observatory.} telescope, which served as targets for a
template study for our large survey. IRAS 12326$-$6245, IRAS 16060$-$5146, and IRAS 16065$-$5158 (Table \ref{tab1}, from now on we
omit the IRAS in the name for brevity) were chosen to be analyzed 
in this pilot study,
because they are the most luminous and line-rich sources in the sample 
of our study based on their {\hbox{${\rm HCO}^+$}} and {\hbox{${\rm SO}_2$}} spectra, and therefore the most suitable targets to test the feasibility of the
project. A fourth very strong hot molecular
core of the sample, IRAS 17233$-$3606, has been analyzed and the results have
been published separately \citep{2008A&A...485..167L}.
Based on the
results obtained in these three sources, the most promising (and feasible) frequency setups for the remaining 44 sources were selected.

\subsection{Source details}

The three sources had already been observed in the course of several
maser studies
\citep{1997MNRAS.291..261W,1998MNRAS.297..215C,1998AJ....116.1897M,1998MNRAS.301..640W,2001MNRAS.326..805C},
as well as in the mm continuum survey of \citet{2004A&A...426...97F}, the {\hbox{\rm CS}} survey of \citet{1996A&AS..115...81B} and the {\hbox{\rm HNCO}} survey of
\citet{2000A&A...361.1079Z}. \\
We derived the kinematic distances to the sources from the center velocities
of the {\hbox{${\rm C}^{17}{\rm O}$}}(3--2) line by using the Galactic
velocity field description of
\citet{2007A&A...469..511K}. 
In the following paragraph, the characteristics of the
sources are summarized.\\

\subsubsection*{IRAS\, 12326$-$6245}

The kinematic distance estimated for 12326$-$6245 from its systemic velocity of
$-39.3$~{\hbox{${\rm km\, s}^{-1}$}}\/ is 4.4~kpc. 
12326$-$6245 features class II {\hbox{${\rm CH}_3{\rm OH}$}}\/ masers at 6.7~GHz, OH masers at 1.6~GHz \citep{1998MNRAS.297..215C} and {\hbox{${\rm H}_2{\rm O}$}} masers \citep{1998AJ....116.1897M}.
Observed at 1.2~mm, 12326-6245 shows compact dust continuum emission
\citep{2000A&A...364..613H,2004A&A...426...97F,2005MNRAS.363..405H,2006A&A...460..721M},
which coincides with the location of two peaks of cm continuum emission
\citep{1998MNRAS.301..640W,2007A&A...461...11U} observed at high
resolution. These peaks correspond to the location of two strong MIR peaks at the
Q~(20~{$\mu\hbox{m}$}) and N~(10~{$\mu\hbox{m}$}) bands \citep{2000A&A...364..613H}. Molecular line
observations in the {\hbox{${\rm CH}_3{\rm CN}$}}(5--4)\footnote{Here and below we
 designate series of lines close together in frequency. The first and second
 numbers are upper and lower values of the total angular momentum quantum
 number J of the individual transitions of the series that have
 different projected angular momentum ($K$) values.}, {\hbox{${\rm CH}_3{\rm CN}$}}(6--5),
{\hbox{${\rm CH}_3{\rm CN}$}}(8--7) and {\hbox{${\rm CH}_3{\rm CN}$}}(12--11) series reveal a two-component temperature structure
hinting at a hot core embedded in a cooler envelope
\citep{2005ApJS..157..279A}. Maps of molecular line data show that 12326$-$6245 seems to have one of the largest bipolar
outflows detected in the southern hemisphere \citep{2000A&A...364..613H,2004A&A...426..503W}.
\citet{2006A&A...460..721M} derive a rotation temperature of 34~K from
observations of {\hbox{${\rm CH}_3{\rm CCH}$}} bands.

\subsubsection*{IRAS\, 16060$-$5146}

The source has class II {\hbox{${\rm CH}_3{\rm OH}$}}\/ masers at 6.7~GHz and  {\hbox{\rm OH}}\/ masers at
1665~MHz, 1667~MHz \citep{1998MNRAS.297..215C} and 6035~MHz \citep{2001MNRAS.326..805C}. 
It shows strong, spherical 1.2~mm dust continuum emission
\citep{2004A&A...426...97F,2005MNRAS.363..405H,2006A&A...460..721M} that peaks at the location of the maser sources. High-resolution radio continuum
observations at 3.6 and 6~cm \citep{2007A&A...461...11U} show two  {\hbox{UCH{\sc ii}}}~ regions at the location of the mm peak.
We derived 5.5~kpc and 9.4~kpc for the near and far distances, respectively, from the systemic velocity of 16060$-$5146
($-90.95$~{\hbox{${\rm km\, s}^{-1}$}}) found in our data.
The kinematic distance estimates for this source found in the literature differ somewhat between 5.3 -- 6.1~kpc for the near distance and 9.6~kpc for the far
distance, which can be explained with the spread found in the systemic velocities assumed for this source, which range from $-88 $~{\hbox{${\rm km\, s}^{-1}$}} to $-91$~{\hbox{${\rm km\, s}^{-1}$}}.

\subsubsection*{IRAS\, 16065$-$5158}

In 16065$-$5158,  {\hbox{\rm OH}}\/ maser emission at 1665~MHz, 1667~MHz, and 1612~MHz and a
{\hbox{${\rm CH}_3{\rm OH}$}}\/ 6.7~GHz maser has been
observed \citep{1998MNRAS.297..215C}. 16065-5158 shows extended 1.2~mm
continuum emission \citep{2004A&A...426...97F} and a  {\hbox{UCH{\sc ii}}}~ region \citep{2007A&A...461...11U}.   
We derive a kinematic distance of 4~kpc and 10.8~kpc for the near
and far values of 16065-5158, using a systemic velocity of
$-62.17$~{\hbox{${\rm km\, s}^{-1}$}}. 

\begin{table*}    
\caption{Peak coordinates of the 1.2~mm dust continuum data, distances and IRAS luminosities of
 the three hot core sources (see text).}
\label{tab1}                     
\centering                                                                          
\begin{tabular}{lccccc}                                                                                            
\hline                                                                         
\hline            
Source & R.A. & Dec. & l,b  & d (near/far)&L (near/far) \\                 
& (J2000) & (J2000) &  & (kpc)  & $(10^5$\Lsun) \\
\hline                            
IRAS 12326$-$6245 & 12:35:35.90 &  -63:02:29.00 & 301.13783,-0.22483 & 4.4 & 2.7 \\
IRAS 16065$-$5158 & 16:10:20.01 & -52:06:13.26 &  330.87748,-0.36821 &  4.0 / 10.8 & 3.0/21.2 \\
IRAS 16060$-$5146 & 16:09:51.40 & -51:55:06.98 &  330.94914,-0.18246 & 5.5 / 9.4 &  7.8/23.8\\
\hline                                                                                               
\end{tabular} 

\end{table*}

\section{Observations and data reduction}
\label{obs}
\subsection{Atacama Pathfinder EXperiment}
\subsubsection{Line data}
The sub-mm data were taken between November 2005 and November 2006 with the
Atacama Pathfinder Experiment \citet[APEX,][]{2006A&A...454L..13G} telescope on Chajnantor in Chile.\\ 
We used both the APEX 2a heterodyne receiver \citep{2006A&A...454L..17R} and
the dual frequency 460/810 GHz First Light APEX Submillimeter Heterodyne (FLASH) receiver
\citep{2006A&A...454L..21H} for our observations (see Table \ref{tab:apexsetup} for
details on the receiver setups). As backend, the MPIfR-built Fast Fourier Transform
Spectrometer (FFTS) \citep{2006A&A...454L..29K} was used, which consists of two
units with 1~GHz bandwidth each. 
The pointing accuracy was checked on G327.3$-$0.6 \citep{2006A&A...454L..91W},
and was found to be accurate within
2\farcs. Most of the data were observed as single pointings. When mapping was
done, raster maps with a map size of $40\farcs \times 40\farcs$ and beam
spacing were observed. The observations were done in the position-switch
mode. The off-positions were chosen with an
offset of (600\farcs,0\farcs) in azimuth and elevation from the pointing
position and were checked to be free of {\hbox{\rm CO}} emission. All observations were obtained in double
sideband mode. Calibration errors were
estimated to be on the order of 30\%. These errors are influenced by pointing
and atmospheric fluctuations, but the largest part of the uncertainty stems
from uncertain knowledge of the sideband gain ratios. The data were reduced with the CLASS
software package\footnote{http://www.iram.fr/IRAMFR/GILDAS}.

\begin{table*}  
\caption{\label{tab:apexsetup} Frequency setup for the molecular line observations. }
\centering
\begin{tabular}{lccccccc}\hline\noalign{\smallskip}
Transition & Center Frequency  & Receiver & $\Theta_{\rm {HPBW}}$  & $F_{\rm eff}$
&$B_{\rm eff}$ & Tracer  &
$T_{\rm sys}$ \\
& (GHz) & & (\farcs) & & & & (K) \\
\noalign{\smallskip}\hline\noalign{\smallskip}
{\hbox{${\rm CH}_3{\rm OH}$}}(6--5) & 290.8  & APEX2a & 20 & 0.95 &0.73 & T & 170 - 190 \\
{\hbox{${\rm CH}_3{\rm OH}$}}(7--6) & 337.8 & APEX2a & 17.5 & 0.95& 0.73 & T & 180 - 210\\
{\hbox{${\rm CH}_3{\rm CN}$}}(16--15) & 294.5  & APEX2a & 20 & 0.95 & 0.73 & T & 140 - 200  \\
{\hbox{${\rm H}_2{\rm CO}$}}(4--3) & 290.0  & APEX2a & 20 & 0.95 & 0.73 &  T & 214 - 291 \\
{\hbox{${\rm H}_2{\rm CO}$}}(6--5)  & 436.9  & FLASH460 & 14 & 0.95& 0.6 & T & 730 - 839  \\
 {\hbox{\rm CO}} (3--2) & 345.8  & APEX2a & 17 & 0.95& 0.73 & o & 550  \\
 {\hbox{\rm CO}} (4--3) & 461.0  & FLASH460 & 13 & 0.95& 0.6 & o & 1000 -1100 \\
 {\hbox{\rm CO}} (7--6) & 806.7  & FLASH810 & 7 & 0.95& 0.43 & o &  5100 - 5700  \\
{\hbox{${\rm C}^{17}{\rm O}$}}(3--2) & 337.8 & APEX2a & 17.5 & 0.95& 0.73 &  cd & 180 - 210  \\ 
{\hbox{${\rm HCO}^+$}}(4--3) & 357.0  & APEX2a & 17 & 0.95& 0.73 & o& 240 - 324\\
\noalign{\smallskip}\hline
\end{tabular}
\tablefoot{
In the column
 marked 'Tracer', T stands for temperature, o for outflow and
 cd for molecular column density.
}

\end{table*}

Table \ref{tab:apexsetup} lists the frequencies of the bands of the molecular lines observed with APEX that are subject of
the pilot study. These lines were the targets of the frequency setups, but because up
to 1.8~GHz bandwidth could be used, lines from many other molecules could be
observed simultaneously.

To analyze the chemical composition and physical properties of the three
sources, five major setups were used, {\hbox{${\rm CH}_3{\rm OH}$}}(6--5) at
291~GHz (together with {\hbox{${\rm H}_2{\rm CO}$}}(4--3) for 16060--5146 and 16065--5158), {\hbox{${\rm CH}_3{\rm OH}$}}(7--6) at 338~GHz, {\hbox{${\rm CH}_3{\rm CN}$}}(16--15) at 294~GHz, {\hbox{${\rm H}_2{\rm CO}$}}(4--3)
at 291~GHz and {\hbox{${\rm H}_2{\rm CO}$}}(6--5) at 436~GHz. 
The positions of the hot cores were first defined as the peak positions of {\hbox{${\rm HCO}^+$}}(4--3) maps at 357~GHz and later refined by APEX continuum scans.\\
On-source integration times for APEX2a were around 5--8~min per single
pointing setup, and 12--20~min per pointing for
FLASH observations, depending on the weather conditions. 

\subsubsection{Continuum data}

In summer 2007, we obtained continuum data for our sources from the Large Apex
BOlometer Camera (LABOCA).\\
LABOCA is a 295 channel bolometer at 870 {$\mu\hbox{m}$}\/ with a field of view of 11.4~\arcmin~ and a
beam FWHM of 18.6~\farcs \citep{2009A&A...497..945S}. The pixel separation is 36~\farcs. To produce a fully
sampled map, the data were
observed in the raster-spiral mode, with 35~s integration time on source, a
spacing of 27~\farcs\/ and 4 subscans. This sets the scanning velocity to about
4\arcmin/s and the total integration time on source was on average 2.3~min. 12326$-$6245 was observed in two scans.\\
The atmospheric opacity was determined every hour through skydips. The zenith
opacity was 0.3 for 12326$-$6245 and 0.09 for 16060$-$5146 and 16065$-$5158. 
The rms noise for 12326$-$6245 is 100~mJy/beam, while the rms noise for 16060$-$5146 and
16065$-$5158 is 50~mJy/beam.\\ IRAS 16293$-$2422,
IRAS 13134$-$6242 and $\eta$ Carinae were used as flux calibrators and the
pointing was checked on $\eta$ Carinae for 12326$-$6245 and on IRAS 16293$-$2422
for 16060$-$5146 and 16065$-$5158. The pointing was good within 3\farcs\/.\\
The data were reduced with the BoA software package (Schuller et al., in prep.).

\subsection{ATCA}

High spatial resolution data at 3~mm were taken in October 2006 with the
Australia Telescope Compact Array (ATCA) interferometer at
Narrabri, Australia. The array was in the H75 configuration with a synthesized
beam of  $6\farcs\times 4\farcs$ for 16060$-$5146, $5\farcs\times 5\farcs$ for 16065$-$5158
and $6\farcs\times 6\farcs$
for 12326$-$6245 and in the FULL$\_$16$\_$256$-$128 correlator configuration. \\
The continuum was observed at 87.89~GHz with a bandwidth of 128~MHz
and a resolution of 8.8~MHz. \\
The sources were observed during two nights, PKS 1253$-$055 was used as bandpass
calibrator, Mars was observed on the second night as flux calibrator, the gain
calibration for 12326$-$6245 was done on PKS 1147$-$6753, and on PKS 1613$-$586 for 16060$-$5146 and
16065$-$5158.
The theoretical continuum rms noise level was 3~mJy/beam for the continuum and 130~mJy/beam for
the line observations, respectively. 
The data were reduced with MIRIAD\footnote{http://www.atnf.csiro.au/computing/software/miriad} and the dirty images were
de-convolved with the CLEAN algorithm \citep{1974A&AS...15..417H}.

\section{Observational results}

\subsection{Line data}
\begin{figure} 
\includegraphics[angle=-90,width=7.2cm]{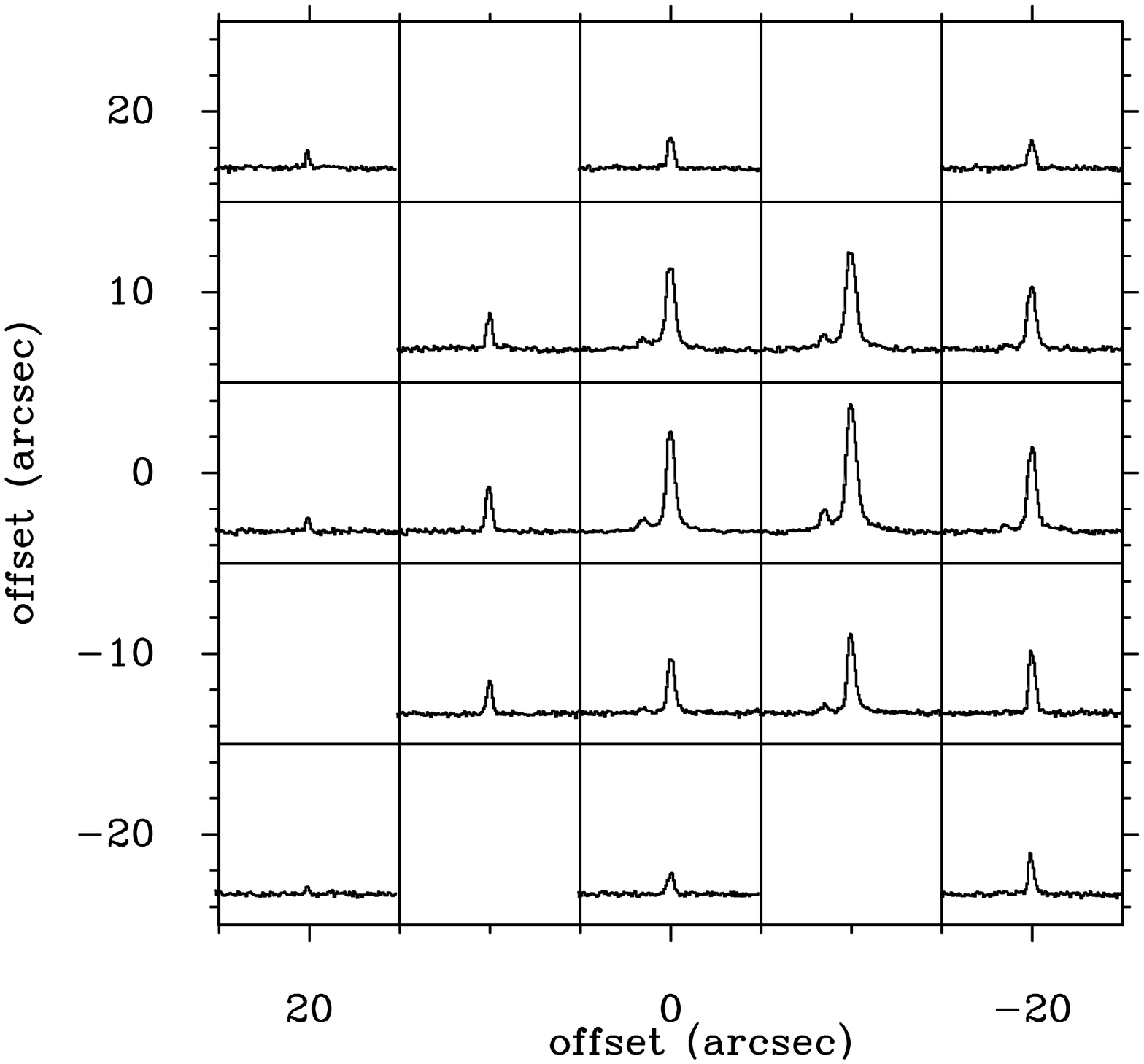}
\includegraphics[angle=-90,width=7.2cm]{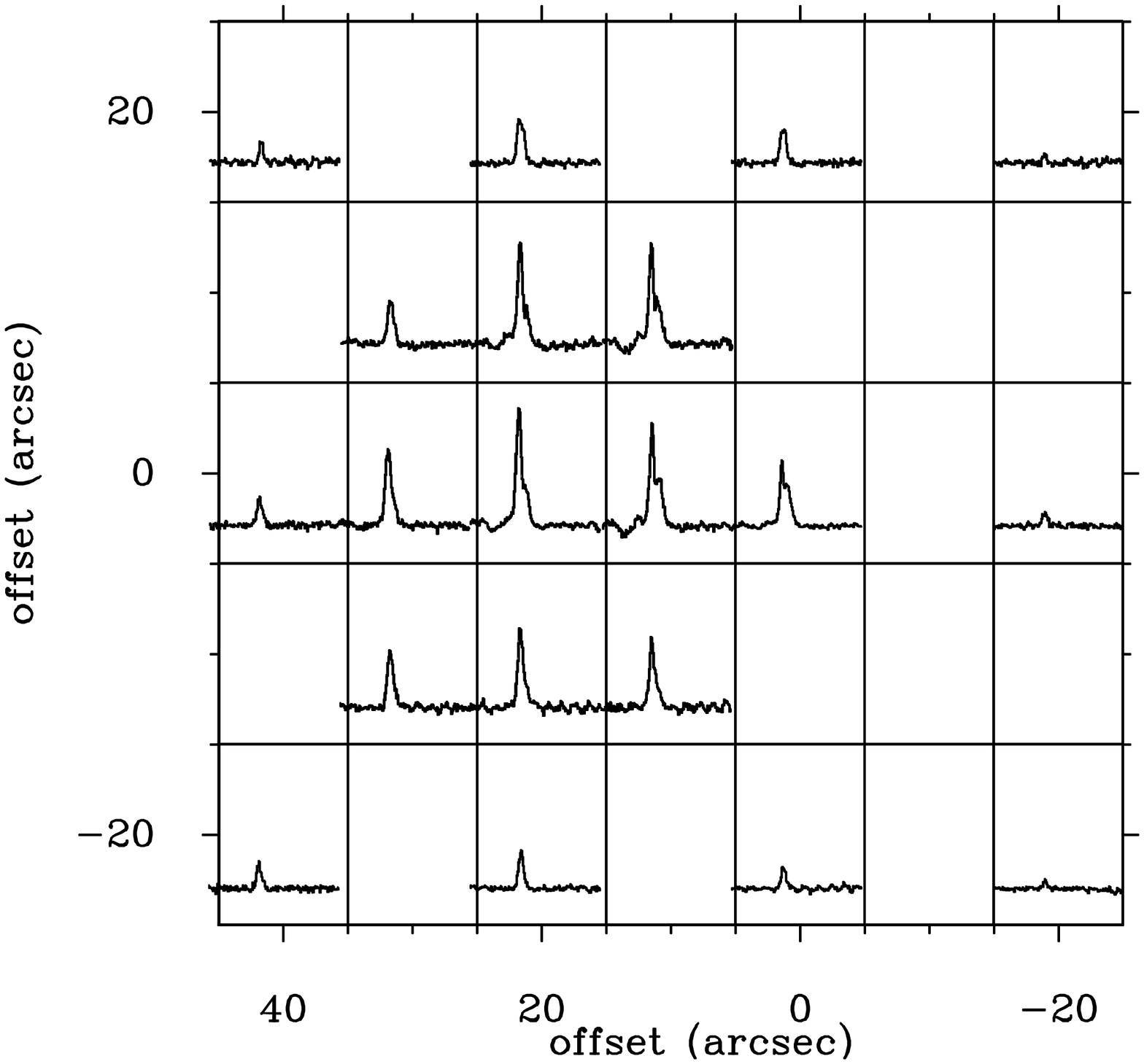}
\includegraphics[angle=-90,width=7.3cm]{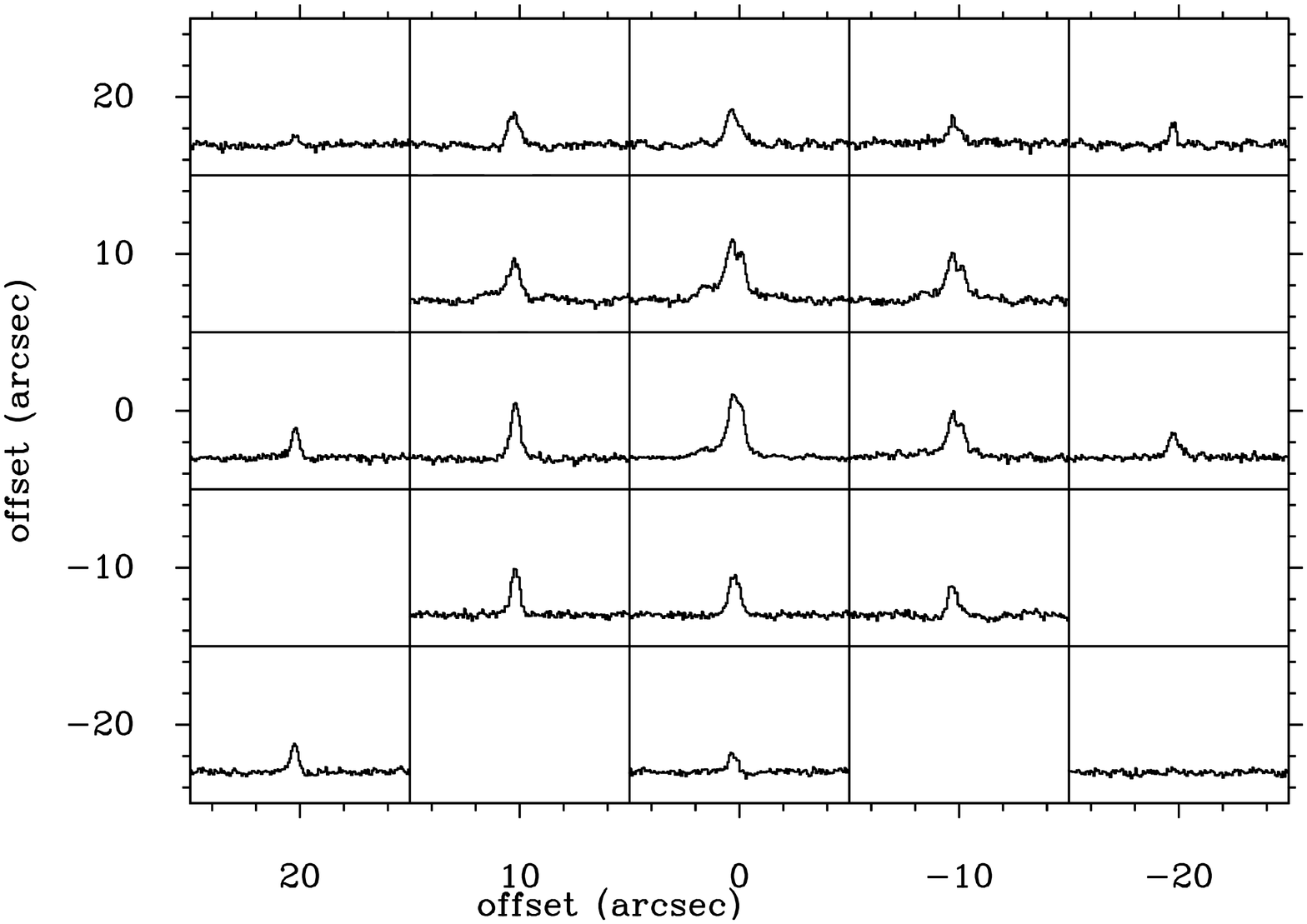}
\caption{\label{hcop} Maps of the {\hbox{${\rm HCO}^+$}}(4--3) line for 12326$-$6245 (top), 16060$-$5146
 (center) and 16065$-$5158 (bottom). The (0,0) positions correspond to
 the 1.2~mm peak positions of Fa{\'u}ndez (2004) listed in
 Tab. \ref{tab1}. At each offset position, the velocity and $T_{\rm mb}$ temperature scales go from -100 --
 20~{\hbox{${\rm km\, s}^{-1}$}}\/ and -5 -- 25~K, -150 -- 0~{\hbox{${\rm km\, s}^{-1}$}} and -5 -- 20~K and -120 -- 0~{\hbox{${\rm km\, s}^{-1}$}}
 and -5 -- 20~K respectively for 12326$-$6245, 16060$-$5146 and 16065$-$5158.}
\end{figure}

Because in the beginning LABOCA data were not available, the three
sources were mapped in {\hbox{${\rm HCO}^+$}}(4--3) at 357~GHz to
verify the peak position observed in the 1.2~mm continuum maps of
\citet{2004A&A...426...97F}. First, the regions were mapped with beam
spacing to locate the peaks, then the coverage was refined with fully sampled maps
around the peak positions (see Fig. \ref{hcop}). The peak positions in
12326$-$6245 and 16060$-$5146 are significantly offset from the 1.2~mm continuum peak positions as determined by
\citet{2004A&A...426...97F}. This is mostly likely owing to a pointing
problem in the 1.2~mm continuum data, because in 12326$-$6245 and
16060$-$5146, respectively, the peak positions derived by
\citet{2004A&A...426...97F} are offset by 6 and 15~\farcs~ from the
LABOCA and ATCA peak positions, which agree well for all three
sources.  The final pointing positions of the line observations can be seen in Table
\ref{tab2a}. The offset relative to the ATCA positions is less than a FWHM in all cases. When comparing results of different frequency setups later on, note that the offsets between the different positions are less than $1/3$ of the FWHM in all but one cases. The extent
 of the sources in {\hbox{${\rm HCO}^+$}}(4--3) is larger than the
 beam, which justifies the determination of beam-averaged column
 density later on.

\begin{table*}    
\caption{Coordinates of the line observations of
 the three sources.}
\label{tab2a}                     
\centering                                                                          
\begin{tabular}{lccl}                                                                                            
\hline                                                                         
\hline            
Source & R.A. & Dec. & Frequency Setup \\                 
& (J2000) & (J2000) &   \\
\hline                            
IRAS 12326$-$6245 & 12:35:33.7 &  -63:02:19.0 & {\hbox{${\rm H}_2{\rm CO}$}}(4--3)  \\
IRAS 12326$-$6245 & 12:35:34.4 &  -63:02:29.0 &   {\hbox{${\rm CH}_3{\rm CN}$}}(16--15),{\hbox{${\rm CH}_3{\rm OH}$}}(6--5) \\
IRAS 12326$-$6245 & 12:35:35.9 &  -63:02:29.0 & {\hbox{${\rm CH}_3{\rm OH}$}}(7--6) \\
\\
IRAS 16065$-$5158 & 16:10:20.0 & -52:06:03.3 & {\hbox{${\rm CH}_3{\rm CN}$}}(16--15),{\hbox{${\rm CH}_3{\rm OH}$}}(6--5)  \\
IRAS 16065$-$5158 & 16:10:20.6 & -52:06:08.3 &  {\hbox{${\rm H}_2{\rm CO}$}}(4--3), {\hbox{${\rm H}_2{\rm CO}$}}(6--5), {\hbox{${\rm CH}_3{\rm OH}$}}(7--6), high-$J$ CO  \\
\\
IRAS 16060$-$5146 & 16:09:53.6 & -51:55:07.0 & {\hbox{${\rm CH}_3{\rm CN}$}}(16--15),{\hbox{${\rm CH}_3{\rm OH}$}}(6--5) \\
IRAS 16060$-$5146 & 16:09:53.9 & -51:54:59.5 &  {\hbox{${\rm H}_2{\rm CO}$}}(4--3), {\hbox{${\rm H}_2{\rm CO}$}}(6--5), {\hbox{${\rm CH}_3{\rm OH}$}}(7--6), high-$J$ CO \\
\hline                                                                                               
\end{tabular} 

\end{table*}

\begin{figure} 
\includegraphics[angle=-90,width=6.8cm]{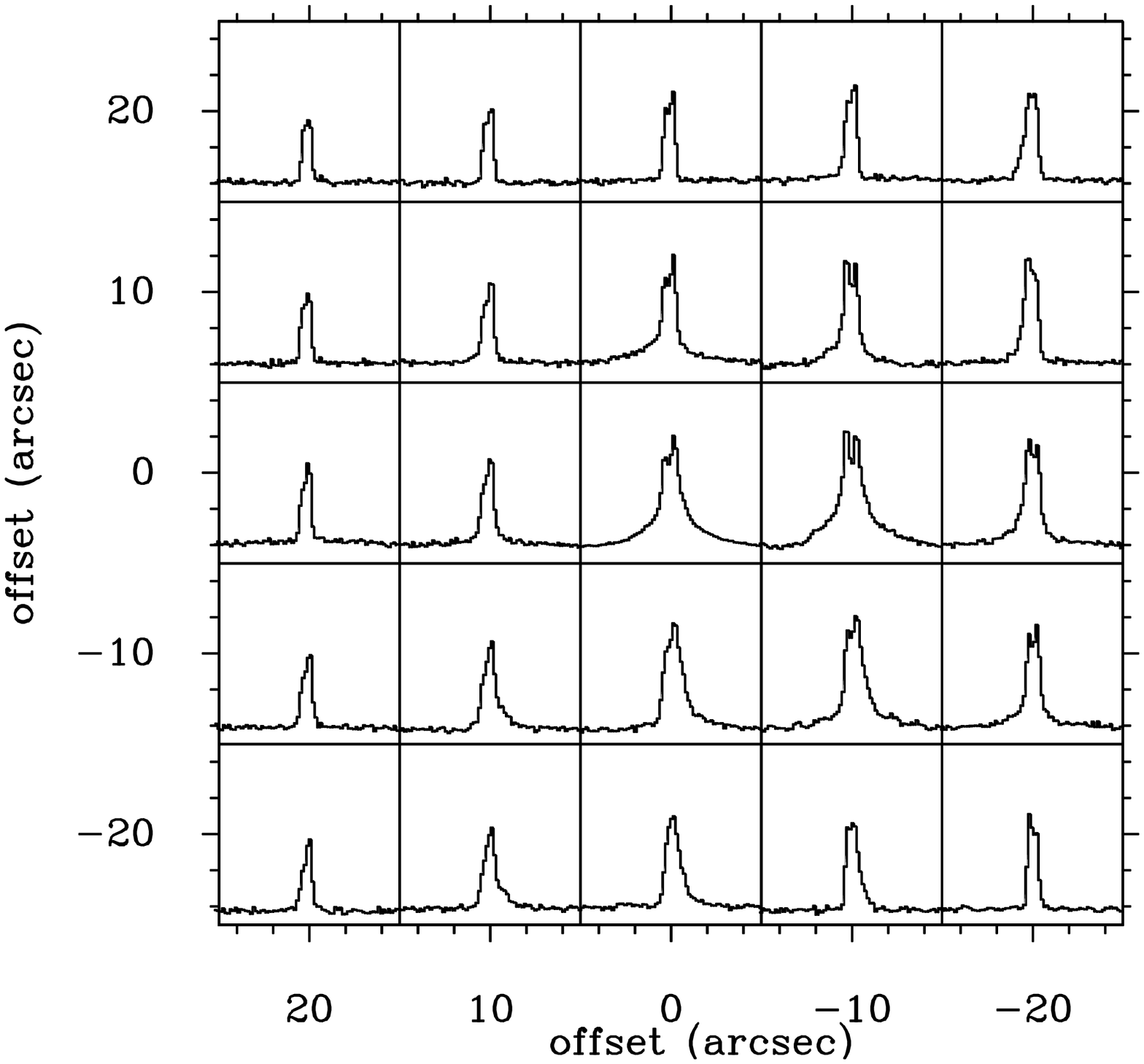}
\includegraphics[angle=-90,width=6.8cm]{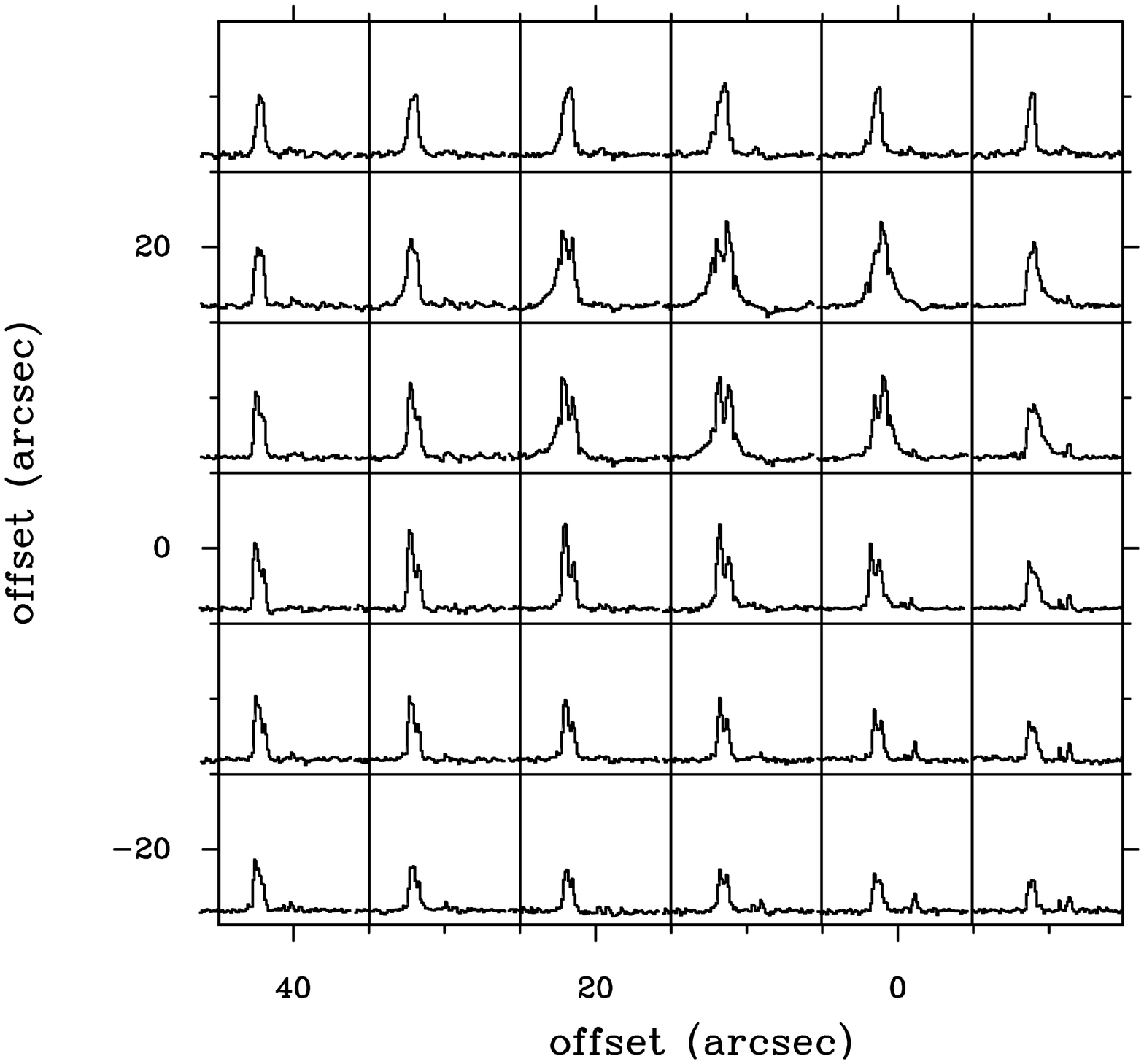}
\includegraphics[angle=-90,width=6.8cm]{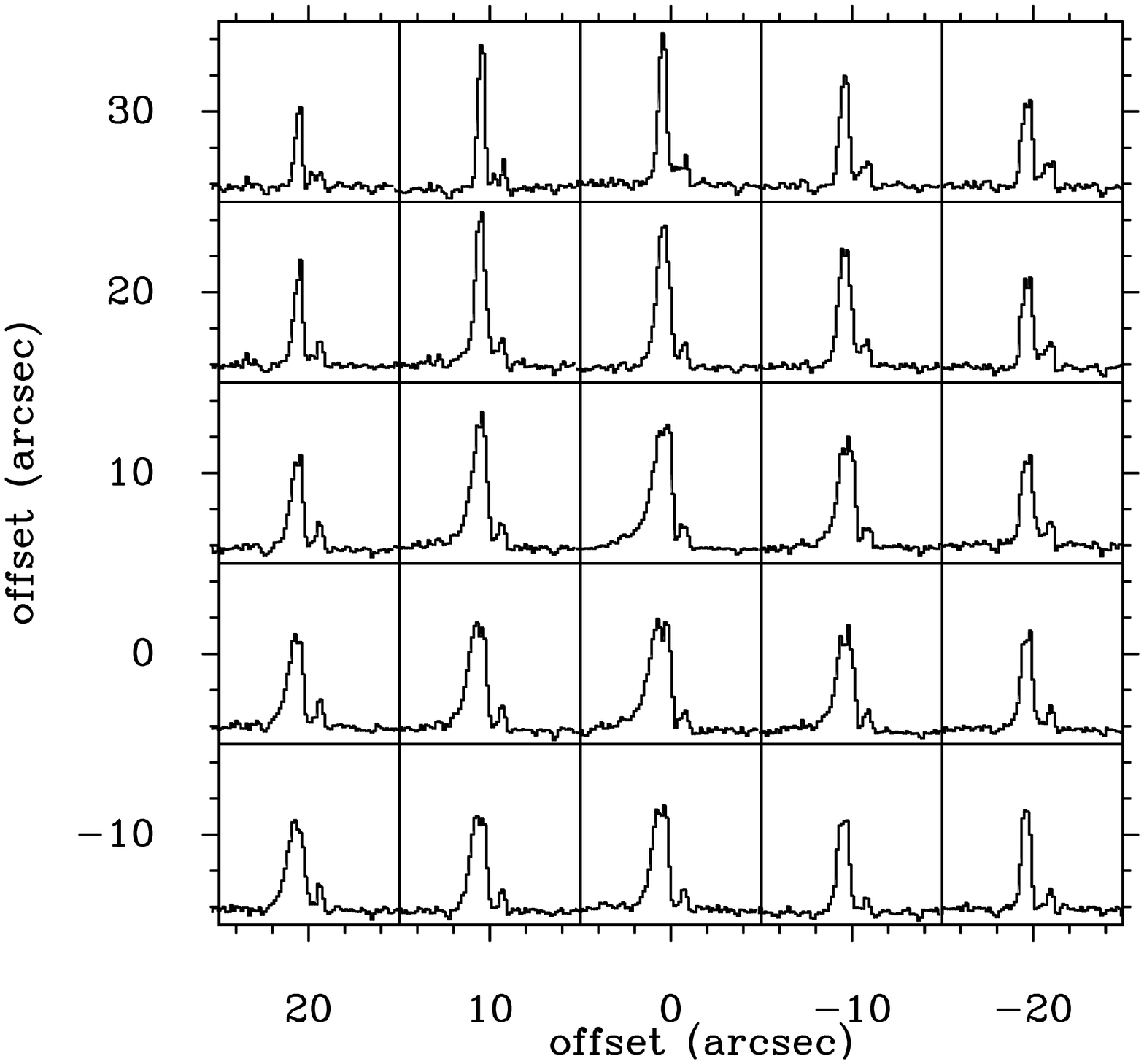}
\caption{\label{pilot:co} Maps of the {\hbox{\rm CO}}(3--2) line for 12326$-$6245 (top), 16060$-$5146
 (middle) and 16065$-$5158 (bottom). The (0,0) positions correspond to
 the 1.2~mm peak positions of Fa{\'u}ndez (2004) listed in
 Tab. \ref{tab1}. The velocity and $T_{\rm mb}$ temperature scales go from -100 --
 20~{\hbox{${\rm km\, s}^{-1}$}}\/ and -5 -- 60~K, -150 -- 0~{\hbox{${\rm km\, s}^{-1}$}} and -5 -- 50~K and -120 -- 0~{\hbox{${\rm km\, s}^{-1}$}}
 and -5 -- 60~K respectively for 12326$-$6245, 16060$-$5146 and 16065$-$5158.}
\end{figure}

\begin{figure} 
\includegraphics[angle=-90,width=7.1cm]{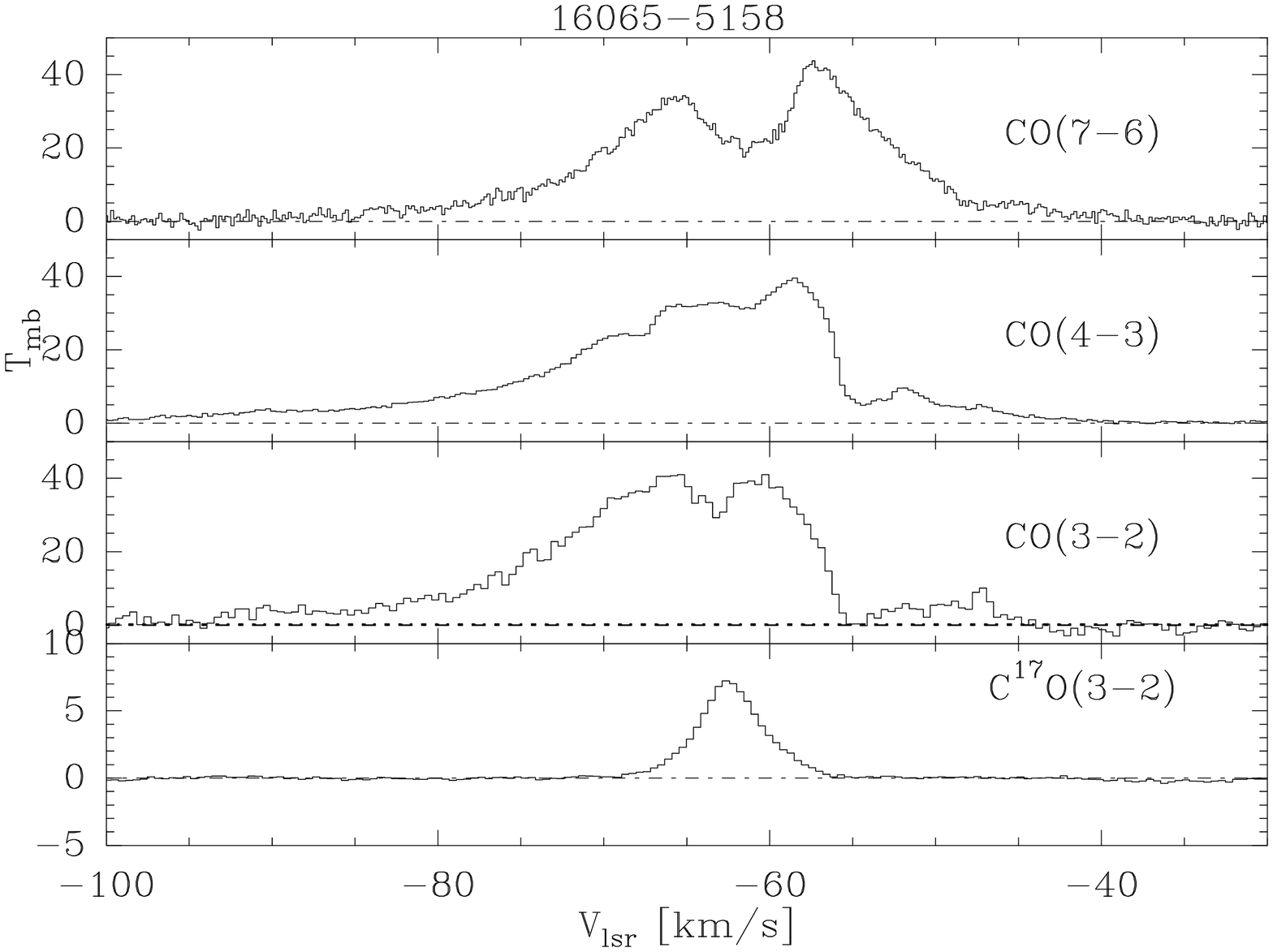}
\includegraphics[angle=-90,width=7.1cm]{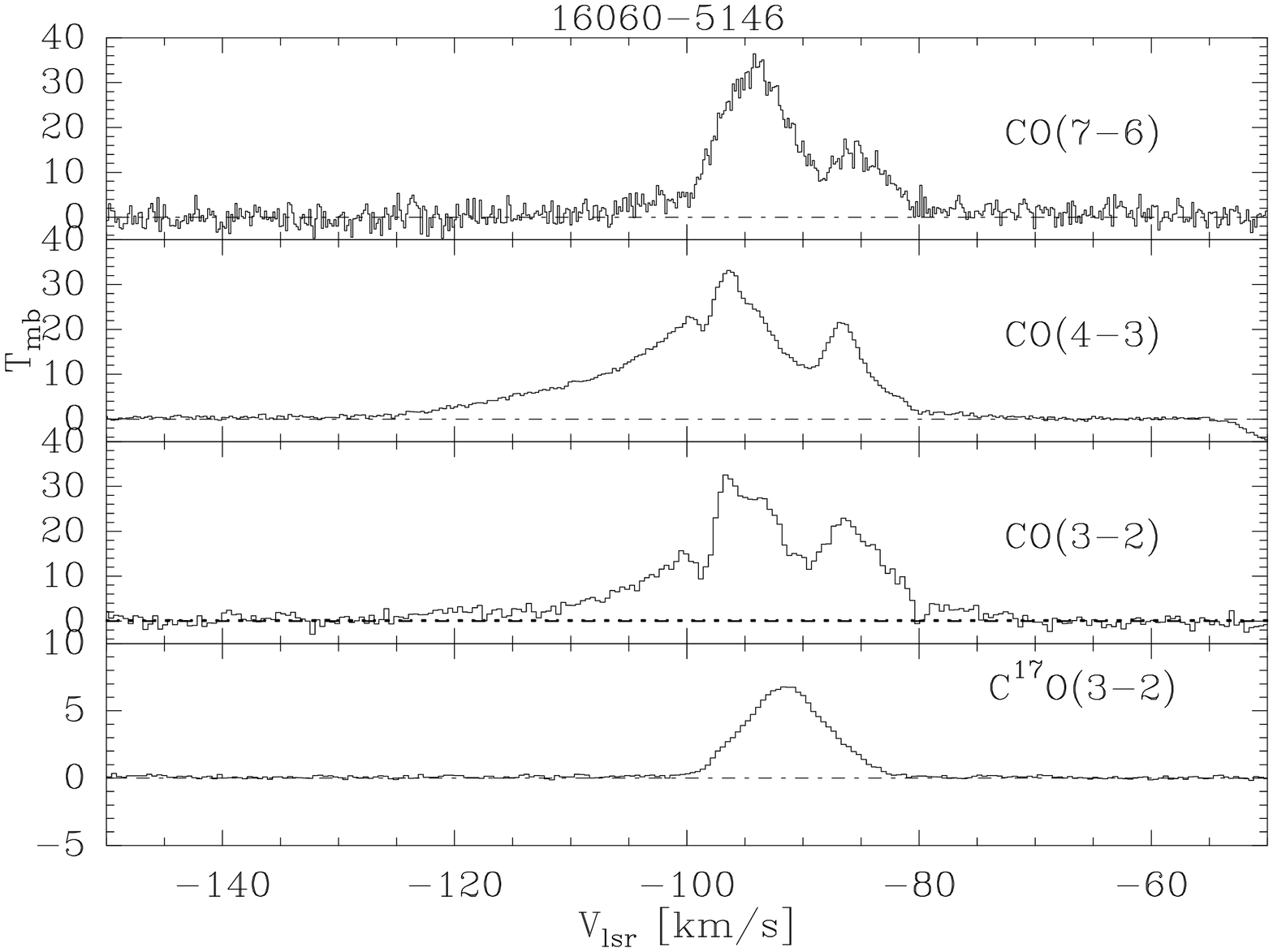}
\includegraphics[angle=-90,width=7.1cm]{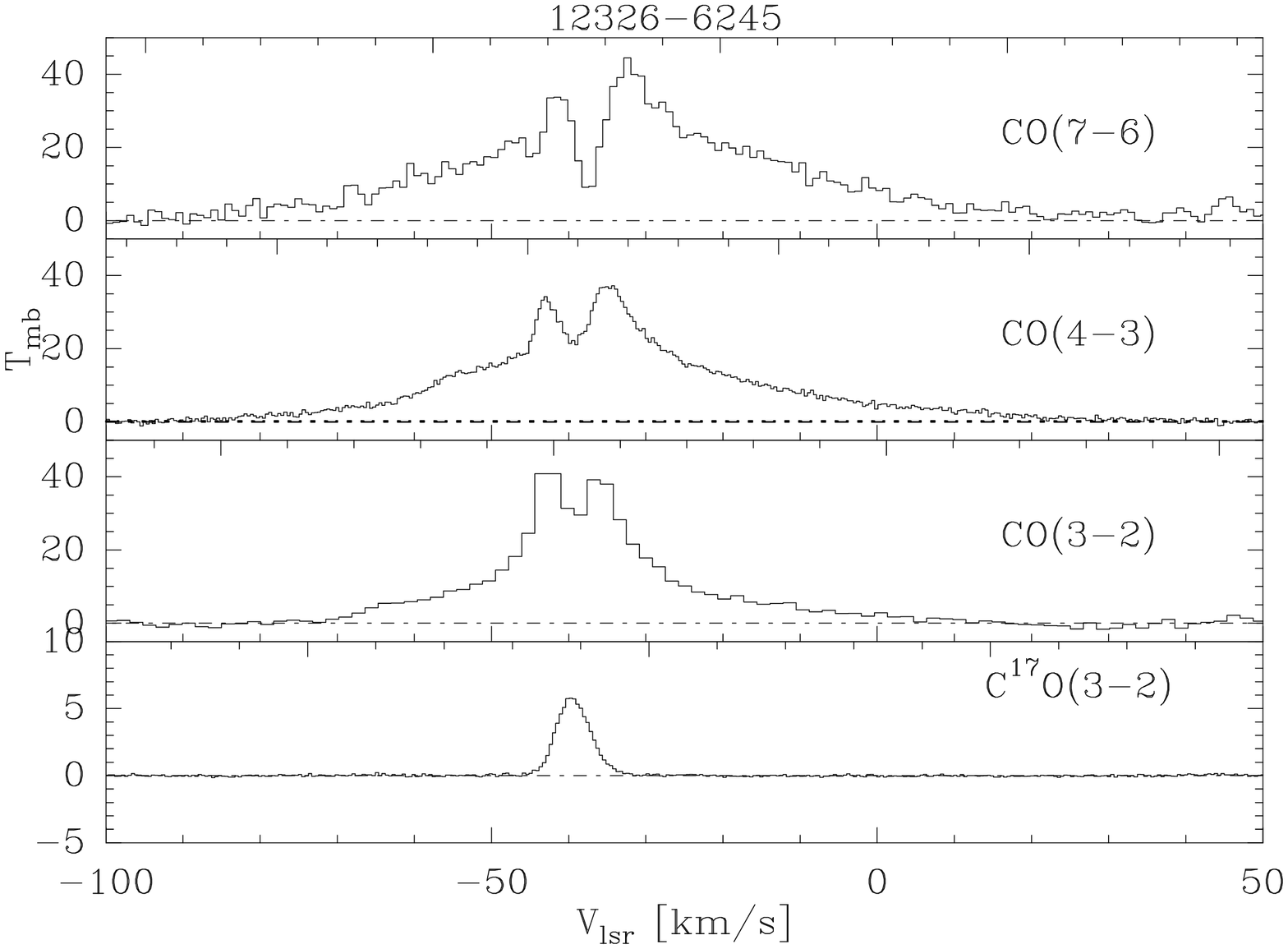}
\caption{\label{obs:fig2a} {\hbox{${\rm C}^{17}{\rm O}$}}(3--2), {\hbox{\rm CO}}(3--2), {\hbox{\rm CO}}(4--3) and {\hbox{\rm CO}}(7--6)
 transitions. While {\hbox{${\rm C}^{17}{\rm O}$}}(3--2) is observed on the peak positions in 16065$-$5158 (top),
 16060$-$5146 (middle) and 12326$-$6245 (bottom), the spectra of the remaining
 transitions were produced by averaging all the spectra taken at the map positions. The dashed line marks
 the zero level.}
\end{figure}

In order to determine the total column density and the properties of
the envelopes, the sources were also mapped in {\hbox{\rm CO}}(3--2)
(Fig. \ref{pilot:co}) and observed in {\hbox{\rm CO}}(4--3),
{\hbox{\rm CO}}(7--6) and {\hbox{${\rm C}^{17}{\rm O}$}}(3--2) on the
peak positions (see Fig.  \ref{obs:fig2a}).  Both the {\hbox{\rm CO}}\/ lines and the {\hbox{${\rm
     HCO}^+$}}(4--3) line show clear signs of outflow activity in the
wings of all sources and self-absorption at the velocity of
{\hbox{${\rm C}^{17}{\rm O}$}. All three sources have very
 high-velocity outflows with a width of the blue wing, $\Delta v_{\rm blue}$, of $30${\hbox{${\rm km\, s}^{-1}$}}\/  and a red wing width, 
 $\Delta v_{\rm red}$, of $20${\hbox{${\rm km\, s}^{-1}$}}\/ in both 16060$-$5146 and 16065$-$5158. In
 12326$-$6245, which was already described as one of the most massive
 and energetic outflows by \citet{2000A&A...364..613H}, we
   found $\Delta v_{\rm blue}=50${\hbox{${\rm km\, s}^{-1}$}}\/ and $\Delta v_{\rm
   red}=60${\hbox{${\rm km\, s}^{-1}$}}. 16060$-$5146 and 16065$-$5158 show an
asymmetric line profile with a stronger blue shoulder at the peak position of
the {\hbox{${\rm HCO}^+$}}(4--3) emission, which is a sign of motion in
the core. At the resolution of the current study, it is not possible to
distinguish between infall and rotation as possible origin of the large scale
motions.
Note that in both 12326--6245 and 16065-5158, the
 high-$J$ CO the intensity in the outflow wings is stronger than in the
 lower-$J$ lines and one sees stronger
 self-absorption. In 16060--5146, however, the outflow wings are less
 strong in the CO(7--6) line, indication of a lower CO abundance in the
 high-velocity gas.}

\begin{figure} 
\includegraphics[angle=0,width=8.2cm]{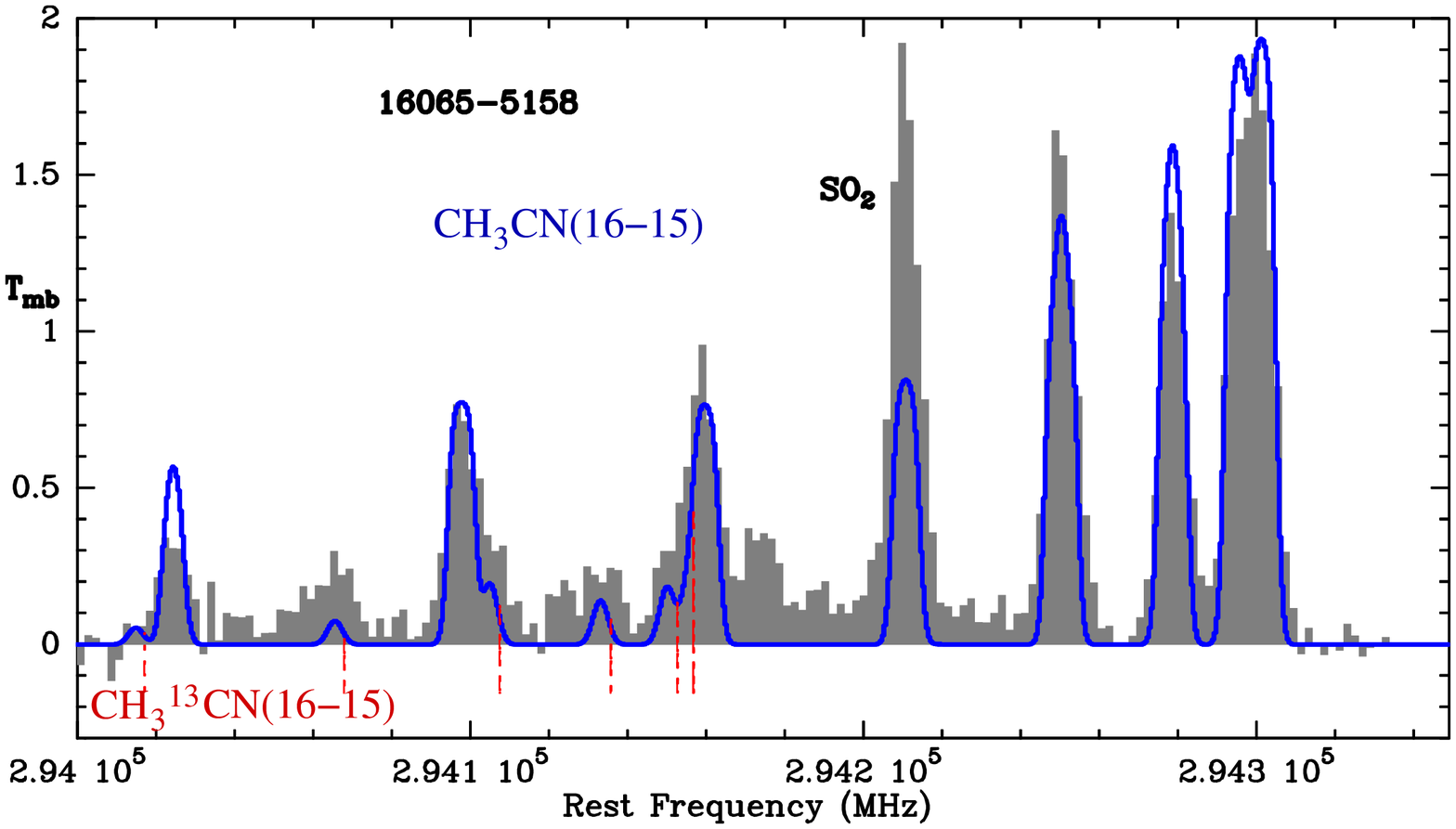}
\includegraphics[angle=0,width=8.2cm]{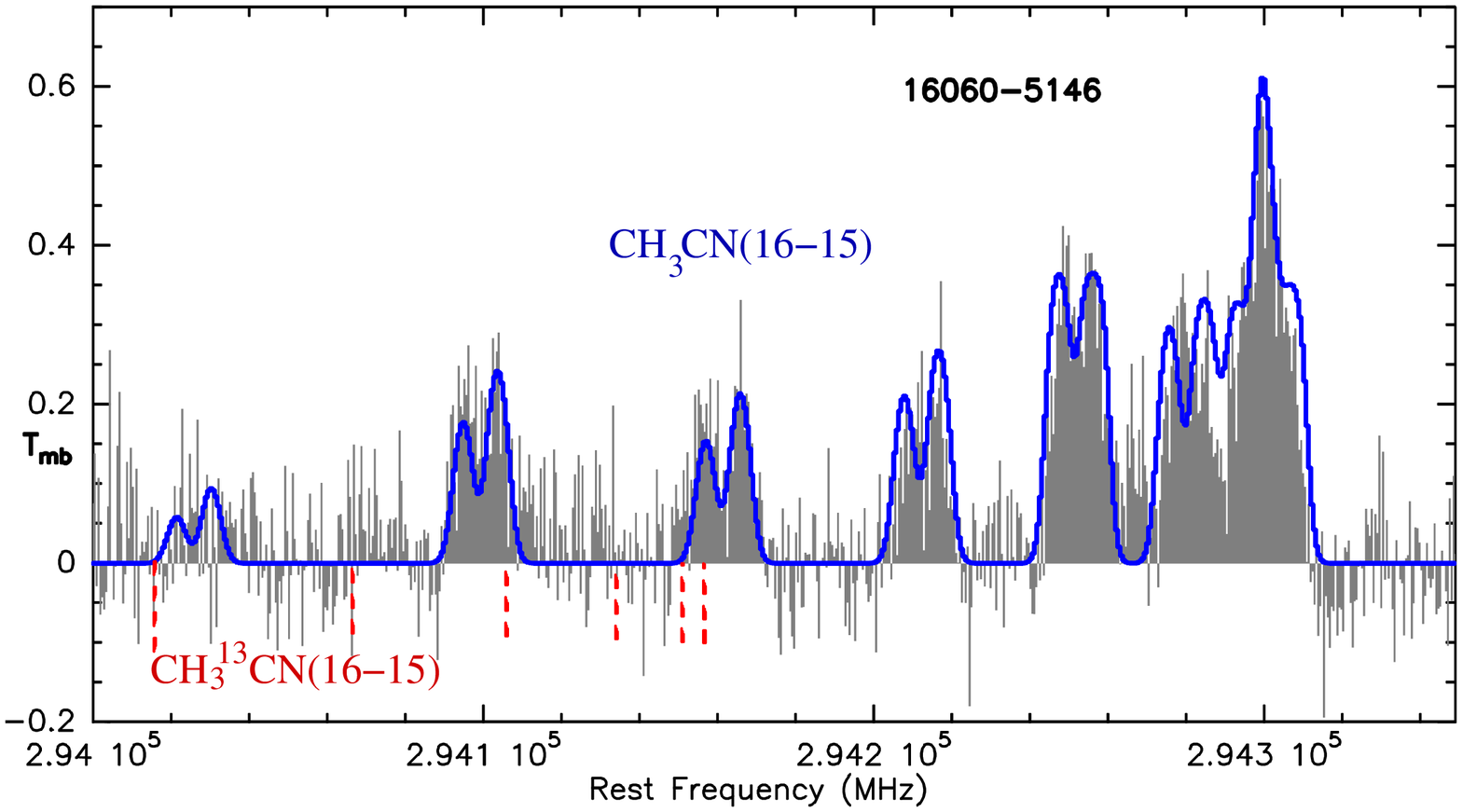} 
\includegraphics[angle=0,width=8.2cm]{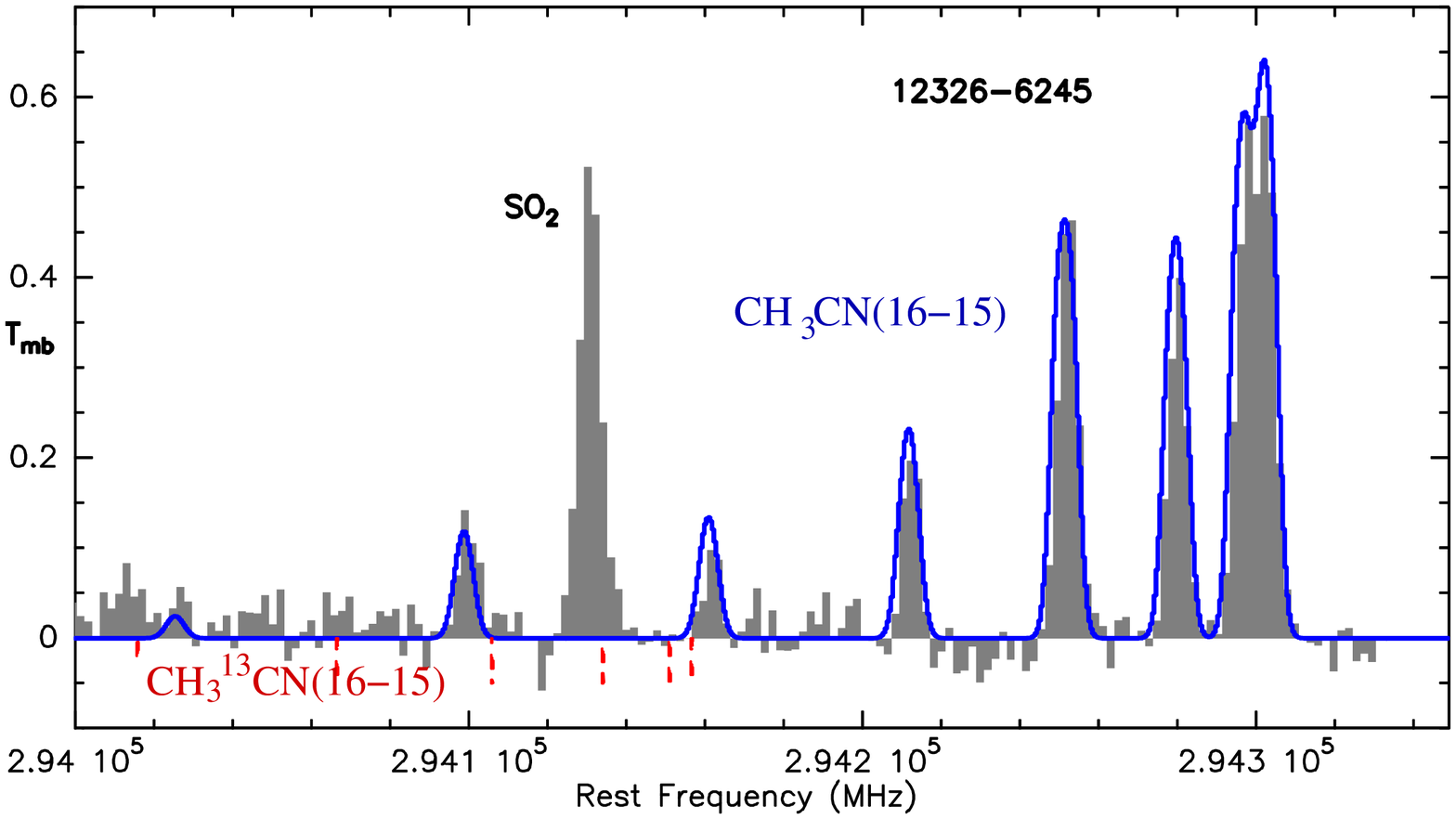} 
\caption{\label{ch3cn_plots} {\hbox{${\rm CH}_3{\rm CN}$}}(16--15) data at 294~GHz. We show in blue the synthetic
 model spectrum (see Tables \ref{tab_mol_16065} -- \ref{tab_mol_16060}). Note that the spectra for the three sources have been
 observed with different frequency setups, therefore the SO$_2$ line from the
image sidebands appears at different frequencies. The rest frequencies of the
{\hbox{${\rm CH}_3^{13}{\rm CN}$}}\/ lines are marked in red. } 
\end{figure}

The three sources were observed in 
{\hbox{${\rm CH}_3{\rm OH}$}}(6--5), {\hbox{${\rm CH}_3{\rm OH}$}}(7--6), {\hbox{${\rm CH}_3{\rm CN}$}}(16--15),
{\hbox{${\rm H}_2{\rm CO}$}}(4--3), and {\hbox{${\rm H}_2{\rm CO}$}}(6--5).   
As will be further discussed in the next section, all three sources exhibit a
rich spectrum of molecular lines. In 12326$-$6245 and 16065$-$5158, we found 18 different
species, while in 16060$-$5146, we found 16 species. As is typical
for hot core sources, high-excitation lines up to energies of 346~K above the ground
state and lines from vibrationally excited states were observed. Especially the setup
around 338~GHz is very rich in lines, with detections of up to 40 lines per GHz.
The lines were identified with the programme XCLASS
\citep{1999pcim.conf..330S,2005ApJS..156..127C} and the rest frequencies
provided by the Cologne database for Molecular Spectroscopy (CDMS)\footnote{http://www.cdms.de} \citep{2001A&A...370L..49M,2005IAUS..235P..62M} and JPL
molecular spectroscopy database\footnote{http://spec.jpl.nasa.gov/}
\citep{1998JQSRT..60..883P}. Because
the line identification went hand in hand with the LTE modeling of the lines,
it is described in more detail in Sec. \ref{lte:modeling}. To give an example
of the chemical complexity of the sources, the identified
lines that were included in the XCLASS modeling are listed for the source
16065-5158 in the online material. Note that it was not possible
to perform individual Gaussian fits to all the
{\hbox{${\rm CH}_3{\rm OH}$}} and {\hbox{${\rm CH}_3{\rm CN}$}} lines included
in this list, because some of them were heavily blended with lines from other species in
all three sources. 

For {\hbox{${\rm CH}_3{\rm OH}$}}, {\hbox{${\rm CH}_3{\rm CN}$}}, and 
{\hbox{${\rm H}_2{\rm CO}$}}, the parameters of the lines (determined from Gaussian fits) can be found in Tables
\ref{ch3oh_16060_tab} to \ref{h2co_16060_tab} in the online material. 
The average line widths of the {\hbox{${\rm CH}_3{\rm OH}$}\/ lines in the three sources are 6~{\hbox{${\rm km\, s}^{-1}$}}\/, 10~{\hbox{${\rm km\, s}^{-1}$}}\/ and
7~{\hbox{${\rm km\, s}^{-1}$}}\/ and of the {\hbox{${\rm CH}_3{\rm CN}$}}(16-15) lines 6~{\hbox{${\rm km\, s}^{-1}$}}\/, 11~{\hbox{${\rm km\, s}^{-1}$}}\/ and 8~{\hbox{${\rm km\, s}^{-1}$}}\/ for 12326$-$6245, 16060$-$5146 and 16065$-$5158 respectively.} 

16060$-$5146 has a slightly broader profile than the
other two sources.  In both {\hbox{${\rm CH}_3{\rm CN}$}}(16--15) and the high-excitation CH$_3$OH(v$_t$=1) lines  in 16060$-$5146, one can see a double-peaked profile
(Fig.~\ref{ch3cn_plots} and Fig.~\ref{fig:molem2b} ),
which might suggest that the larger line widths seen in this source in the other
species are caused by the superposition of two velocity components.  

In 16065$-$5158, a similar double-peaked profile can be found for the highly (torsionally)
excited CH$_3$OH(v$_t$=1) lines. 16065$-$5158 also displays strong
{\hbox{${\rm CH}_3{\rm CN}$}} lines from the v$_8=1$ bending mode, which is not
seen toward the other two sources. Looking at the distribution of peak velocities versus
energy of the lines, one can see no trend in all three sources and
molecules. The peak velocities of 16060$-$5146 are scattered over a broader field than those of
the other two sources.

\onllongtab{4}{

\begin{longtable}   {cccccc}                                                                               
\caption{Line parameters for CH$_3$OH in 16060-5146. }
\label{ch3oh_16060_tab} \\

\hline                                                                                               
\hline                                                                                               
\multicolumn{1}{c}{Frequency}  &  \multicolumn{1}{c}{Transition}  &  \multicolumn{1}{c}{v} & \multicolumn{1}{c}{$\delta$ (v)} & \multicolumn{1}{c}{$\int T_{mB}\Delta v$} & \multicolumn{1}{c}{$T_{mB}$} \\         
(MHz)  &            &   (km/s) & ($\times 10^{-1}$km/s)               &  (K km/s) & (K)  \\                
\hline                                                                          
\endfirsthead 
\caption{continued.}\\                 

\hline
\hline 

\multicolumn{6}{c}{{\bfseries \tablename\ \thetable{} -- continued from
    previous page}} \\
\hline
Frequency  &  Transition  &   v & $\delta$ (v) & $\int T_{mB}\Delta v$ & $T_{mB}$ \\         
(MHz)  &            &   (km/s) & ($\times 10^{-1}$km/s)               &  (K km/s) & (K)  \\     
\hline                                                                                               
\endhead                                                                                             
\hline                                                                                               
\multicolumn{6}{r}{{Continued on next page. v and $\delta$ (v) are the peak velocity and its error as derived from the Gauss fit.}} \\\hline  
\endfoot                                                                                             
\endlastfoot

289939.5  &   6 $_{0,6 }E$ --   5 $_{0,5 }E$        &  -90.70           &  0.9              &  17.1             &  1.5             \\   
290069.8  &   6 $_{1,6 }E$ --   5 $_{1,5 }E$        &  -90.96           &  0.4              &  21.0             &  2.1             \\   
\\
290110.7  &   6 $_{0,6 }A^+$ --   5 $_{0,5 }A^+$    &  -91.09           &  1.0              &  22.1             &  2.3             \\   
290117.8  &   6 $_{5,1 }E$ --   5 $_{5,0 }E$        &  & & & \\   
\\
290183.2  &   6 $_{4,2 }E$ --   5 $_{4,1 }E$        &  -96.78           &  2.5              &  8.1              &  0.7             \\   
290184.7  &    6 $_{2,5 }A^-$ --   5 $_{2,4 }A^-$   & & & & \\
\end{longtable} }                               

\onllongtab{5}{
\begin{longtable}{cccccc}                                                                        
\caption{Line parameters for CH$_3$OH in 12326$-$6245.} \\

\hline                                                                                               
\hline                                                                                               
\multicolumn{1}{c}{Frequency}  &  \multicolumn{1}{c}{Transition}  &  \multicolumn{1}{c}{v} & \multicolumn{1}{c}{$\delta$ (v)} & \multicolumn{1}{c}{$\int T_{mB}\Delta v$} & \multicolumn{1}{c}{$T_{mB}$} \\         
(MHz)  &            &   (km/s) & ($\times 10^{-1}$km/s)               &  (K km/s) & (K)  \\                
\hline                                                                          
\endfirsthead 
\caption{continued.}\\                 

\hline
\hline 

\multicolumn{6}{c}{{\bfseries \tablename\ \thetable{} -- continued from
    previous page}} \\
\hline
Frequency  &  Transition  &   v & $\delta$ (v) & $\int T_{mB}\Delta v$ & $T_{mB}$ \\         
(MHz)  &            &   (km/s) & ($\times 10^{-1}$km/s)               &  (K km/s) & (K)  \\ 
\hline                                                                                               
\endhead                                                                                             
\hline                                                                                               
\multicolumn{6}{r}{{Continued on next page. v and $\delta$ (v) are the peak velocity and its error as derived from the Gauss fit.}} \\\hline  

\endfoot                                                                                             
\endlastfoot              

289939.5  &   6 $_{0,6 }E$ --   5 $_{0,5 }E$        &  -39.11           &  0.4              &  12.0             &  1.9             \\   
290069.8  &   6 $_{1,6 }E$ --   5 $_{1,5 }E$        &  -39.21           &  0.2              &  17.1             &  2.7             \\   
\\
290110.7  &   6 $_{0,6 }A^+$ --   5 $_{0,5 }A^+$    &  -39.19           &  0.1              &  18.0             &  2.9             \\   
290117.8  &   6 $_{5,1 }E$ --   5 $_{5,0 }E$        &   & & &     \\   
\\
290138.9  &   6 $_{5,2 }E$ --   5 $_{5,1 }E$        &  -39.8           &  36.3             &  3.4            &  0.6            \\   
\\
\hline

\hline                                                                                                                                                                 
\end{longtable}}

\onllongtab{6}{
\begin{longtable} {cccccc}                                                         
\caption{Line parameters for CH$_3$OH in 16065-5158.}   \\                                                                                                    
\hline                                                                                               
\hline                                                                                   

\multicolumn{1}{c}{Frequency}  &  \multicolumn{1}{c}{Transition}  &  \multicolumn{1}{c}{v} & \multicolumn{1}{c}{$\delta$ (v)} & \multicolumn{1}{c}{$\int T_{mB}\Delta v$} & \multicolumn{1}{c}{$T_{mB}$} \\         
(MHz)  &            &   (km/s) & ($\times 10^{-1}$km/s)               &  (K km/s) & (K)  \\ 
\hline                                                                          
\endfirsthead 
\caption{continued.}\\                 

\hline
\hline 

\multicolumn{6}{c}{{\bfseries \tablename\ \thetable{} -- continued from
    previous page}} \\
\hline
Frequency  &  Transition  &   v & $\delta$ (v) & $\int T_{mB}\Delta v$ & $T_{mB}$ \\         
(MHz)  &            &   (km/s) & ($\times 10^{-1}$km/s)               &  (K km/s) & (K)  \\ 
\hline                                                                                               
\endhead                                                                                             
\hline                                                                                               
\multicolumn{6}{r}{{Continued on next page.  v and $\delta$ (v) are the peak velocity and its error as derived from the Gauss fit.}} \\\hline  

\endfoot                                                                                             
\endlastfoot

290161.3  &   6 $_{4,2 }A^+$ --   5 $_{4,1 }A^+$    &  -61.66           &  0.6              &  11.4             &  1.4             \\   
\\
290183.2       &    6 $_{4,2 }E$ --   5 $_{4,1 }E$       &  -62.01	     &  0.4		 &  14.1	     &  1.9	        \\
290184.7  &   6 $_{2,5 }A^-$ --   5 $_{2,4 }A^-$    & & & &  \\   
\\
290209.7  &   6 $_{3,4 }E$ --   5 $_{3,3 }E$        &  -62.13           &  0.3              &  15.1             &  2.1             \\   
290213.2       &    6 $_{3,3 }E$ --   5 $_{3,2 }E$       &  & & & \\
\\
\hline                                                                        

\end{longtable}  }

\onltab{7}{                                                                                           
\begin{table*}                                                                             
\begin{center}                                                                                       
\caption{Line parameters for CH$_3$CN in 16065-5158. }                                                  
\begin{tabular}{cccccc}                                                                              
\hline                                                                                               
\hline                                                                                               
Frequency  &  Transition  &  Velocity v & $\delta$ (v) & $\int T_{mB}\Delta v$ & $T_{mB}$ \\         
(MHz)  &            &   (km/s) & ($\times 10^{-1}$km/s)               &  (K km/s) & (K)  \\ 
\hline                                                                                               
294025.5  &  16$_{7}$--15$_{7}$    &  -61.06           &  5.7              &  3.7              &  0.3             \\   
294098.9  &  16$_{6}$--15$_{6}$    &  -62.40           &  2.9              &  7.9              &  0.6             \\   
294161.0  &  16$_{5}$--15$_{5}$    &  -59.56           &  2.5              &  9.0              &  0.7             \\   
294251.5  &  16$_{3}$--15$_{3}$    &  -61.20           &  1.0              &  11.9             &  1.3             \\   
294279.8  &  16$_{2}$--15$_{2}$    &  -61.27           &  3.7              &  9.5              &  1.1             \\   
294302.4  &  16$_{0}$--15$_{0}$    &  -58.46           &  0.6              &  20.1             &  1.6             \\
\hline

\end{tabular}                                                                                        
\tablefoot{
v and $\delta$ (v) are the peak velocity and its error as derived from the Gauss fit.}

\end{center}                                                                                         
\end{table*} }

\onltab{8}{                                                                                        
\begin{table*}                                                                                 
\begin{center}                                                                                       
\caption{Line parameters for CH$_3$CN in 12326-6245. }                                                  
\begin{tabular}{cccccc}                                                                              
\hline                                                                                               
\hline                                                                                               
Frequency  &  Transition  &  Velocity v & $\delta$ (v) & $\int T_{mB}\Delta v$ & $T_{mB}$ \\         
(MHz)  &            &   (km/s) & ($\times 10^{-1}$km/s)               &  (K km/s) & (K)  \\ 
\hline                                                                                               
294098.9  &  16$_{6}$--15$_{6}$    &  -40.11           &  4.0              &  1.4              &  0.2             \\   
294161.0  &  16$_{5}$--15$_{5}$    &  -40.22           &  9.4              &  1.1              &  0.2             \\   
294251.5  &  16$_{3}$--15$_{3}$    &  -39.81           &  1.1              &  2.3              &  0.4             \\   
294279.8  &  16$_{2}$--15$_{2}$    &  -40.04           &  0.6              &  4.3              &  0.7             \\   
294302.4  &  16$_{0}$--15$_{0}$    &  -39.88           &  3.3              &  3.6              &  0.6             \\   
349212.3  &  19$_{6}$--18$_{6}$    &  -41.06           &  0.6              &  10.4             &  1.4             \\
\hline

\end{tabular}  
\tablefoot{
v and $\delta$ (v) are the peak velocity and its error as derived from the Gauss fit.}
                                                                                      
\end{center}                                                                                         
\end{table*} }

\onltab{9}{
\begin{table*}                                                                                  
\begin{center}                                                                                       
\caption{Line parameters for CH$_3$CN in 16060-5146.}                                                  
\begin{tabular}{cccccc}                                                                              
\hline                                                                                               
\hline                                                                                               
Frequency  &  Transition  &  Velocity v & $\delta$ (v) & $\int T_{mB}\Delta v$ & $T_{mB}$ \\         
(MHz)  &            &   (km/s) & ($\times 10^{-1}$km/s)               &  (K km/s) & (K)  \\ 
\hline                                                                                               
294098.9         &  16$_{6}$--15$_{6}$    &  -95.02           &  10.1             &  2.9              &  0.2             \\   
                &                   &  -86.50           &  6.4              &  0.8              &  0.1             \\   
294161.0         &  16$_{5}$--15$_{5}$    &  -96.57           &  3.1              &  1.3              &  0.2             \\   
                &                   &  -86.67           &  5.1              &  1.8              &  0.2             \\   
294211.9         &  16$_{4}$--15$_{4}$    &  -96.28           &  6.3              &  0.9              &  0.2             \\   
                &                   &  -89.23           &  13.8             &  1.5              &  0.2             \\
\hline

\end{tabular}                                                                   \label{ch3cn_16060_tab}                
\tablefoot{
v and $\delta$ (v) are the peak velocity and its error as derived from the Gauss fit.}

\end{center}                                                                                         
\end{table*} }

\onltab{10}{                                                                                           
\begin{table*}                                                                                
\begin{center}                                                                                       
\caption{Line parameters for H$_2$CO in 16065-5158. }                                                  
\begin{tabular}{cccccc}                                                                              
\hline                                                                                               
\hline                                                                                               
Frequency  &  Transition  &  Velocity v & $\delta$ (v) & $\int T_{mB}\Delta v$ & $T_{mB}$ \\         
(MHz)  &            &   (km/s) & ($\times 10^{-1}$km/s)               &  (K km/s) & (K)  \\ 
\hline                                                                                               
290623.4  &  4$_{0,4}$ --  3$_{0,3}$  &  -62.59           &  0.9              &  27.3             &  4.3             \\   
291237.8  &  4$_{2,3}$ --  3$_{2,2}$  &  -61.96           &  1.3              &  11.7             &  1.3             \\   
\\
291380.5       &   4$_{3,2}$ --  3$_{3,1}$ & -57.40	      &  5.0		  &  10.2	      &  1.3	      \\
291384.3  &  4$_{3,1}$ --  3$_{3,0}$  &  & & &   \\   
\\
436586.5  &  6$_{2,5}$ --  5$_{2,4}$  &  -61.09           &  2.0              &  23.3             &  2.4             \\   
\\
436751.0  &  6$_{5,1}$ --  5$_{5,0}$  &  -60.16           &  3.2              &  13.1             &  1.2             \\   
436751.0       &   6$_{5,2}$ --  5$_{5,1}$ & & & &\\ 
\\
436957.3  &  6$_{4,3}$ --  5$_{4,2}$  &  -61.92           &  3.6              &  22.1             &  1.8             \\   
436957.5       &   6$_{4,2}$ --  5$_{4,1}$ & & & & \\ 
\\
437199.6  &  6$_{3,4}$ --  5$_{3,3}$  &  -61.92           &  3.5              &  32.6             &  2.8             \\   
437236.0  &  6$_{3,3}$ --  5$_{3,2}$  &  -61.51           &  4.0              &  36.4             &  3.1             \\   
\hline

\end{tabular}                                                                                        
\tablefoot{
v and $\delta$ (v) are the peak velocity and its error as derived from the Gauss fit.}

\end{center}                                                                                         
\end{table*} }

\onltab{11}{                                                                                      
\begin{table*}                                                                                 
\begin{center}                                                                                       
\caption{Line parameters for H$_2$CO in 12326-6245. }                                                  
 \begin{tabular}{cccccc}                                                                              
\hline                                                                                               
\hline                                                                                               
Frequency  &  Transition  &  Velocity v & $\delta$ (v) & $\int T_{mB}\Delta v$ & $T_{mB}$ \\         
(MHz)  &            &   (km/s) & ($\times 10^{-1}$km/s)               &  (K km/s) & (K)  \\ 
\hline                                                                                               
290623.4  &  4$_{0,4}$ --  3$_{0,3}$  &  -39.56           &  0.2              &  23.0             &  3.5             \\   
291237.8  &  4$_{2,3}$ --  3$_{2,2}$  &  -39.80           &  0.3              &  12.6             &  1.9             \\   
\\
291380.5       &   4$_{3,2}$ --  3$_{3,1}$ & -37.89	      &  2.3		  &  14.3	      &  1.5	    \\
291384.3  &  4$_{3,1}$ --  3$_{3,0}$  &   & & &    \\   
\hline                                                                      
\end{tabular}                                                                                        
\tablefoot{
v and $\delta$ (v) are the peak velocity and its error as derived from the Gauss fit.}

\end{center}                                                                                         
\end{table*} }

\onltab{12}{
\begin{table*}                                                                                
\begin{center}                                                                                       
\caption{Line parameters for H$_2$CO in 16060-5146.}                                                  
\begin{tabular}{cccccc}                                                                              
\hline                                                                                               
\hline                                                                                               
Frequency  &  Transition  &  Velocity v & $\delta$ (v) & $\int T_{mB}\Delta v$ & $T_{mB}$ \\         
(MHz)  &            &   (km/s) & ($\times 10^{-1}$km/s)               &  (K km/s) & (K)  \\ 
\hline                                                                                               
290623.4  &  4$_{0,4}$ --  3$_{0,3}$  &  -92.07           &  0.4              &  26.7             &  2.7             \\   
291237.8  &  4$_{2,3}$ --  3$_{2,2}$  &  -92.00           &  0.7              &  15.8             &  1.4             \\   
\\
291380.5       &   4$_{3,2}$ --  3$_{3,1}$ & -91.22	      &  1.2		  &  14.8	      &  1.3	      \\
291384.3  &  4$_{3,1}$ --  3$_{3,0}$  &   & & &  \\   
\\
436586.5  &  6$_{2,5}$ --  5$_{2,4}$  &  -91.65           &  2.6              &  15.6             &  1.2             \\   
\\
436751.0  &  6$_{5,1}$ --  5$_{5,0}$  &  -92.69           &  10.2             &  27.1             &  1.6             \\   
436751.0       &   6$_{5,2}$ --  5$_{5,1}$ & & & &\\   
\\
436957.3  &  6$_{4,3}$ --  5$_{4,2}$  &  -93.81           &  3.0              &  12.1             &  0.9             \\   
436957.5       &   6$_{4,2}$ --  5$_{4,1}$ & & & & \\ 
\\
437199.6  &  6$_{3,4}$ --  5$_{3,3}$  &  -93.36           &  4.7              &  15.7             &  1.3             \\   
437236.0  &  6$_{3,3}$ --  5$_{3,2}$  &  -93.71           &  3.0              &  14.1             &  1.2             \\   
\hline                                                                                               
\end{tabular}                                                                 
\label{h2co_16060_tab}                  
\tablefoot{
v and $\delta$ (v) are the peak velocity and its error as derived from the Gauss fit.}

\end{center}                                                                                         
\end{table*} }

\subsection{Continuum data}

\subsubsection{LABOCA}\label{lab:obsres}
{\bf}

\begin{figure} 
\includegraphics[clip=true, trim= 0 0 0 20,angle=0,scale=0.4]{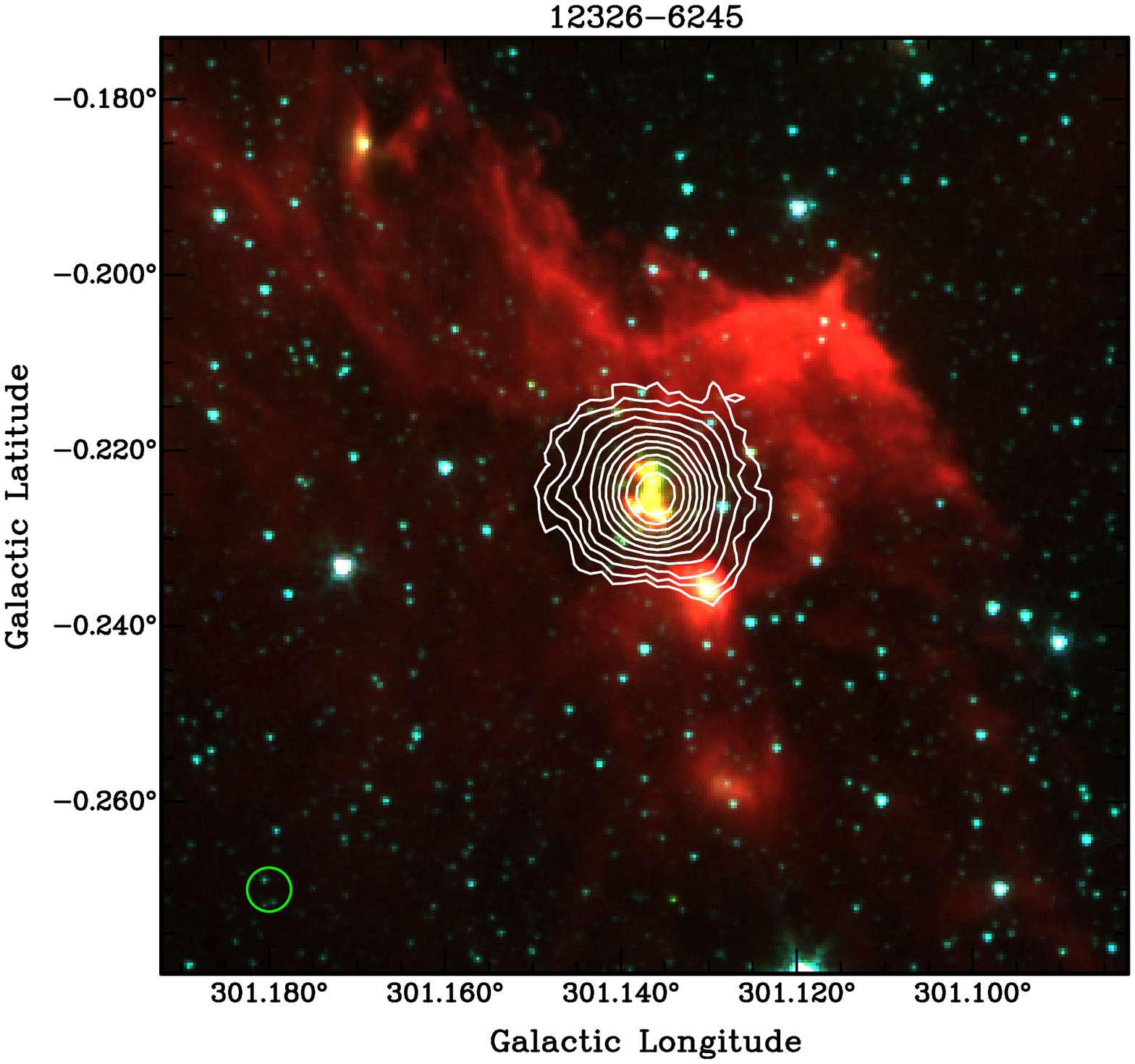}
\includegraphics[angle=0,scale=0.4]{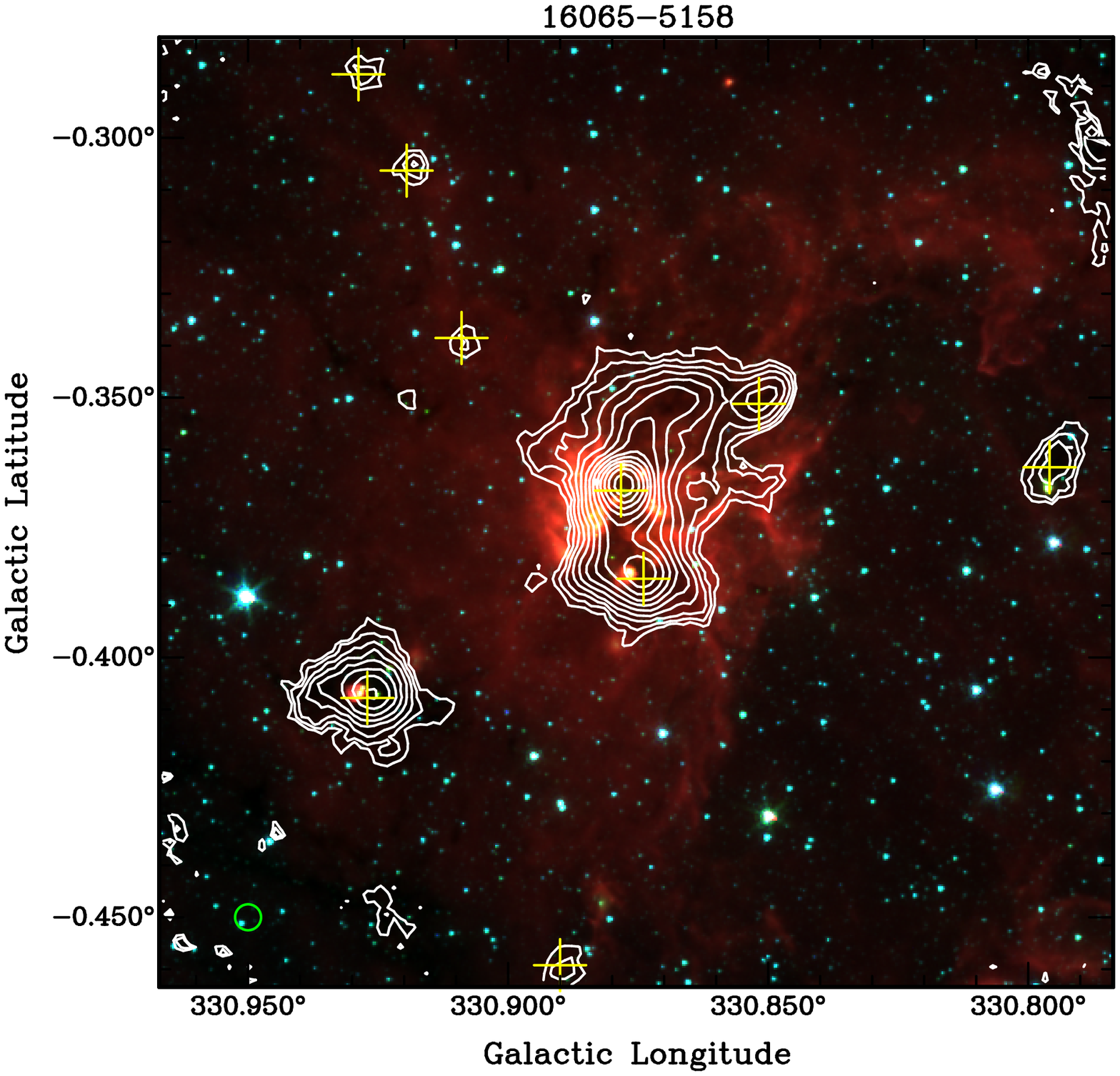}
\includegraphics[clip=true, trim= 0 40 0 60,angle=0,scale=0.4]{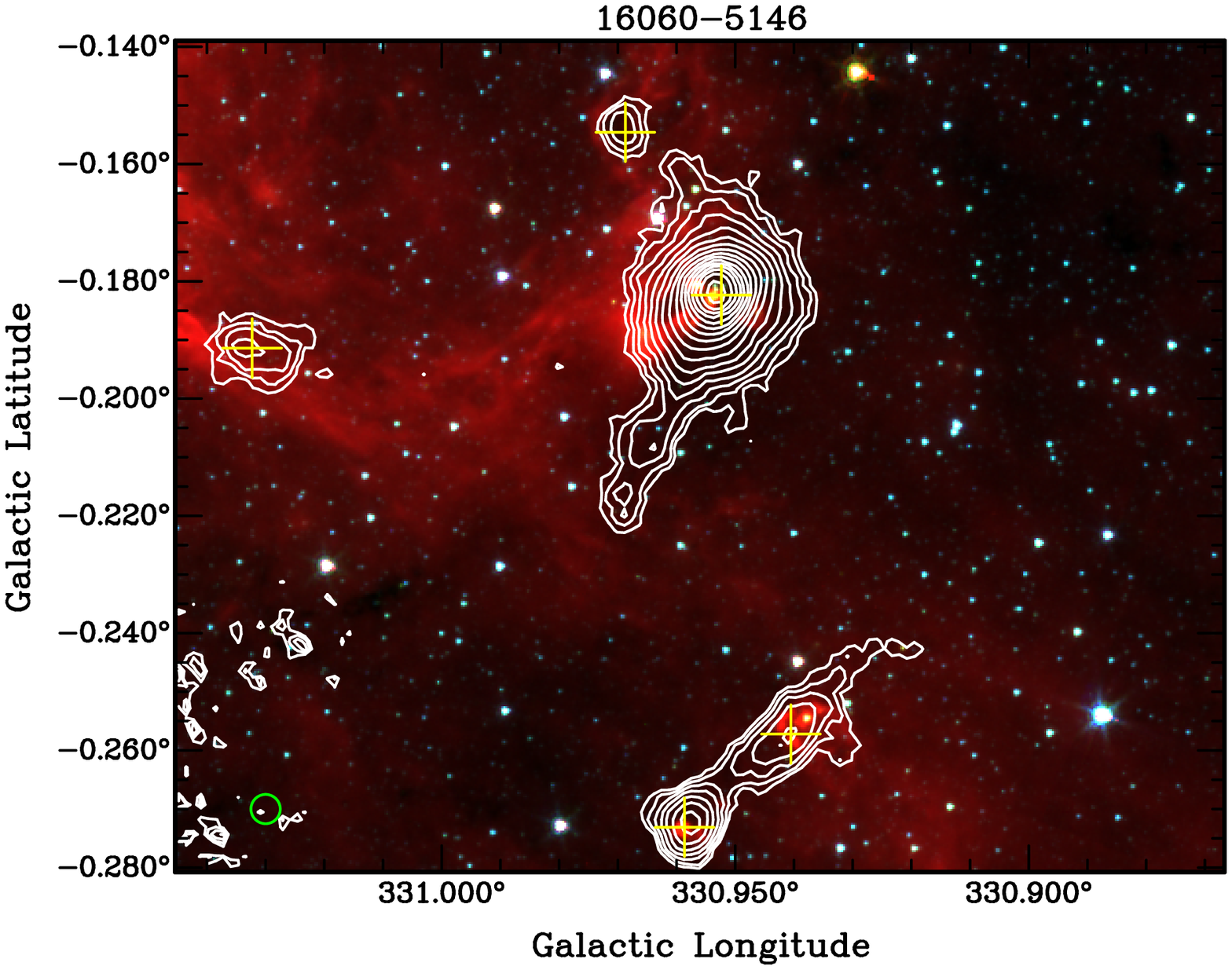} 
\caption{\label{ir_over} LABOCA 870~{$\mu\hbox{m}$}\/ emission (shown as contours),
 starting from 5~$\sigma$ and continuing in multiples of 1.4~$\sigma$ (see
 Table \ref{tab_lab}). The yellow crosses mark the sub-cores in 16060$-$5146
 and 16065$-$5158. The LABOCA beam is shown as green circle in the lower left
 of the image. The three-color image shows 3.6 (blue),
 4.5 (green) and 8~{$\mu\hbox{m}$}\/ (red) emission from the {\it Spitzer} GLIMPSE survey.} 
\end{figure}

While 12326$-$6245 (see Fig. \ref{ir_over} for the three sources) 
shows a spherical, centrally peaked morphology at 870~{$\mu\hbox{m}$}\/,
16060$-$5146 and 16065$-$5158 show an extended morphology with one and three
peaks, respectively. The latter two sources are also
surrounded by secondary peaks, unlike 12326$-$6245, which is isolated within the
LABOCA field of view. Here, the bright MIR cluster seen in the {\it Spitzer} Space Telescope Galactic Legacy Infrared Midplane Survey
Extraordinaire (GLIMPSE) \citep{2003PASP..115..953B} emission, on
which 12326$-$6245 is centered is the only sign of activity in the region, while
the region immediately around the dust emission traced by LABOCA is dominated by enhanced
8~{$\mu\hbox{m}$}\/ emission, a tracer of photon-dominated regions (PDRs). 

In 16060$-$5146, the dominant feature
in the LABOCA field of view is a large bubble that is visible in the MIR emission. The
870~{$\mu\hbox{m}$}\/ emission of 16060$-$5146 corresponds to an active star-forming site
located, at least in projection, at the edge of the bubble. The other cores
traced by 870~{$\mu\hbox{m}$}\/ emission can be found tracing the rim of the bubble and in another
region to the south of 16060$-$5146. 

16065$-$5158, which is only 12\arcmin\/ away from
16060$-$5146, is located in an environment with very active star formation. One finds
a very extended region of 8~{$\mu\hbox{m}$}\/ emission, interspersed with infrared dark
clouds and an association of bright
young stars at the center. 16065$-$5158 is located right at the center of the
association, in a deeply embedded region free of MIR emission. An elongated
structure at 4.5~{$\mu\hbox{m}$}\/ (color-coded green) seems to be coming from this deeply embedded
region. Because 4.5~{$\mu\hbox{m}$}\/ is associated with hot shocked gas
\citep{2008arXiv0810.0530C}, this situation is very suggestive of a massive
outflow stemming from 16065$-$5158. The very extended 870~{$\mu\hbox{m}$}\/ emission found
in the LABOCA field of view for 16065$-$5158 traces nearly the whole extent of
the active star-forming region visible in the GLIMPSE data.  

To obtain source positions, sizes, and peak fluxes, the LABOCA maps (see Fig. \ref{ir_over}) were analyzed with the
\it{sfind}\rm\/ routine in the MIRIAD software package. The integrated
fluxes were obtained over the area inside the contour line representing 10\% of
the peak flux, which corresponds to an average signal--to--noise ratio of 120. The
size of the sources was determined fitting a circularly symmetric two dimensional Gaussian source model. The source positions, fluxes, sizes, and the calculated {\hbox{\rm H$_2$}}\/ column density can be found in
Tables \ref{tab_lab1a} and \ref{tab_lab}. The positions agree within the pointing
uncertainties with the 3~mm peak positions derived with ATCA. The column density was obtained from
Eq. \ref{nh2}, following \citet{1998A&A...336..150M,2006A&A...460..721M}

\begin{equation}
N({\rm H_2})=1.67\times 10^{-22}\frac{I_\nu^{\rm dust}}{B_\nu({\hbox{$T_{\rm d}$}} )\mu m_{\rm H}\kappa_{\rm
   d}R_{\rm d}} ~~[{\rm cm}^{-2}]
\label{nh2}
\end{equation}

where $I_{\nu}$ is the peak flux in Jy/beam,  $B_{\nu}$(T$_{\rm d}$) the Planck
function of a blackbody at dust temperature {\hbox{$T_{\rm d}$}} , $\mu$ is the mean
molecular mass assuming 10\% contribution of helium,  $\kappa_d$ the dust absorption
coefficient of 0.176 m$^2$/kg linearly approximated at 870~{$\mu\hbox{m}$}\/ from model V
(thin ice mantles, $n=10^6~{\hbox{{\rm cm}$^{-3}$}}$, $\beta=1.8$) of
\citet{1994A&A...291..943O} and a dust-to-gas mass ratio $R_{\rm d} =
\frac{1}{100}$. Model V is suited for sources in which considerable depletion
on ice occurs. Unlike the models with thick ice mantles, which apply for
dark clouds, the conditions in model V include the influence of a heating
source that is starting to evaporate the ices. \citet{1994A&A...291..943O}
give the uncertainties for the dust absorption coefficient to be a factor of 2
for ice covered dust and state that it can be up to a factor of 5 higher in
disk regions, where the ice mantles are already evaporated off the grains. As
provided by model V, we
use $\beta=1.8$ ($\kappa \sim (\nu/\nu_0)^{\beta}$) in this work, which is
consistent with values of $\beta=1.5-2.0$ found in massive star-forming
regions \citep{2000A&A...355..617M}. Using the dust opacities derived by
\citet{1983QJRAS..24..267H}, albeit for grains without ice mantles, as has been done by \citet{2002ApJ...566..945B},
leads to masses and column densities that are about a factor of 4 higher. \\
We also list the source multiplicity, in this case all the
sources above 3~$\sigma$ found
within the 11.4~\arcmin\/ field of view.

\begin{table} 
\caption{Properties of the 870~{$\mu\hbox{m}$}\/ continuum as derived from the LABOCA observations. }
\label{tab_lab1a}

\centering                                                                          
\begin{tabular}{lccc}                                                                                            
\hline                                                                          \hline                     
Source & $S_{\rm peak}$ & $S_{\rm int}$ & $N({\rm H_2})$    \\     
& (Jy/beam)  & (Jy) & (x10$^{23}$cm$^{-2}$)  \\
\hline                           
12326$-$6245 &   19.9  & 42.6 & 4.2    \\
16065$-$5158 &   17.9 &  41.8 & 3.8    \\
16060$-$5146 &   46.2 &  81.0 & 8.2        \\ 
\hline                                                                                               
\end{tabular} 

\end{table}

\begin{table*}  
\caption{Positions of the LABOCA 870~{$\mu\hbox{m}$}\/ continuum emission peaks. }
\label{tab_lab}

\centering                                                                          
\begin{tabular}{lcccc}                                                                                            
\hline                                                                        
\hline                
Source & R.A. & Dec. & l,b & Rms   \\     
& (J2000) & J(2000) &  & (mJy/beam) \\
\hline                           
12326$-$6245 & 12:35:35.0 & -63:02:28.9& 301.14,-0.22   & 100   \\
16065$-$5158 & 16:10:19.8 & -52:06:09.7 & 330.88,-0.37  &  50  \\
16065$-$5158 LABOCA-A & 16:10:23.4 & -52:07:06.1 & 330.87,-0.38 &  \\
16065$-$5158 LABOCA-B & 16:10:07.8 & -52:06:31.4 & 330.85,-0.35 &  \\
16065$-$5158 LABOCA-C & 16:09:55.1 & -52:09:20.8 & 330.80,-0.36 &  \\
16065$-$5158 LABOCA-D & 16:10:44.7 & -52:05:56.2 & 330.93,-0.41 &  \\
16065$-$5158 LABOCA-E & 16:10:47.9 & -52:09:43.0 & 330.89,-0.46 &  \\
16065$-$5158 LABOCA-F & 16:10:21.2 & -52:03:35.9 & 330.91,-0.34 &  \\
16065$-$5158 LABOCA-G & 16:10:15.6 & -52:01:47.4 & 330.92,-0.31 &  \\
16065$-$5158 LABOCA-H & 16:10:13.2 & -52:00:35.5 & 330.93,-0.29 &  \\
16060$-$5146 & 16:09:52.5 & -51:54:55.3 & 330.95,-0.18  &  50      \\ 
16060$-$5146 LABOCA-A &	16:09:49.6 & -51:53:03.7 & 330.97,-0.15 & \\  
16060$-$5146 LABOCA-B & 16:10:08.7 & -51:58:43.0 & 330.94,-0.26 & \\
16060$-$5146 LABOCA-C & 16:10:18.1 & -51:58:43.1 & 330.96,-0.27 & \\
16060$-$5146 LABOCA-D & 16:10:17.3 & -51:52:06.2 & 331.03,-0.19 & \\

\hline                                                                                               
\end{tabular} 

\tablefoot{
The
 sub-cores 16065-5158 A-H and 16060-5146 A-D are marked in Fig. \ref{ir_over}.
}

\end{table*}

\subsubsection{ATCA}

\begin{figure} 
\includegraphics[scale=0.4,angle=-90]{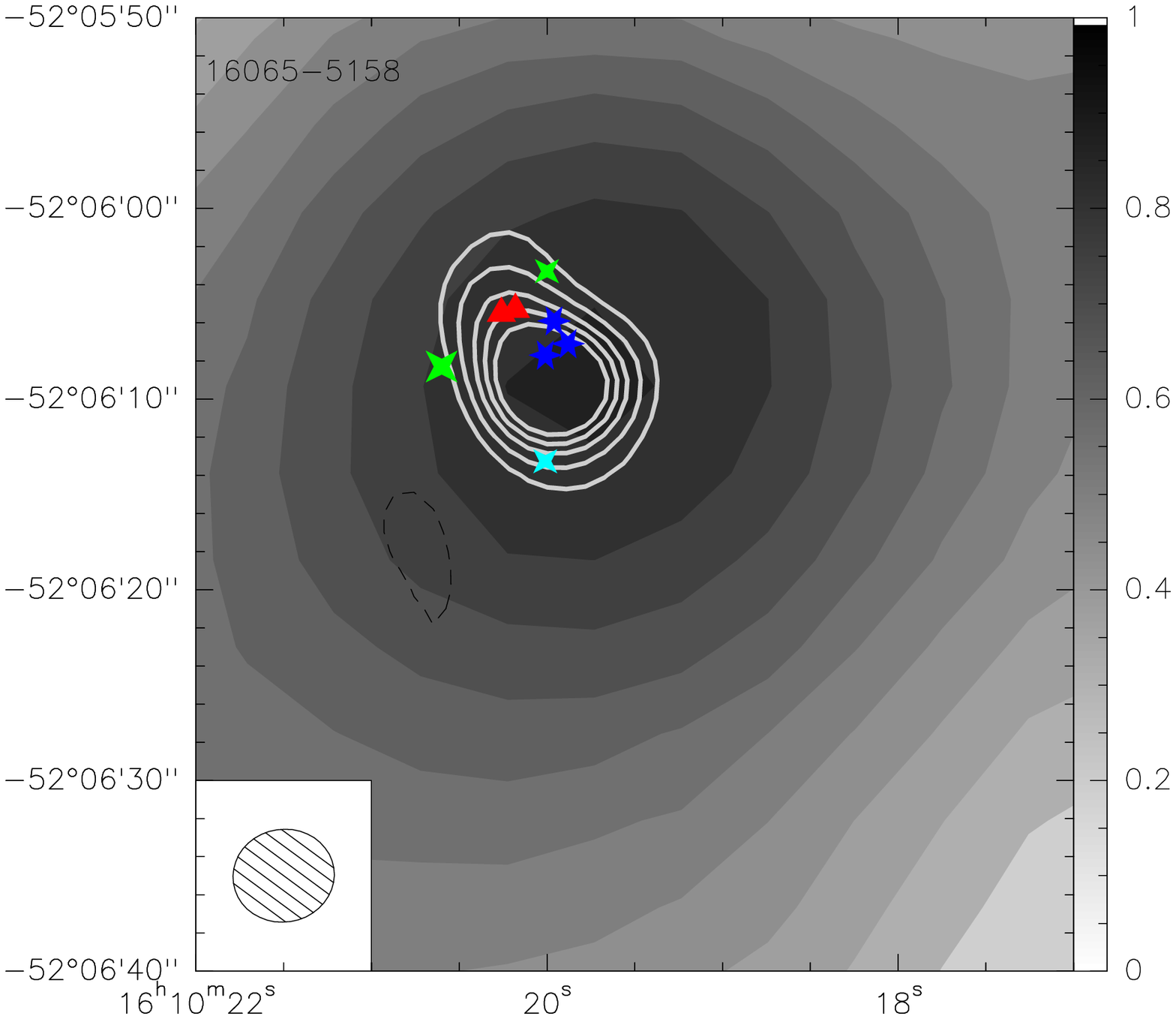}
\includegraphics[clip=true, trim= 250 900 0 0 ,angle=-90,scale=0.4]{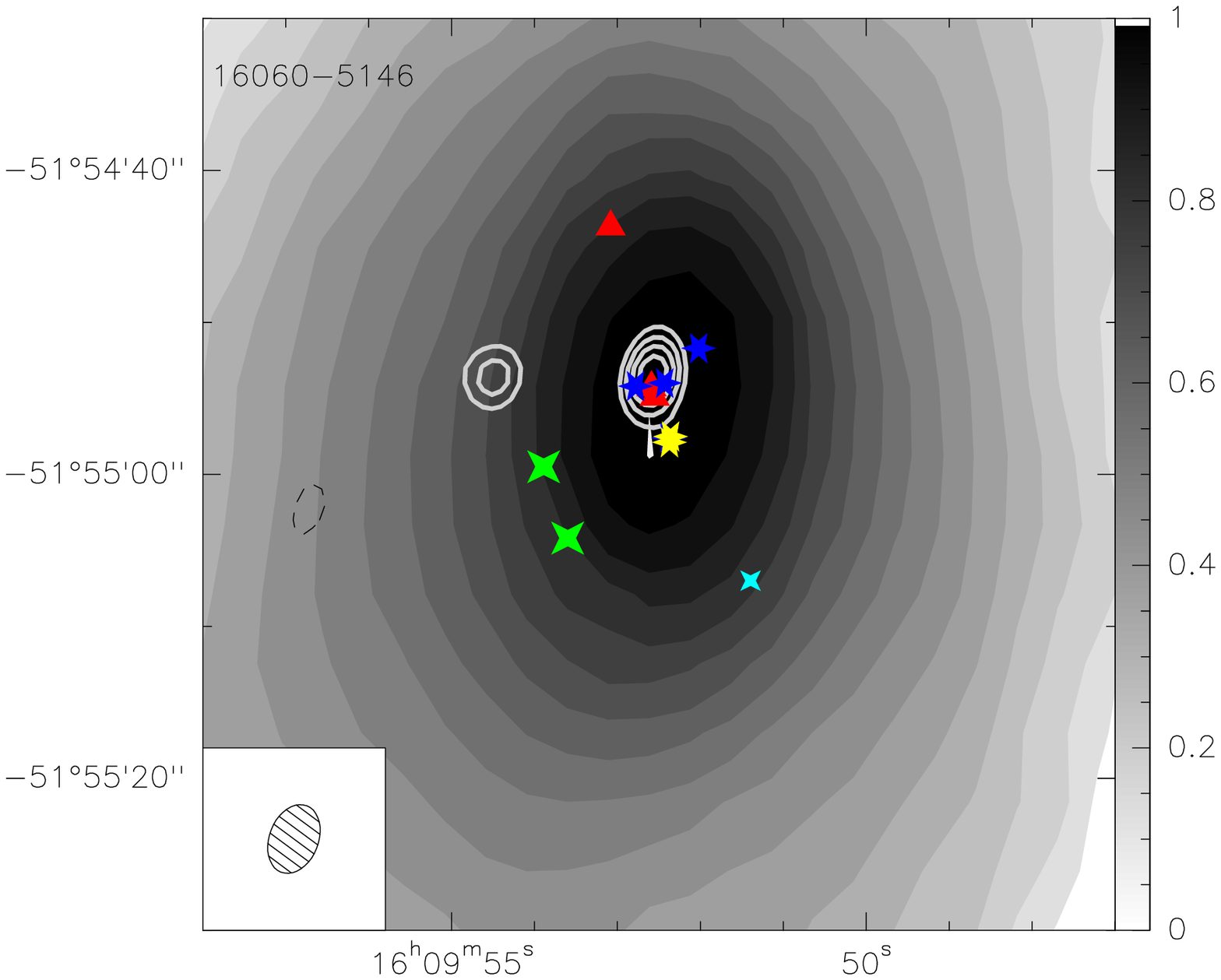}
\includegraphics[scale=0.34,angle=-90]{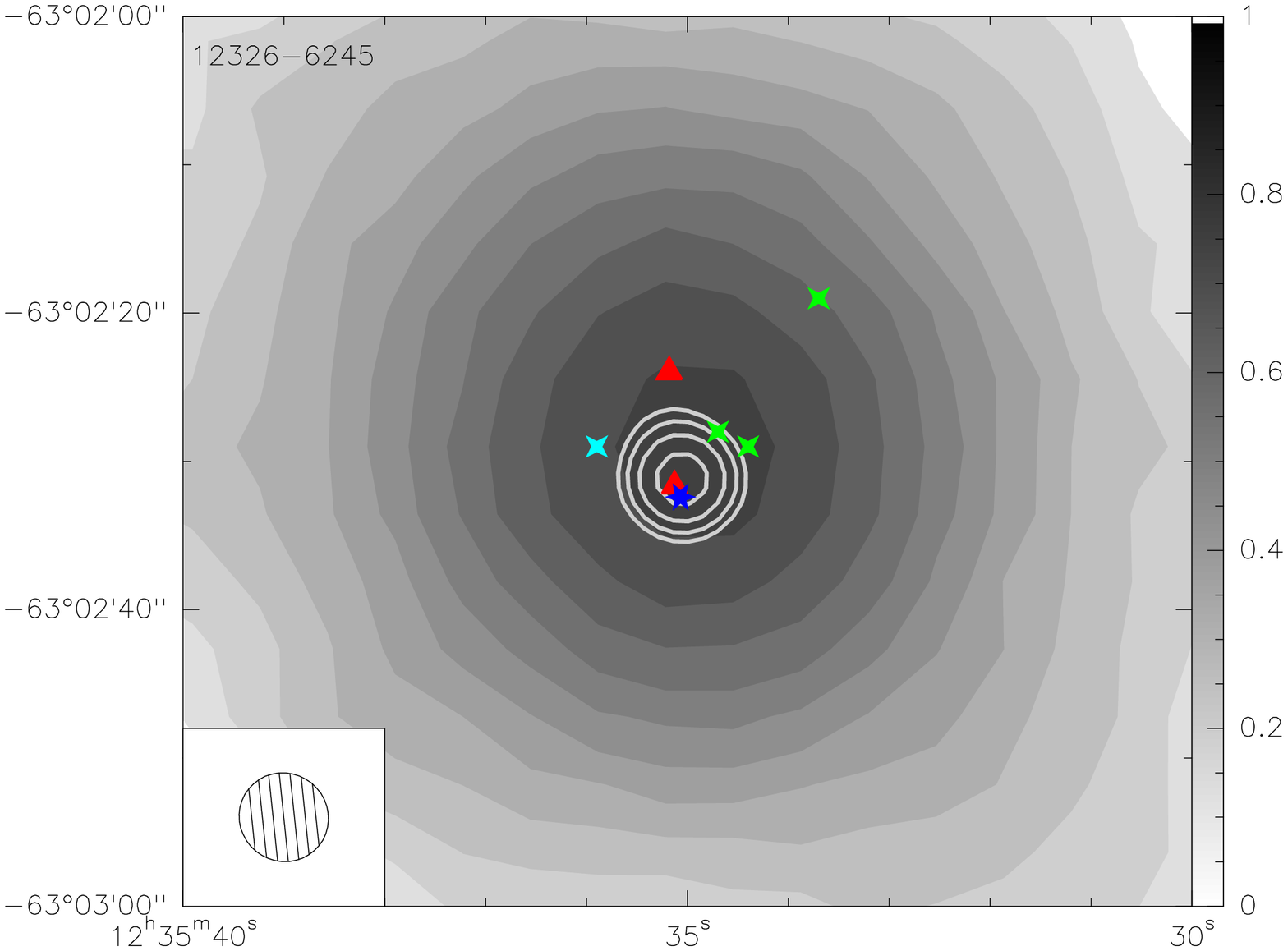}
\caption{\label{fig4} ATCA 3~mm continuum data (solid lines) top
 16060$-$5146 (contour steps -3,3,5,7...$\sigma$, with $\sigma$=0.1~Jy/beam), middle
 16065$-$5158 (contour steps -3,3,5,7...$\sigma$, with $\sigma$=8~mJy/beam) and
 bottom 12326$-$6245 (contour steps -3,3,5,7...$\sigma$, with
 $\sigma$=0.14~Jy/beam). The synthesized ATCA beam is shown in the lower left
 corner. LABOCA 870~{$\mu\hbox{m}$}\/ continuum is shown in gray-scales. Symbols: red: radio continuum (Urquhart et al. 2007), blue: OH maser positions, green: offsets of the APEX line observations, turquoise: IRAS position, yellow: {\hbox{${\rm CH}_3{\rm OH}$}}\/ maser  (Caswell et al. 1998).} 
\end{figure}

Figure \ref{fig4} shows the mm continuum emission imaged with ATCA. One
can see that its distribution agrees well with the peak of the LABOCA
870~{$\mu\hbox{m}$}\/ emission. The ATCA continuum emission is still unresolved
in 16060$-$5146 and 12326$-$6245. In 16060$-$5146, two cores can be
found in the 3~mm emission, and one of them, core ATCA-a, is
associated with the LABOCA peak. The other core, ATCA-b, is too far
away from the APEX pointing position to be picked up with the APEX
beam of about 19\farcs\/. Therefore, these two cores cannot be
responsible for the double peaked structure seen in the spectra of
16060$-$5146. Figure \ref{fig4} also shows the location of the offset
positions determined for the molecular line observations. While the
somewhat crude determination of the position via the {\hbox{${\rm HCO}^+$}}\/ peaks
meant that we may have missed the position of the hot cores by a few
seconds of arc, we were still close enough to pick up the hot core
emission in the beams.

In Fig. \ref{fig4}, overlaid on the data are the interferometric positions of associated  {\hbox{\rm OH}}\/
(Caswell et al. 1999, 2001, 2004) and
{\hbox{${\rm CH}_3{\rm OH}$}}\/ masers (J. Caswell, private communication), as well as the 3~cm and 6~cm
peaks of radio continuum emission \citep{2007A&A...461...11U}. It is
remarkable that in 16060$-$5146, only core A (position 16060$-$5146 ATCA-a) has associated masers and radio
continuum, suggesting that core 16060$-$5146 ATCA-b might be in a very early
evolutionary stage. This core is neither evident in the LABOCA continuum data nor in the
{\it Spitzer} GLIMPSE MIR data.

The ATCA data provide a precise position of the core (see Table \ref{atca_cont}).
The beam de-convolved source sizes, positions, and fluxes for the sources were
determined from the image plane with the {\it imfit}\rm\/ task in the MIRIAD software package.

\begin{table*}[htbp]  

\caption{\label{atca_cont} ATCA 3mm continuum parameters}                                                 
\begin{center} 
\begin{tabular}{lcccc}                                                                     
\hline                                                                                               
\hline                                   

Source & Position  & Source Size & $S_{\nu}$ & $S_{\rm int}$\\
& (J2000) & (\farcs) & (Jy/beam) & (Jy) \\
\hline
12326-6245 & 12:35:35.06 -63:02:31.00 &  $3.3\times2.0$ &  1.96 & 2.36 \\
16060-5146 ATCA-a& 16:09:52.57  -51:54:53.69 & $4.6\times2.2$ & 1.24 & 1.95\\
16060-5146 ATCA-b & 16:09:54.52 -51:54:53.56 & $6.5\times1.7$ & 0.78 & 1.33 \\
16065-5158 & 16:10:20.00 -52:06:08.79 & $6.4\times3.1$ & 0.15 & 0.28 \\
\hline                                                                                               
\end{tabular}                                                                                        
\end{center}                                                                                         
\end{table*}                 

\subsubsection{IR and radio emission}

In addition to the 870~{$\mu\hbox{m}$}\/ continuum observed with LABOCA and the 3~mm continuum
observed with ATCA (this work), the associated fluxes of point sources found
in the {\it Spitzer} GLIMPSE survey in the 3.6,
4.5, 5.6 and 8~{$\mu\hbox{m}$}\/ bands observed with the Infrared Array Camera (IRAC) and
in the MSX mission at 8.3, 12.1, 14.7 and
21.3~{$\mu\hbox{m}$}\/ \citep{2003AAS...203.5708E} were obtained. Photometry on the
{\it Spitzer} MIPSGAL 24 and 70~{$\mu\hbox{m}$}\/ images \citep{2006AAS...209.8801C} could not be performed, because
the sources are saturated in all cases. All the three regions show extended
8~{$\mu\hbox{m}$}\/ emission, which is commonly associated with emission of
photon-dominated regions (PDRs) owing to the IR features of polycyclic aromatic
hydrocarbons (PAHs).

In order to find the infrared point sources associated with the 870~{$\mu\hbox{m}$}\/ dust peaks, we
searched the catalogs within a radius of 9\arcsec\/, the LABOCA beam, around the dust peak. When studying environments of high-mass star formation, one frequently encounters a situation where many infrared
sources cluster in one single dish beam -- in this case the LABOCA beam --
because massive stars are thought to form exclusively in clusters (see Fig. \ref{ir_over}). Because many of the young
high-mass proto-stellar candidates are very red and bright, they are saturated in
the 8~{$\mu\hbox{m}$}\/ band and therefore appear with null values in the
GLIMPSE point source catalog \citep{2007A&A...472..155K}, so when a null value at 8~{$\mu\hbox{m}$}\/ was found, we did a manual check on the image to
distinguish between sources with no detectable emission at 8~$\mu m$ and those that are
saturated. If the GLIMPSE data was saturated at 8~$\mu m$, the MSX value was used.\\
For 16060$-$5146 and 12326$-$6245, both MSX and GLIMPSE emission is 
associated with the hot core traced by the 3~mm continuum observed with ATCA.
In 16065$-$5158 however, there is neither an MSX nor a GLIMPSE source at the
position of the hot core, yet both types of sources can be found a few seconds
of arc offset from
the hot core, which might indicate an external heating source or a deeply
embedded  {\hbox{UCH{\sc ii}}}\/ region.\\

\begin{figure} 
\includegraphics[clip=true, trim= 0 0 0 40 , angle=0,width=7.8cm]{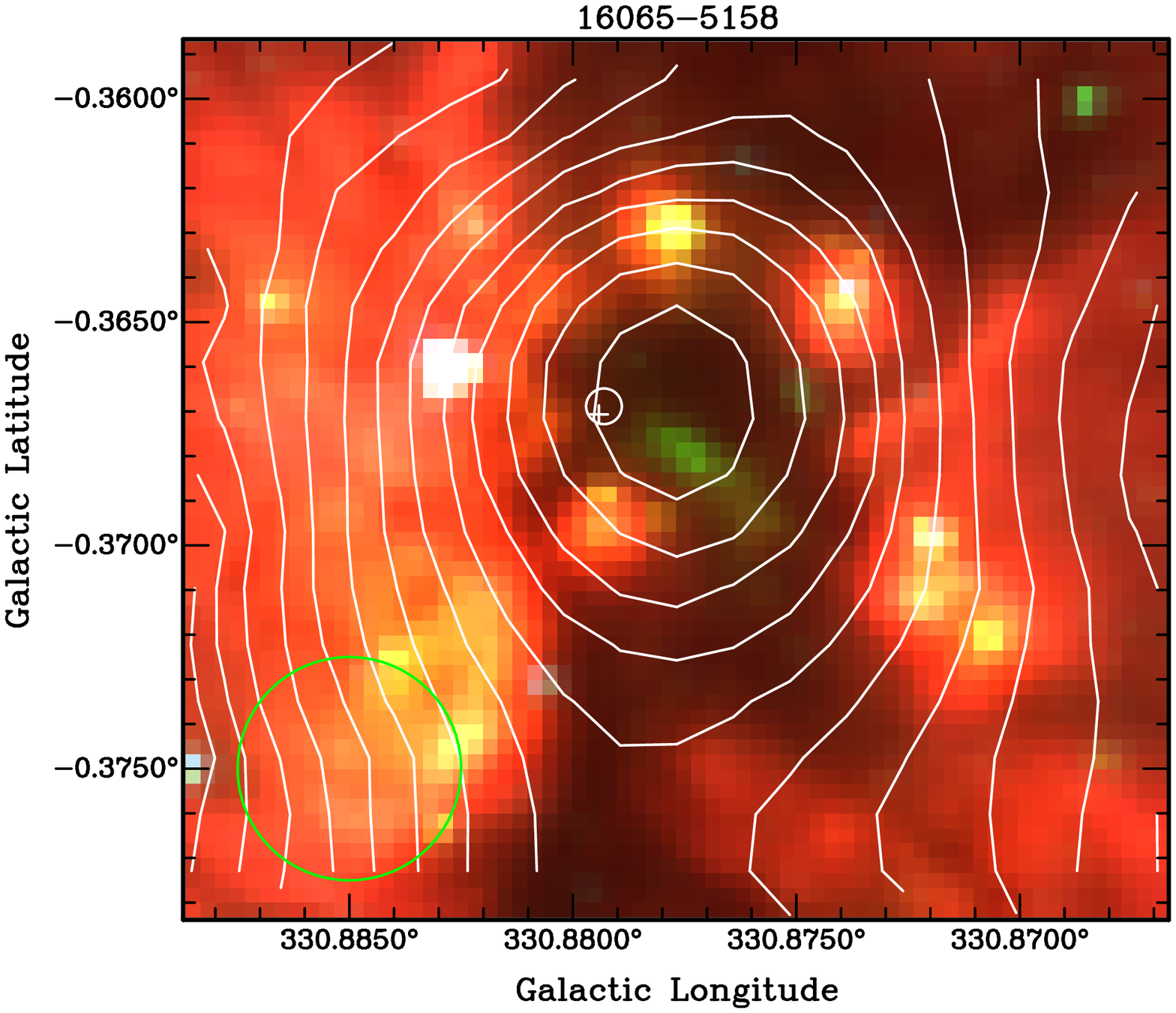}
\includegraphics[clip=true, trim= 0 15 0 50 ,angle=0,width=7.8cm]{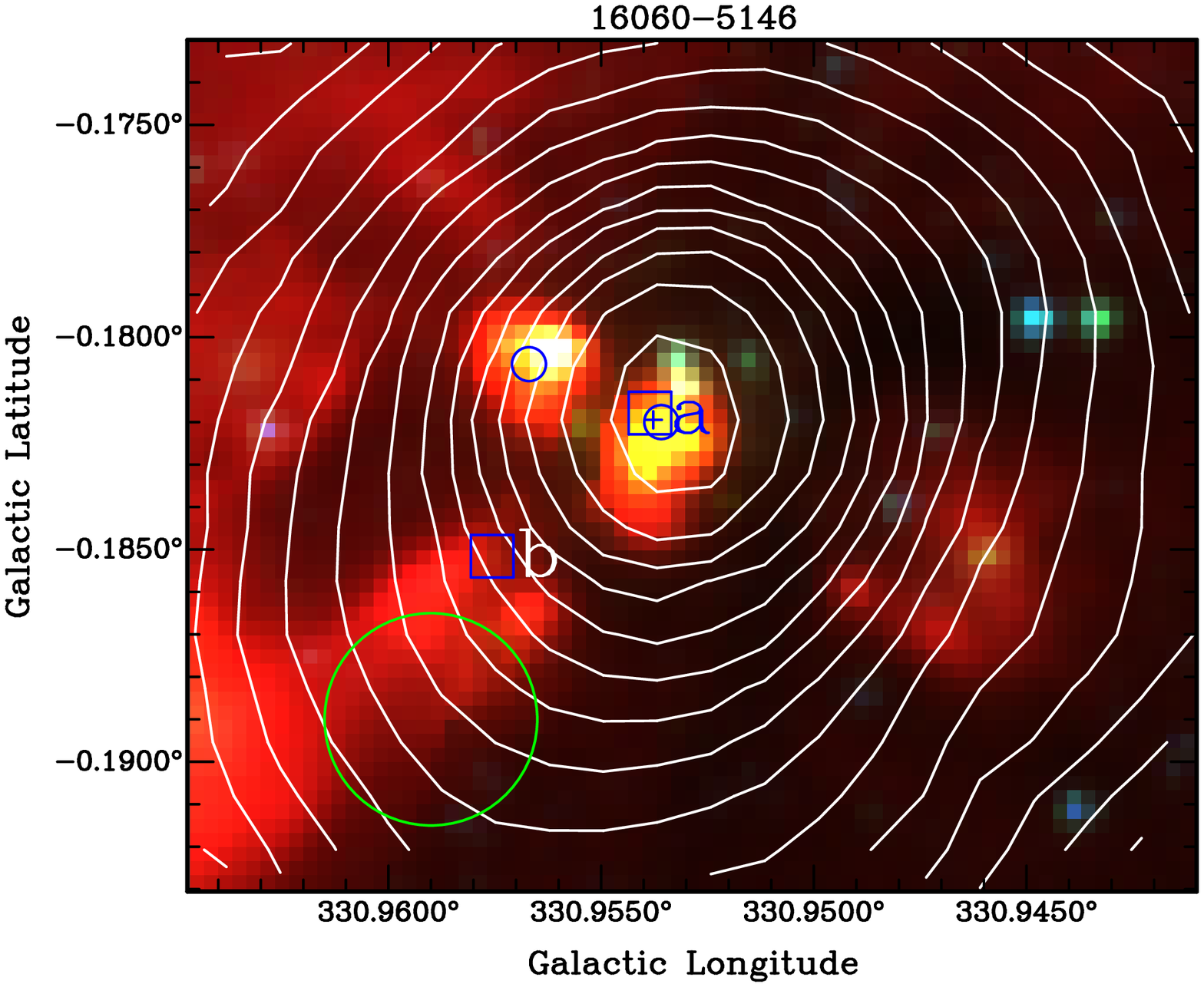}
\includegraphics[clip=true, trim= 0 0 0 0 ,angle=0,width=7.8cm]{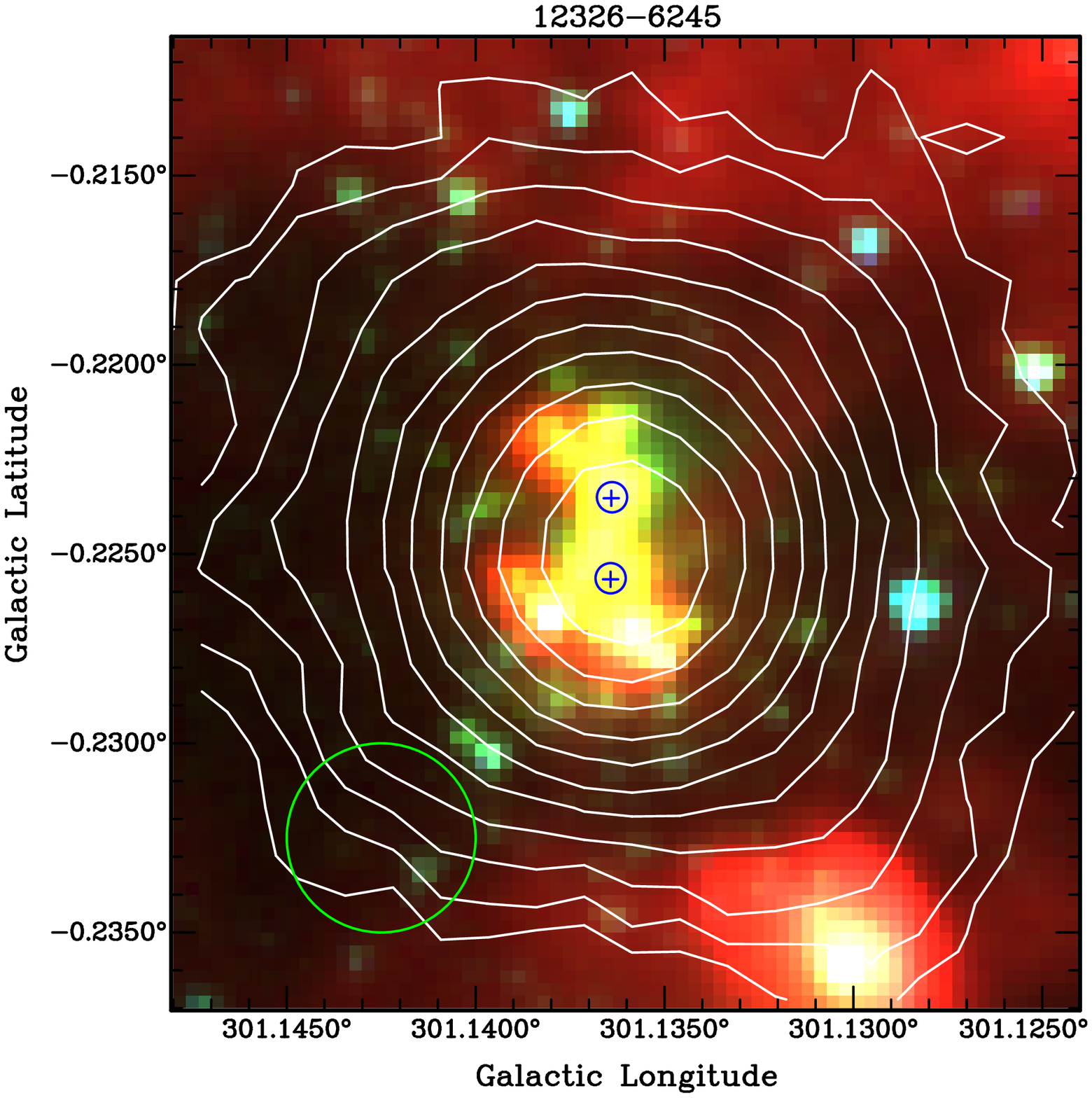}
\caption{\label{ir_over1} Zoom-in of the {\it Spitzer} GLIMPSE composite image of 3.6,
 4.5 and 8~{$\mu\hbox{m}$}\/ emission. LABOCA 870~{$\mu\hbox{m}$}\/ emission is shown as contours,
 starting from 5$\sigma$. The small blue circles and crosses
 mark the position of the 6~cm and 3.6~cm continuum (Urquhart et al. 2007). 16060-5146: the
 blue boxes show the location of the ATCA 3~mm continuum emission (marked $"$a$"$
 and $"$b$"$). 16065$-$5158:
 the cross and circle marking the radio continuum emission are shown in white.}
\end{figure}

For cm wavelengths, archival 3~cm and 6~cm ATCA data published by
\citet{2007A&A...461...11U} were obtained for the hot cores in 12326$-$6245
and 16060$-$5146. In
16065$-$5158, there is an offset of 4.3\farcs\/ between the mm position obtained from ATCA and the
ATCA cm position (see Fig. \ref{fig4}). The cm emission in this source
has a negative spectral index of $-0.9$, hinting at synchrotron emission. No extragalactic object could
be found at this position in the NED database. This could be similar to the
situation in W3({\hbox{${\rm H}_2{\rm O}$}}),
where a synchrotron source \citep{1995ApJ...443..238R} can be found in a hot
core. Note that \citet{2007A&A...461...11U} state that the
spectral indices of individual sources might not be reliable in all cases, because the ATCA
is not a scaled array at the observed frequencies (which would provide comparable
resolution at the different wavelengths) and the measurements may
sample different spatial scales. \\
Figure \ref{ir_over1} shows a zoom-in on the region of the 870~{$\mu\hbox{m}$}\/ peak. In
12326$-$6245, the IR flux is coming from a bright cluster of objects. One can
see emission at 4.5{$\mu\hbox{m}$}\/ (color-coded green) surrounding the cluster, which is a sign of
shocked gas \citep{2008arXiv0810.0530C} and can point to outflow action. In 16060$-$5146, the
peak of the 870~{$\mu\hbox{m}$}\/ emission can be associated with a single object. There is
weak 4.5~{$\mu\hbox{m}$}\/ emission at the edge of the IR source. 16065$-$5158 has no
associated IR source at the position of the 870~{$\mu\hbox{m}$}\/ peak. There is however
a strong elongated 4.5~{$\mu\hbox{m}$}\/ emission. This situation is suggestive of a
deeply embedded young object with a strong outflow.

\section{Derivation of physical parameters}

\subsection{LTE modeling}\label{lte:modeling}

\begin{figure} 
\includegraphics[clip=true, trim= 0 0 0 0,angle=-90,width=8cm]{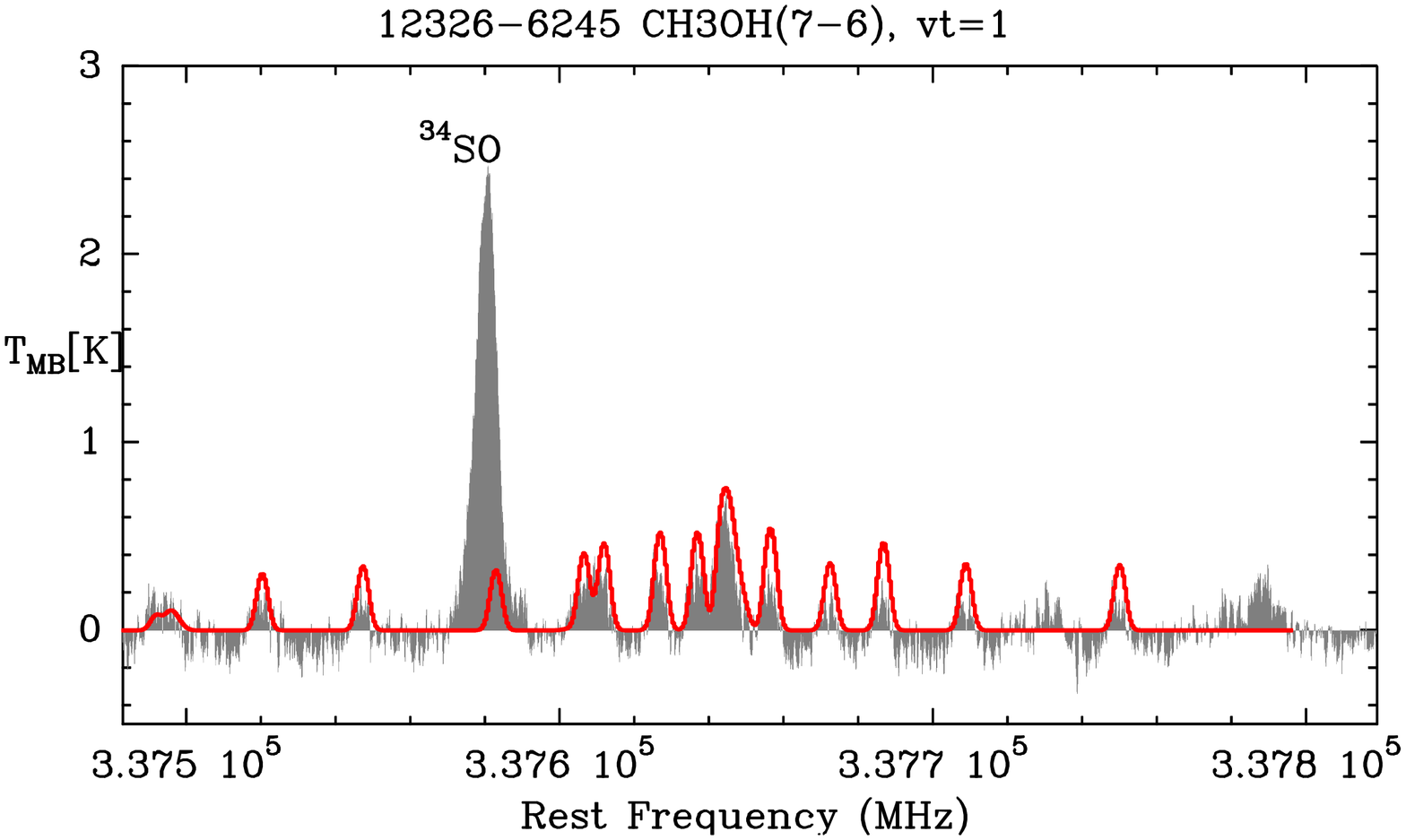}
\includegraphics[clip=true, trim= 0 0 0 0,angle=-90,width=8cm]{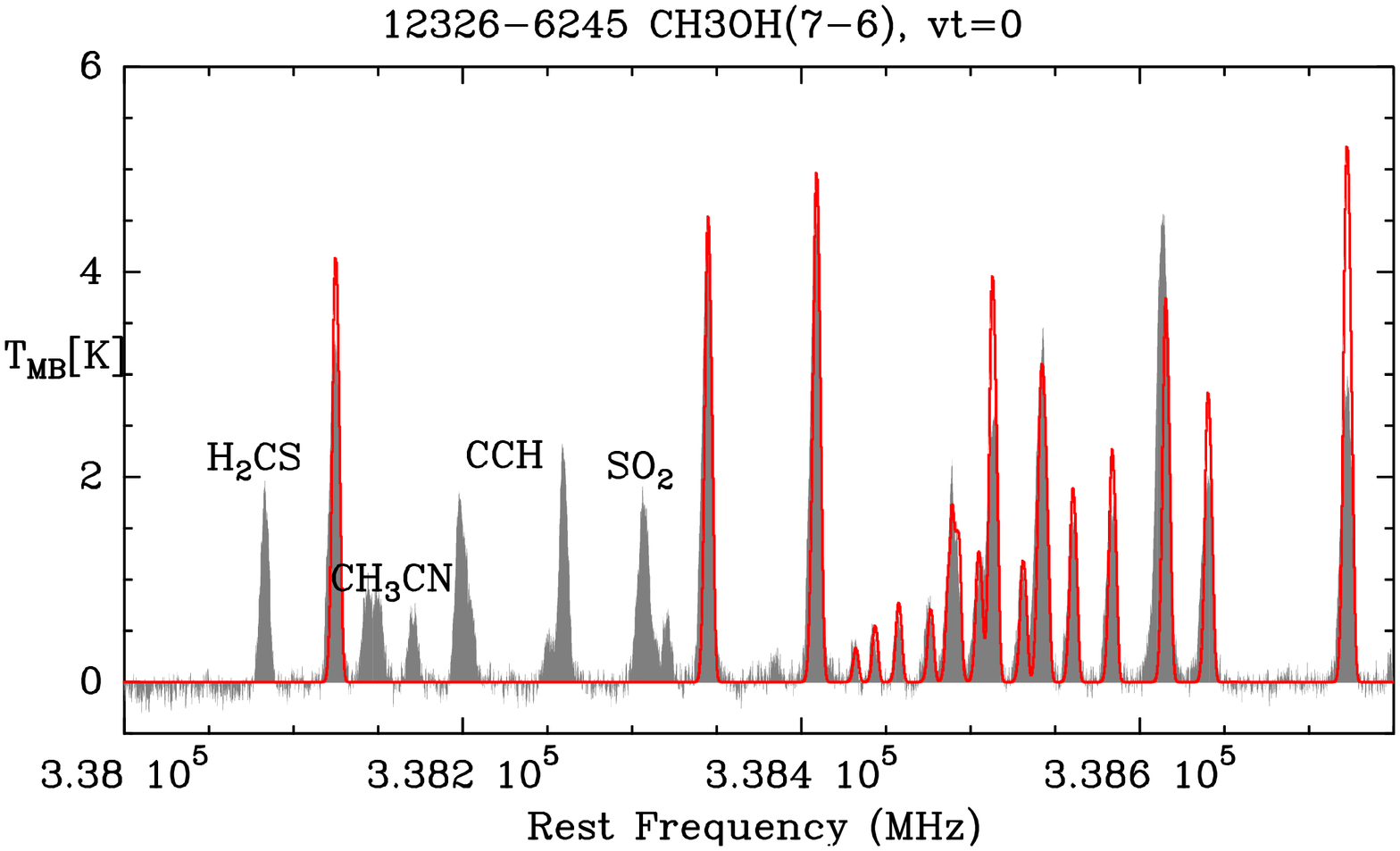}
\caption{\label{ch3oh1_plots} Spectrum of 12326$-$6145. The synthetic model
 spectrum of {\hbox{${\rm CH}_3{\rm OH}$}}\/(7--6) is shown in red. In the upper plot ($v_{\rm t}=1$),
 the synthetic spectrum of the torsionally excited {\hbox{${\rm CH}_3{\rm OH}$}}\/ transitions (red)
 is shown. } 
\end{figure}

\begin{figure} 
\includegraphics[clip=true, trim= 0 0 0 0,angle=-90,width=8cm]{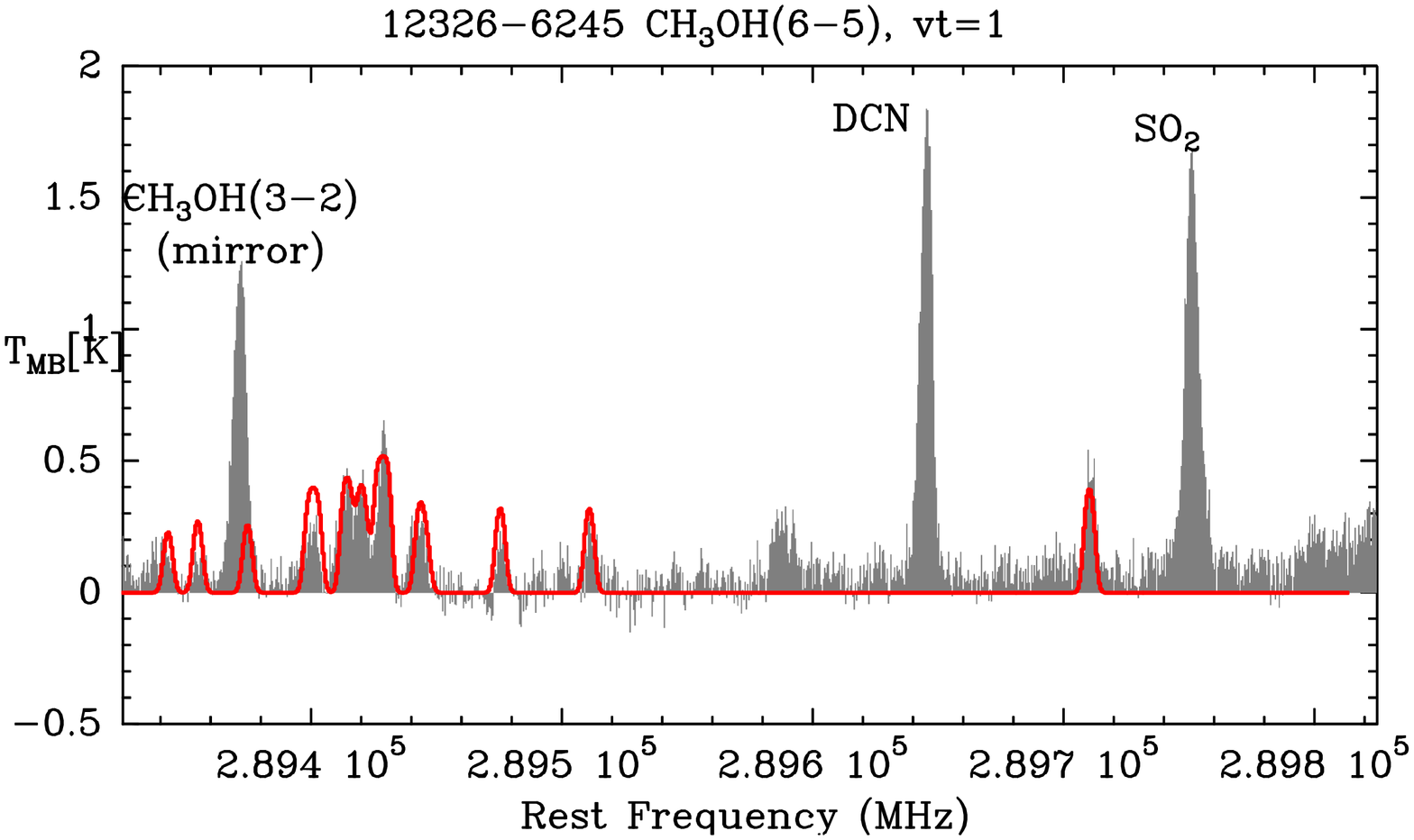}
\includegraphics[clip=true, trim= 0 0 0 0,angle=-90,width=8cm]{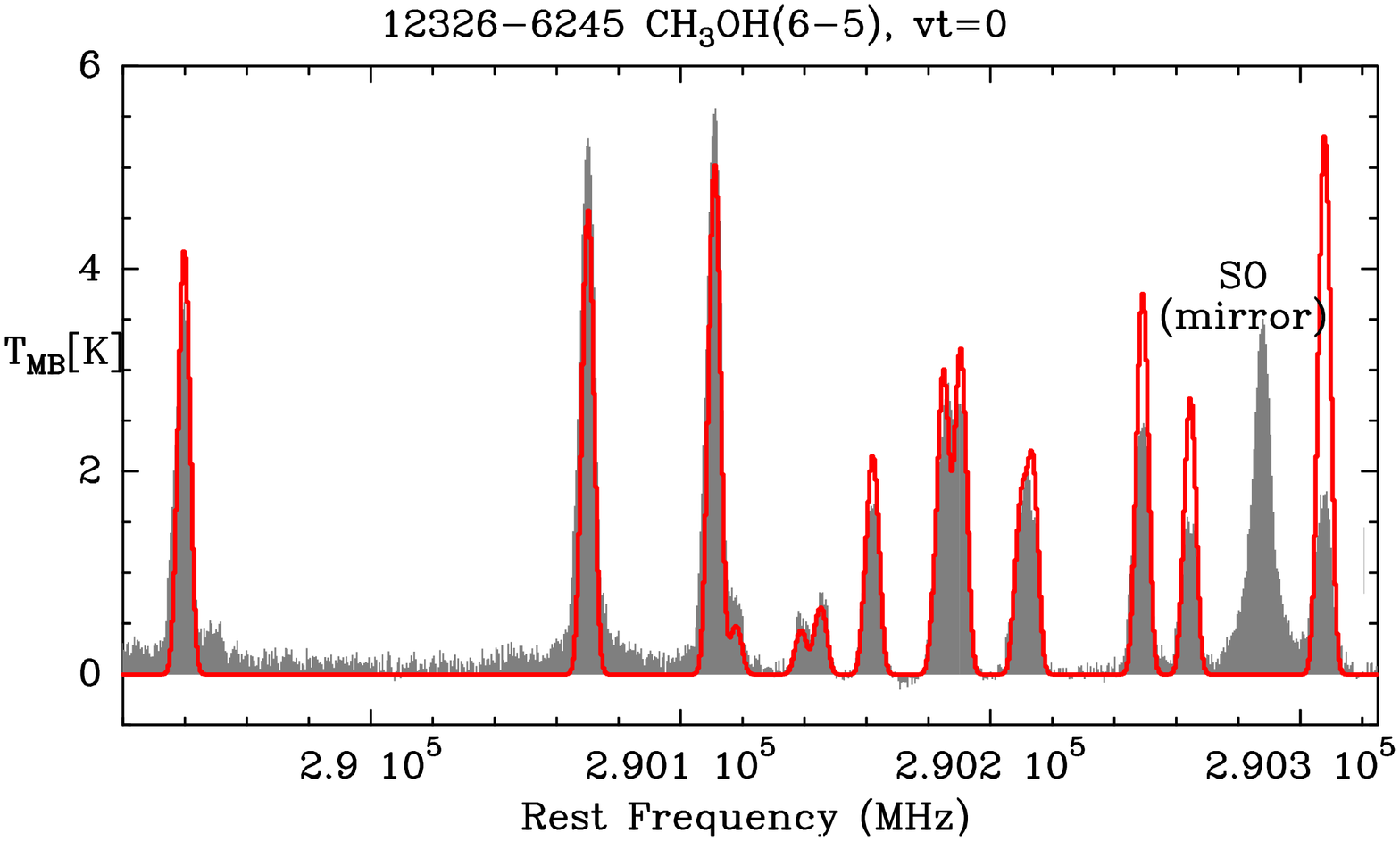}
\caption{\label{ch3oh2_plots} Spectrum of 12326$-$6245. The synthetic model
 spectrum of {\hbox{${\rm CH}_3{\rm OH}$}}\/(6--5) is shown in red. In the upper plot ($v_{\rm t}=1$),
 the synthetic spectrum of the torsionally excited {\hbox{${\rm CH}_3{\rm OH}$}}\/ transitions (red)
 is shown. The two lines marked (mirror) in both plots show lines that lie
 outside the shown frequency range, but were folded into the spectrum owing to
 imperfect filter edges. } 
\end{figure}

We observed our sources in lines from {\hbox{${\rm CH}_3{\rm CN}$}}, {\hbox{${\rm CH}_3{\rm OH}$}}\/, and {\hbox{${\rm H}_2{\rm CO}$}}, which are useful
tracers of temperature. 
While classically rotational diagrams have been used to determine rotational
temperature, $T_{\rm rot}$, and column density, $N$, of a molecule (see
\citet{1993ApJS...89..123M,1993A&A...276..489O}), we
make use of an LTE approach developed by \citet{1999pcim.conf..330S} and
further improved by
\citet{2005ApJS..156..127C}. Implemented into the program XCLASS, it allows us to model synthetic
spectra that take line blends and optical depth effects into account and
furthermore allows us to simulate a double sideband (DSB) spectrum.
Input parameters for the simulation are source size, temperature, column density,
line widths, and offset from systemic velocity. It is possible to
simultaneously model
transitions in several frequency bands and include more than one component,
i.e. a core and an envelope component and/or several velocity
components.  If there is a hot core and cold envelope component, both components need to be modeled independently though, i.e. the cold component is not absorbing the hot component, because we do not know the source geometry. Figures. \ref{ch3cn_plots}, \ref{ch3oh1_plots}, and \ref{ch3oh2_plots} give an example of
a synthetic spectrum for a single molecule overplotted on the measured
sub-spectra, while Figs. \ref{fig:molem1a} -- \ref{fig:molem3b} in the appendix show the synthetic
spectra of all included species. 

The degeneracy between column density and source
size can be solved for species where both optically thin and thick lines are
present, as we did for {\hbox{${\rm CH}_3{\rm CN}$}}\/ and to some degree {\hbox{${\rm CH}_3{\rm OH}$}}.
For species
where only one or two lines were available to model, only the beam-averaged
column density was derived, and for species that we assume to reside in the hot core, a rotation temperature $T_{\rm rot}=$150~K \citep{2006A&A...454L..41S} was assumed while for species  located in the
more extended envelope 50~K was chosen, unless we had both optically thick and
thin lines to derive a temperature from the model. \\

The lines were identified with the catalogues of the CDMS and the JPL
\citep{1998JQSRT..60..883P,2001A&A...370L..49M,2005IAUS..235P..62M} as a reference for the rest frequencies. \\    
The tables with the detected lines for the three sources can be found in the
online material (Tables \ref{appen:co} to \ref{app:trans_etcn1}) -- here we will present the results of the modeling.\\

XCLASS produces a synthetic spectrum (model) of all species and transitions
included in the modeling, which is then overlaid on the data. For {\hbox{${\rm CH}_3{\rm CN}$}}\/,
we performed a ${\rm \chi}^2$ analysis on the synthetic model spectrum to determine the best-fit model, while for
the remaining setups a comparison by eye was done to decide on the best
fit. This was necessary to account for the high complexity when modeling
several often blended transitions in the double sideband spectra. For
sources labeled ``b'' in Tables \ref{tab_mol_16065} -- \ref{tab_mol_12326}, the lines are weak and/or blended with many other
lines, so that the fit is less accurate than a fit marked ``c''.\\
If both temperature and source size were fixed, the column
density could be determined to an accuracy of 10\%, assuming perfect calibration. 
The uncertainties introduced when modeling both source size and temperature
are discussed in the following sections. This modeling is done under the assumption of LTE, which may not necessarily be valid for every molecule and source, although it is a reasonable choice given the high densities. 
Nonetheless, this method gives us a good and fast overview on the
chemical composition of the sources, and it can be applied to a large sample of
sources with considerably less effort than individual non-LTE modeling would
require.

The results of the modeling can be found in Tables \ref{tab_mol_16065} to
\ref{tab_mol_12326}. Tables \ref{tab_16065} to \ref{tab_12326} show the model
results for those species that were modeled in several frequency setups,
namely {\hbox{${\rm CH}_3{\rm OH}$}}, {\hbox{${\rm CH}_3{\rm CN}$}}, {\hbox{${\rm H}_2{\rm CO}$}}\/ and {\hbox{${\rm SO}_2$}}. For some species, marked as $^f$ in the Tables, we modeled isotopologues, using ratios of 
$^{12}$C/$^{13}$C=60, $^{32}$S/$^{34}$S=23, $^{14}$N/$^{15}$N=300 \citep{1994ARA&A..32..191W} and
$^{16}$O/$^{17}$O=1500 \citep{1994ARA&A..32..191W,2008A&A...487..237W}.

\onltab{16}{

\begin{table*}                                                                                 
\begin{center}                                                                                       
\caption{ Identified features of {\hbox{\rm CO}} and {\hbox{${\rm HCO}^+$}}}                                                            
\label{appen:co}
\begin{tabular}{cc}                                                                                  
\hline
\hline                                                                                               
$\nu$  &   Transition\\                                                                            
(MHz)  &  (J)  \\                                                                                      
\hline                                                                                               
{\hbox{\rm C$^{17}$}}O &   \\                                                                   
337061.1       &   3 --  2  \\
{\hbox{${\rm HCO}^+$}} & \\
356734.2 & 4$_{0,0}$  --  3$_{0,0}$  \\

\hline                                                                                               
\end{tabular} \\                                                                                     
\end{center}                                                                                         

\end{table*}          }

\onltab{17}{
\begin{table*}                                                                                 
\begin{center}                                                                                       
\caption{Identified features of {\hbox{\rm CS}}\/ and {\hbox{${\rm HC}_3{\rm N}$}}}                                                  
\label{app:trans_cthfos}
\begin{tabular}{cc}                                                                                  
\hline                                                                                               
\hline 
$\nu$  &   Transition \\                                                                            
(MHz)  &  (J) \\                                                                                      
\hline                                                                                               
{\hbox{${\rm C}^{34}{\rm S}$}}  &   \\                                                                                     
337396.5       &   7 -- 6     \\   
{\hbox{${\rm C}^{33}{\rm S}$}}  &    \\                                                                                     
291485.9       &   6 -- 5     \\   
437165.0  &  9 -- 8   \\
{\hbox{${\rm HC}_3{\rm N}$}} & \\                                                                                       
291068.4       &  32 -- 31         \\
436468.9       &  48 -- 47     \\ 
{\hbox{${\rm HC}_3{\rm N}$}}(v7=1)  & \\  
291782.3       &  32$f$ -- 31$f$   \\
337344.7       &  37$e$ -- 36$e$    \\   
337825.3       &  37$f$ -- 36$f$    \\   
437535.9       &  48$f$ -- 47$f$   \\
\hline                                                                                               
\end{tabular} \\                                                                                     
\end{center}                                                                                         

\end{table*} }

\onltab{18}{                            

\begin{table*}                                                                                  
\begin{center}                                                                                       
\caption{Identified features of {\hbox{\rm HCN}}, {\hbox{\rm DCN}} and {\hbox{\rm HNCO}} }                                                                 
\label{app:dcn}
\begin{tabular}{cc}                                                                                  
\hline
\hline                                                                                               
$\nu$  &   Transition   \\                                                                            
(MHz)  &  (J$_{K_-,K_+}$) \\                                                                          
\hline
{\hbox{${\rm H}^{13}{\rm CN}$}} & \\

431659.8 & 5$_{0,0}$   --  4$_{0,0}$ \\

{\hbox{${\rm HC}^{15}{\rm N}$}} & \\
430235.3  & 5$_{0,0}$  --  4$_{0,0}$ \\

{\hbox{\rm DCN}} &  \\                     
289644.9       &   4$_{0,0}$ --  3$_{0,0}$  \\   
{\hbox{\rm HNCO}}  &  \\                                                                                       
350333.3       &  16$_{1,16}$ -- 15$_{1,15}$  \\

\hline                                                                                               
\end{tabular} \\                                                                                     
\end{center}                                                                                         

\end{table*} }

\onltab{19}{

\begin{table*}                                                                                  
\begin{center}                                                                                       
\caption{Identified features of {\hbox{${\rm H}_2{\rm CO}$}}}                                                                
\label{app:h2co}
\begin{tabular}{ccc}                                                                                  
\hline     
\hline                                                                                          
$\nu$  &   Transition &  \\                                                                            
(MHz)  &  (J$_{K_-,K_+}$) &  \\                                                                          
\hline                                                                                               
290623.4       &   4$_{0,4}$ --  3$_{0,3}$ & \\  
291237.8       &   4$_{2,3}$ --  3$_{2,2}$ & \\
\\ 
291380.5       &   4$_{3,2}$ --  3$_{3,1}$ & *\\
291384.3       &   4$_{3,1}$ --  3$_{3,0}$ & *\\   
\\
436586.5       &   6$_{2,5}$ --  5$_{2,4}$ & \\ 
\\
436751.0       &   6$_{5,1}$ --  5$_{5,0}$ & *\\   
436751.0       &   6$_{5,2}$ --  5$_{5,1}$ & *\\   
\\
436957.3       &   6$_{4,3}$ --  5$_{4,2}$ & *\\ 
436957.5       &   6$_{4,2}$ --  5$_{4,1}$ & * \\ 
\\
437199.5       &   6$_{3,4}$ --  5$_{3,3}$ & \\ 
437236.0       &   6$_{3,3}$ --  5$_{3,2}$ & \\ 
\hline                                         						       
\end{tabular} \\                               				
Notes: (*) Blended with each other

\end{center}                                                     

\end{table*} }

\onltab{20}{

\begin{table*}                                                                               
\begin{center}                                                                                       
\caption{ Identified features of {\hbox{${\rm CH}_3{\rm OCHO}$}}.}                                                              
\label{tab:lines}
\begin{tabular}{cccc}                                                                                  
\hline
\hline                                                                                               
$\nu$  &   Transition & Notes &  \\                                                                            
(MHz)  &  (J$_{K_-,K_+}$) & & \\                                                                          
\hline                                                                                               
336889.213       &  26$_{0,26}E$ -- 25$_{0,25}E$  &   &  \\   
336918.095       &  26$_{6,20}A$ -- 25$_{6,19}A$  & a)  &  \\   
338355.771       &  27$_{8,19}A$ -- 26$_{8,18}A$  &   &  \\   
348909.480       &  28$_{9,20}E$ -- 27$_{9,19}E$  & b)  &  \\   
348914.975       &  28$_{9,20}A$ -- 27$_{9,19}A$  &   &  \\   
350442.250       &  28$_{8,20}E$ -- 27$_{8,20}E$  & c)  &  \\   
\hline                                                                                               
\end{tabular} \\                                                                                     
Notes: (a) Blend with CH$_{3}$OH at 350687.8. (b) Blend with CH$_{3}$OH at 337490.5. (c) Blend with CH$_{3}$CN at 350443.3.
\end{center}                                                                                         

\end{table*} }

\onllongtab{21}{              
\begin{longtable}{ccc}                                                                             

\caption{Identified features of {\hbox{${\rm CH}_3{\rm OH}$}}}  
\label{app:tab_meth}\\

\hline
\hline 
\multicolumn{1}{c}{$\nu$}  &    \multicolumn{1}{c}{Transition} & \multicolumn{1}{c}{Notes} \\                                                                             
(MHz)  &  (J$_{K_-,K_+}$) &  \\                                                       \hline                    
\endfirsthead                                                                    
\caption{continued.}\\                 

\hline
\hline 

\multicolumn{3}{c}{{\bfseries \tablename\ \thetable{} -- continued from
    previous page}} \\
\hline
$\nu$ & Transition & Notes   \\
(MHz)  &  (J$_{K_-,K_+}$) &\\  
\hline                                                                                               
\endhead                                                                                             
\hline                                                                                               

\multicolumn{3}{r}{{Continued on next page}} \\\hline  

\endfoot                                                                                             
\endlastfoot                                                  

{\hbox{${\rm CH}_3{\rm OH}$}}(vt=0)  & &  \\                                                                                 
289939.5      &    6 $_{0,6 }E$ --   5 $_{0,5 }E$       &    \\   
290069.8       &    6 $_{1,6 }E$ --   5 $_{1,5 }E$       &    \\   
\\
290110.7       &    6 $_{0,6 }A^+$ --   5 $_{0,5 }A^+$   &  *  \\   
290117.8       &    6 $_{5,1 }E$ --   5 $_{5,0 }E$       &  *  \\   
\\
290138.9  &   6 $_{5,2 }E$ --   5 $_{5,1 }E$        & \\
\end{longtable} }

\onltab{22}{

\begin{table*}                                                                                
\begin{center}                                                                                       
\caption{Identified features of {\hbox{${\rm SO}_2$}}}                                                                
\label{app:trans_sotw}
\begin{tabular}{ccc}                                                                                  
\hline    
\hline                                                                                           
$\nu$  &   Transition & Notes \\                                                                            
(MHz)  &  (J$_{K_-,K_+}$) & \\                                                                          
\hline                                                                                               
{\hbox{${\rm SO}_2$}}  &  & \\                                                                  
301896.6       &  19$_{2,18}$ -- 19$_{1,19}$ &  \\                        
338611.8       &  20$_{1,9}$ -- 19$_{2,8}$  & a) \\   
338306.0       &  18$_{4,4}$ -- 18$_{3,5}$  & b) \\   
430193.7       &  24$_{1,23}$ -- 23$_{2,22}$ & c)\\
\\
430228.6       &  24$_{0,24}$ -- 23$_{1,23}$ & d)* \\   
430232.3       &  21$_{2,20}$ -- 20$_{1,19}$ & * \\   
\\
430347.6       &  26$_{5,21}$ -- 26$_{4,22}$ & e) \\   
{\hbox{$^{34}{\rm SO}_2$}}  & &  \\                                                                                   
338320.356       &  13$_{2,2}$ -- 12$_{1,1}$ & b) \\   
\hline                                                                                               
\end{tabular} \\                                                                                     
Notes: (a) Blend with CH$_3$OH at  338615. (b) Blend. \\
(c) Blend with CH$_3$OH at 430207.1 and OCS at 437624.522. \\
(d) Blend with HC$^{15}$N at 430235.\\ (e) Blend with SO at 430339.5.

\end{center}                                                                                         

\end{table*} }

\onltab{23}{

\begin{table*}                                                                                
\begin{center}                                                                                       
\caption{Identified features of {\hbox{\rm SO}} }                                                                 
\label{app:so}
\begin{tabular}{ccc}                                                                                  
\hline 
\hline                                                                                              
$\nu$  &   Transition & Notes\\                                                                            
(MHz)  &  (N$_J$;J,F) & \\                                                                              
\hline                                                                           {\hbox{\rm SO}}  &   \\                                                                                         
430339.544       &  10$_{10}$ --  9$_{ 9}$  & a) \\                                                                 
{\hbox{$^{34}{\rm SO}$}}  &   & \\                                                                                     
290562.238       &   7$_{ 6}$ --  6$_{ 5}$ & \\ 
337580.147       &   8$_{ 8}$ --  7$_{ 7}$ & b)\\   
{\hbox{$^{33}{\rm SO}$}} & & \\
337198.6      &   8$_{ 7}$ --  7$_{ 6}$ & \\    
\hline                                                                                               
\end{tabular} \\                                                                                     
Notes: (a) Blend with {\hbox{$^{34}{\rm SO}_2$}} at 430347.6. (b) Blend with {\hbox{${\rm CH}_3{\rm OH}$}} at 337581.6

\end{center}                                                                                         

\end{table*} }

\onltab{24}{

\begin{table*}                                                                                 
\begin{center}                                                                                       
\caption{Identified features of {\hbox{${\rm CH}_3{\rm OCH}_3$}} and {\hbox{${\rm CH}_2{\rm CHCN}$}}}                                                              
\label{app:trans_dimeet}
\begin{tabular}{ccc}                                                                                  
\hline 
\hline                                                                                              
$\nu$  &   Transition & Notes  \\                                                                            
(MHz)  &  (J$_{K_-,K_+}$)  & \\                                                                          
\hline      
 {\hbox{${\rm CH}_3{\rm OCH}_3$}} & & \\                                                                  
337420.9       &  21$_{2,19}AA$ -- 20$_{3,18}AA$   & * \\
337421.3       &  21$_{2,19}EE$ -- 20$_{3,18}EE$   & * \\
337421.8       &  21$_{2,19}AE$ -- 20$_{3,18}AE$   & * \\
337421.8       &  21$_{2,19}EA$ -- 20$_{3,18}EA$   & * \\
\\
337722.3       &   7$_{4,4}EE$ --  6$_{3,3}EE$   & * \\
337723.0       &   7$_{4,4}AE$ --  6$_{3,3}AE$   & * \\
\\
337730.7       &   7$_{4,4}AA$ --  6$_{3,3}AA$   & a) * \\
337731.9       &   7$_{4,3}EA$ --  6$_{3,3}EA$   &  *\\
337732.2       &   7$_{4,3}EE$ --  6$_{3,3}EE$   &  *\\
\\
337770.6       &   7$_{4,4}EA$ --  6$_{3,4}EA$   &  b)\\
\\
337778.0       &   7$_{4,4}EE$ --  6$_{3,4}EE$   & * \\
337779.5       &   7$_{4,3}AE$ --  6$_{3,4}AE$   & * \\
\\
337787.2       &   7$_{4,3}AA$ --  6$_{3,4}AA$   & * \\
337787.9       &   7$_{4,3}EE$ --  6$_{3,4}EE$   & * \\
337790.1       &   7$_{4,3}EA$ --  6$_{3,4}EA$   & * \\
{\hbox{${\rm CH}_2{\rm CHCN}$}} & & \\
337039.730	 &  36$_{2,35}$ -- 35$_{2,34}$   & c) \\
\hline                                                                                               
\end{tabular} \\ 
Notes: (*) Blended with each other. (a) Blend with {\hbox{${\rm C}_2{\rm H}_5{\rm OH}$}} at 337727.1. (b) bad baseline \\
c) Blend with wing of {\hbox{\rm C$^{17}$}}O at 337061.1                                                                       
\end{center}                                                                                         

\end{table*}  }

\onltab{25}{
\begin{table*}                                                                                  
\begin{center}                                                                                                                         
\caption{Identified features of {\hbox{${\rm CH}_3{\rm CN}$}}}                                                                                                     
\label{app:trans_methcn}
\begin{tabular}{ccc}                                                                                  

\hline 
\hline                                                                          

$\nu$ & Transition & Notes\\
(MHz)  &  (J$_{K_-,K_+}$) & \\

{\hbox{${\rm CH}_3{\rm CN}$}}  &  &  \\

294025.5  &  16$_{7}$ -- 15$_{7}$  &  \\
294098.9  &  16$_{6}$ -- 15$_{6}$  &  \\
294161.0  & 16$_{5}$ -- 15$_{5}$   &  \\
294211.9  & 16$_{4}$ -- 15$_{4}$   &  \\
294251.4   & 16$_{3}$ -- 15$_{3}$   & \\
294279.7  & 16$_{2}$ -- 15$_{2}$   &  \\
\\
294296.7  & 16$_{1}$ -- 15$_{1}$   & * \\
294302.4   & 16$_{0}$ -- 15$_{0}$   & *\\
\\      
348911.4       &  19$_{9}$ -- 18$_{9}$ &  \\
349125.3       &  19$_{7}$ -- 18$_{7}$ & a) \\
349212.3       &  19$_{6}$ -- 18$_{6}$ &  \\
349286.0       &  19$_{5}$ -- 18$_{5}$ & \\
349346.3       &  19$_{4}$ -- 18$_{4}$ & b)  \\
349393.3       &  19$_{3}$ -- 18$_{3}$ & c) \\
349426.8       &  19$_{2}$ -- 18$_{2}$ &  \\
\\
349448.5       &  19$_{1}$ -- 18$_{1}$ & d) * \\
349452.1       &  19$_{0}$ -- 18$_{0}$ &  *\\
\\
{\hbox{${\rm CH}_3{\rm CN}$}}(v$_8$=1)  & &  \\  
350168.1       &  19$_{1}$ -- 18$_{-1}$  & \\
350423.6       &  19$_{2}$ -- 18$_{2}$  &\\
\\
350444.8       &  19$_{0}$ -- 18$_{0}$ & * \\
350449.6       &  19$_{1}$ -- 18$_{1}$ & *  \\
\\
350465.6       &  19$_{4}$ -- 18$_{4}$ &  \\
350507.1       &  19$_{3}$ -- 18$_{3}$ &  \\
350552.4       &  19$_{2}$ -- 18$_{-2}$ & e) \\
\\
{\hbox{${\rm CH}_3^{13}{\rm CN}$}}  &  & \\                        
294066.6	& 16$_{4}$ -- 15$_{4}$ & \\
294106.1	& 16$_{3}$ -- 15$_{3}$ & \\
294134.2	&16$_{2}$ -- 15$_{2}$  & \\
\\
294151.1	&16$_{1}$ -- 15$_{1}$  & *  \\
294156.7 	& 16$_{0}$ -- 15$_{0}$ & f) *\\
\\
349220.6   &  19$_{3}$ -- 18$_{3}$ &  \\
349254.0   &  19$_{2}$ -- 18$_{2}$ & g)  \\
\\
349274.1   &  19$_{1}$ -- 18$_{1}$  & h) * \\
349280.7   &  19$_{0}$ -- 18$_{0}$  & * \\

\hline        
\end{tabular} \\ 

Notes: (*) Blended with each other. (a) Blend with {\hbox{${\rm CH}_3{\rm OH}$}} at 338475.290.  (b) Blend with CCH at 349337.7. \\ 
(c) Blend with {\hbox{${\rm C}_2{\rm H}_5{\rm CN}$}} at 349392.2 and CCH at 349399.3.\\
(d) Blend with {\hbox{${\rm C}_2{\rm H}_5{\rm CN}$}} at 349442.9.\\
(e) Blend with wing of {\hbox{\rm C$^{17}$}}O at 337061.1. \\
(f) Blend with {\hbox{${\rm CH}_3{\rm CN}$}} at 294161.0. (g) Blend with {\hbox{${\rm CH}_3{\rm OH}$}} at 338344.6.\\
(h) Blend with {\hbox{${\rm CH}_3{\rm CN}$}} at 349286.0.                                                                            
\end{center}                                                                                         

\end{table*}   }

\onltab{26}{

\begin{table*}                                                                                  
\begin{center}                                                                                       
\caption{Identified features of {\hbox{${\rm CH}_3{\rm CCH}$}}}                                                
\label{app:ch3cch}
\begin{tabular}{ccc}                                                                                  
\hline 
\hline 

$\nu$  &   Transition & Notes\\                                                                            
(MHz)  &  (J) &\\                                                                 
\hline                     
290452.2      &  17$_3$--16$_3$     &       \\					 
290479.9       &  17$_2$--16$_2$     &      \\   
290496.5       &  17$_1$--16$_1$     &     \\  
290502.1       &  17$_0$--16$_0$     &     \\  
\hline                                                                                               
\end{tabular} \\                                                                                     
\end{center}                                                                                         

\end{table*} }                           

\onltab{27}{                  
\begin{table*}                                                                                
\begin{center}                                                                                       
\caption{Identified features of {\hbox{{\rm CCH}}}  }                                                                
\label{app:cch}
\begin{tabular}{ccc}                                                                                  
\hline
\hline

$\nu$  &   Transition & Notes  \\                                                                            
(MHz)  &  (N,J,F) & \\                                                                                  
\hline                                                                                               
349337.7      &   4,$\frac{9}{2}$,$\frac{9}{2}$ --  3,$\frac{7}{2}$,$\frac{7}{2}$  &  a)* \\   
349339.1       &   4,$\frac{9}{2}$,$\frac{7}{2}$ --  3,$\frac{7}{2}$,$\frac{5}{2}$  &  * \\   

349399.3       &   4,$\frac{7}{2}$,$\frac{7}{2}$ --  3,$\frac{5}{2}$,$\frac{5}{2}$  & b)*  \\   
349400.7       &   4,$\frac{7}{2}$,$\frac{5}{2}$ --  3,$\frac{5}{2}$,$\frac{3}{2}$  &  * \\   

436661.0       &   5,$\frac{11}{2}$,$\frac{11}{2}$ --  4,$\frac{9}{2}$,$\frac{9}{2}$ &  *  \\   
436661.9       &   5,$\frac{11}{2}$,$\frac{9}{2}$ --  4,$\frac{9}{2}$,$\frac{7}{2}$ &   * \\	

436723.1       &   5,$\frac{9}{2}$,$\frac{9}{2}$ --  4,$\frac{7}{2}$,$\frac{7}{2}$  & *  \\   
436723.9       &   5,$\frac{9}{2}$,$\frac{7}{2}$ --  4,$\frac{7}{2}$,$\frac{5}{2}$  & *  \\   
\hline                                                                                               
\end{tabular} \\   
Notes: (*) Blended with each other. (a) Blend with {\hbox{${\rm CH}_3{\rm CN}$}} at 349346.3.\\(b) Blend with {\hbox{${\rm CH}_3{\rm CN}$}} at 349393.3                                                                               
\end{center}                                                                                         

\end{table*} }

\onltab{28}{

\begin{table*}                                                                              
\begin{center}                                                                                       
\caption{ Identified features of {\hbox{${\rm C}_2{\rm H}_5{\rm CN}$}}}
\label{app:trans_etcn1}
\begin{tabular}{ccc}                                                                                  
\hline
\hline                                                                                               
$\nu$  &   Transition & Notes  \\                                                                            
(MHz)  &  (J$_{K_-,K_+}$) & \\

\hline              
{\hbox{${\rm C}_2{\rm H}_5{\rm CN}$}}(v=0) &  &\\
337441.7      &  13$_{6,17}$ -- 14$_{4,10}$  & * \\
337445.9       &  37$_{4,33}$ -- 36$_{4,32}$ & *  \\
\\
349547.0       &  39$_{9,31}$ -- 38$_{9,30}$ &  *\\
349547.0       &  39$_{9,30}$ -- 38$_{9,39}$  &* \\
\\
349730.8       &  39$_{8,32}$ -- 38$_{8,31}$ & * \\
349731.3       &  39$_{8,31}$ -- 38$_{8,30}$ & * \\
\\
349796.0       &  39$_{4,36}$ -- 38$_{4,35}$ & a) \\

350139.6       &  41$_{1,41}$ -- 40$_{1,40}$  & \\

350145.1       &  41$_{0,41}$ -- 40$_{0,40}$ &  \\

{\hbox{${\rm C}_2{\rm H}_5{\rm CN}$}}(v=1) & &\\
 349790.2      &  30$_{3,8}$ -- 29$_{2,7}$ & a) * \\   
 349791.1       &  30$_{2,8}$ -- 29$_{3,7}$ & * \\   
\hline
\end{tabular} \\                                                                                     
Notes: (*) Blended with each other.  (a) Blend with {\hbox{${\rm CH}_3{\rm OCH}_3$}} at 349796.6

\end{center}                                                                                         

\end{table*} }

\onlfig{10}{
\begin{figure*}
\centerline{
\includegraphics[clip,trim= 0 0 0 0,angle=-90,scale=0.37]{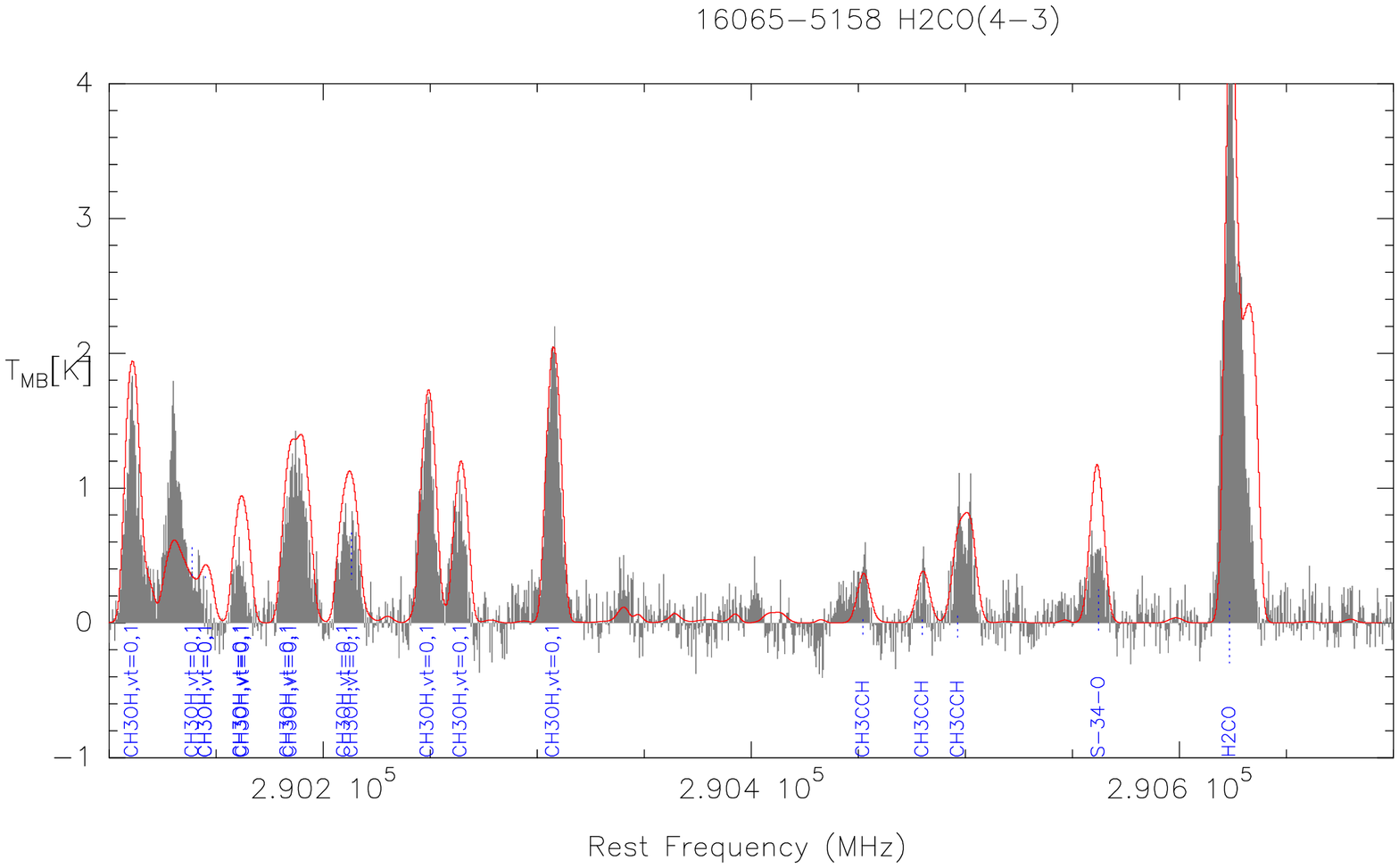}
\includegraphics[clip,trim= 0 0 0 0,angle=-90,scale=0.37]{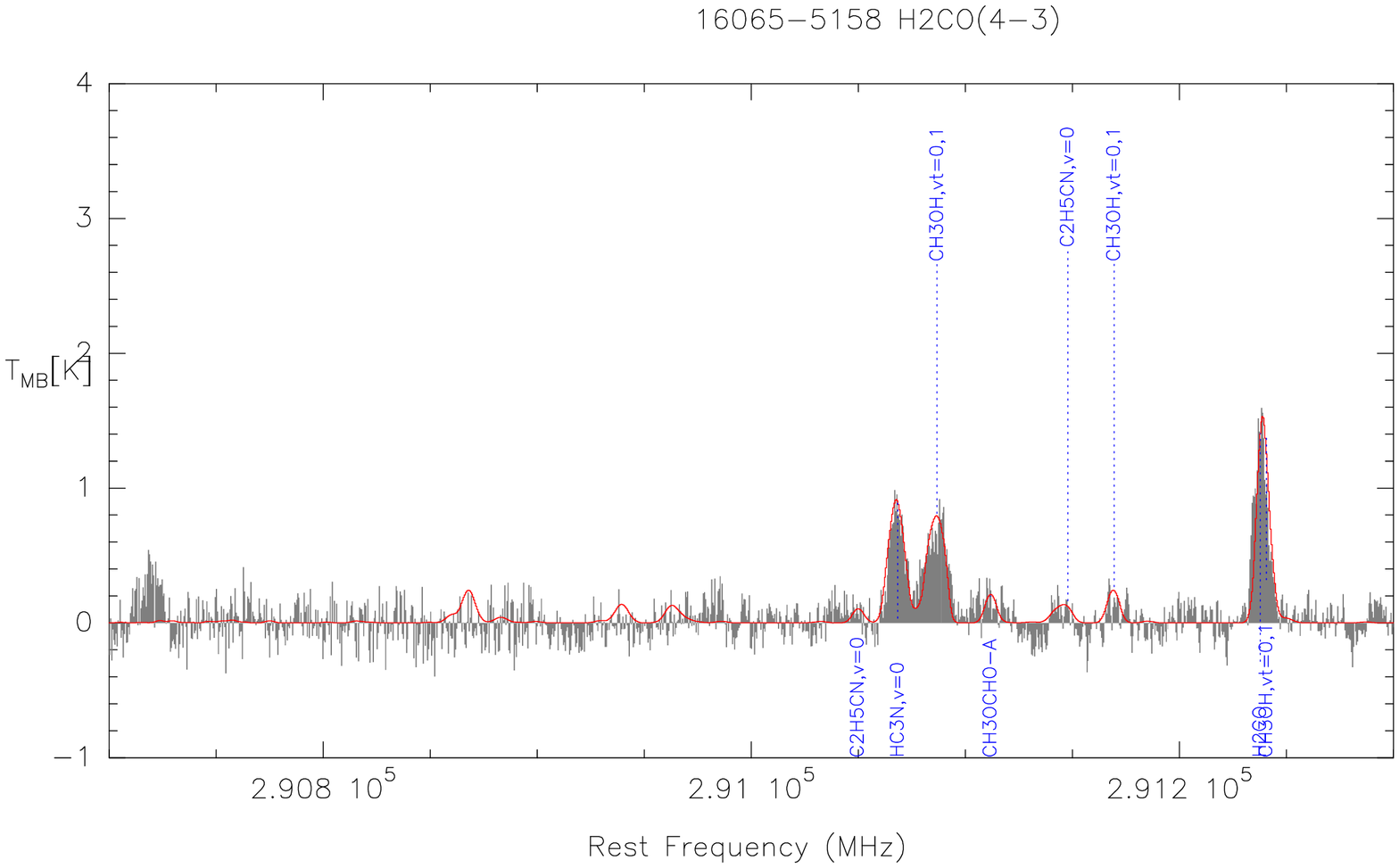}}
\centerline{
\includegraphics[clip,trim= 0 0 0 0,angle=-90,scale=0.37]{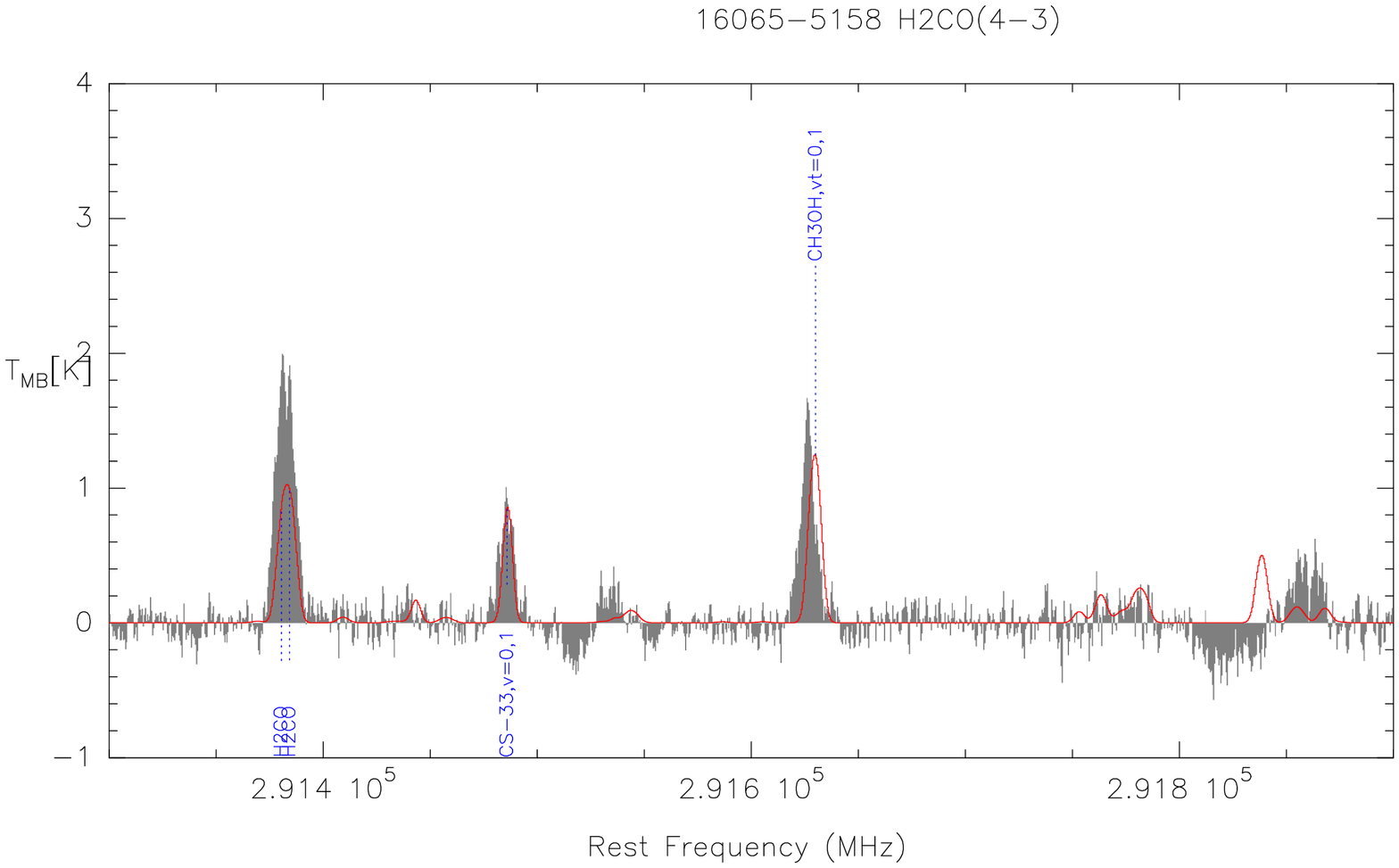}
\includegraphics[clip,trim= 0 0 0 30,angle=-90,scale=0.37]{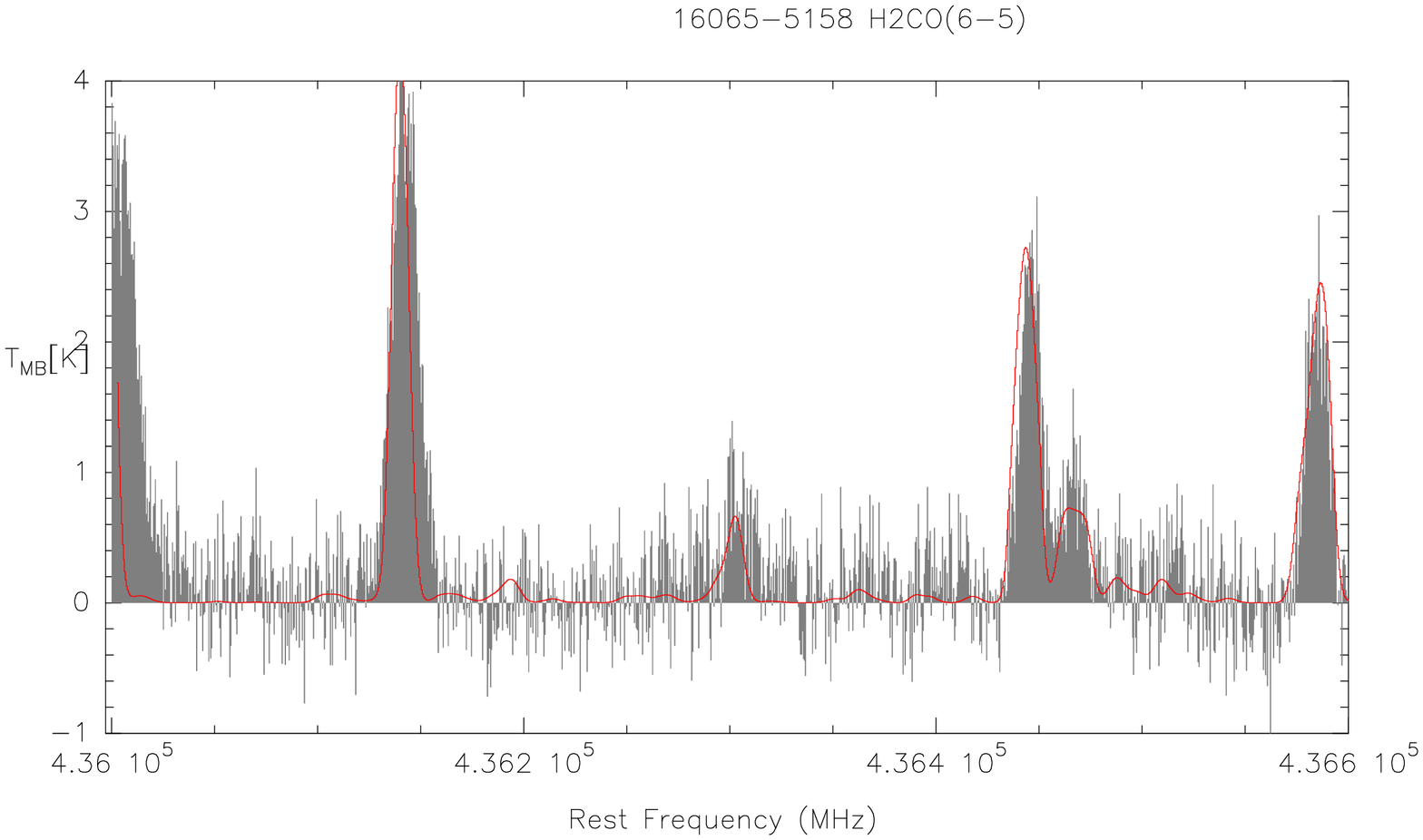}}
\centerline{
\includegraphics[clip,trim= 0 0 0 30,angle=-90,scale=0.37]{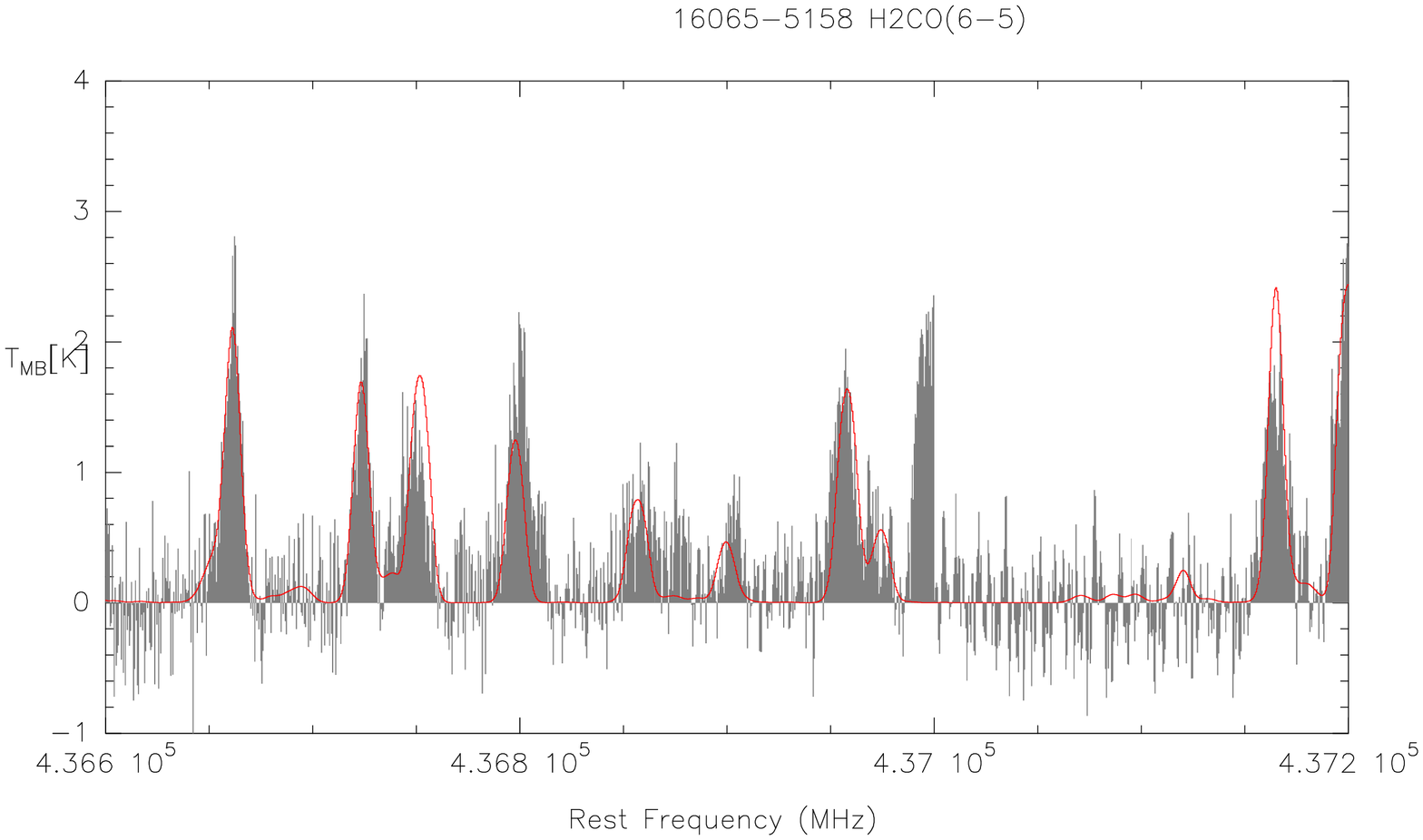}
\includegraphics[clip,trim= 0 0 0 0,angle=-90,scale=0.37]{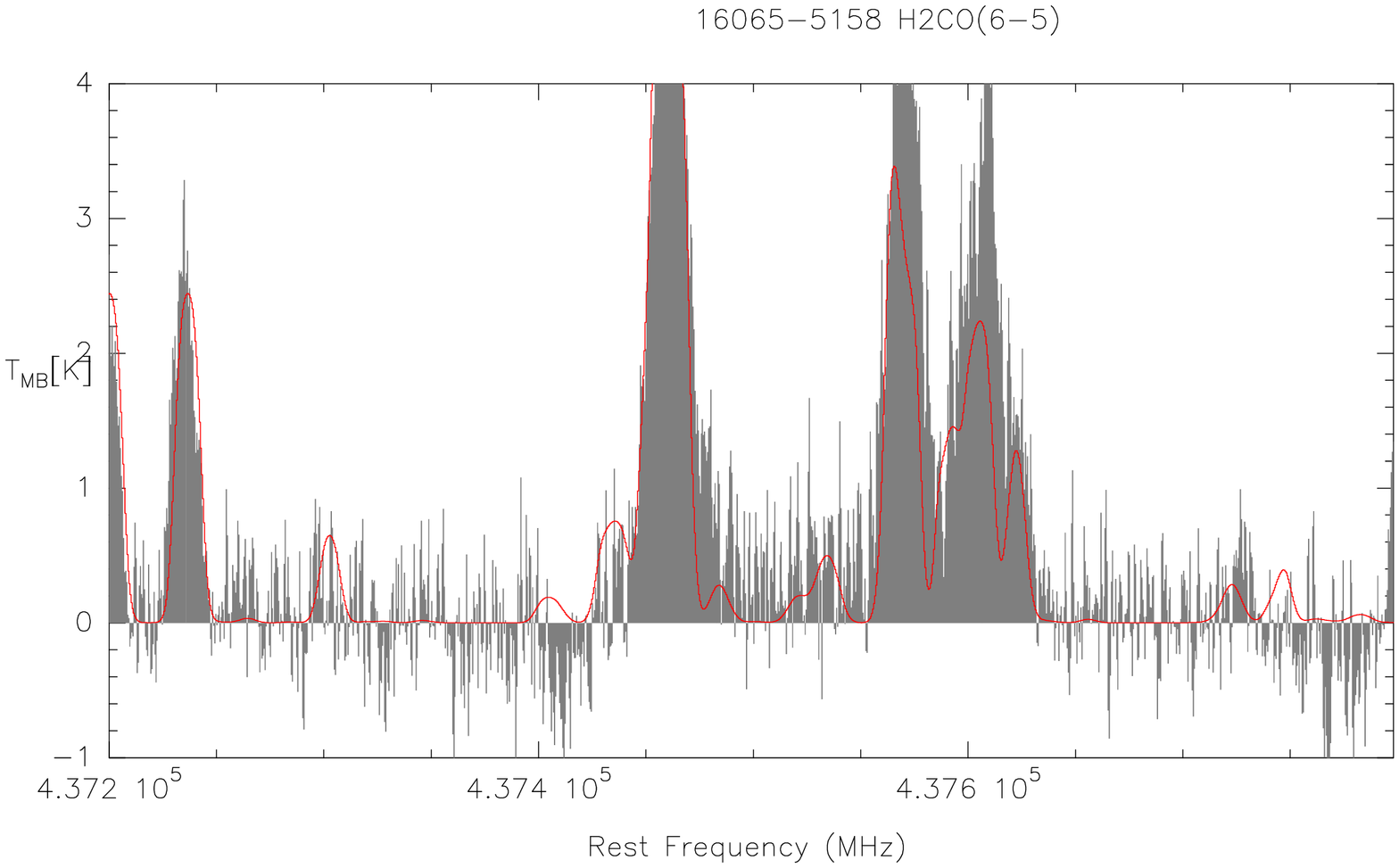}}
\caption{\label{fig:molem1a} Molecular emission of 16065-5158 around 290~GHz
 and 436~GHz. The best-fit synthetic spectrum is overplotted (solid line).  }
\end{figure*} }

\onlfig{11}{
\begin{figure*}
\centerline{
\includegraphics[clip,trim=0 0 0 30,angle=-90,scale=0.37]{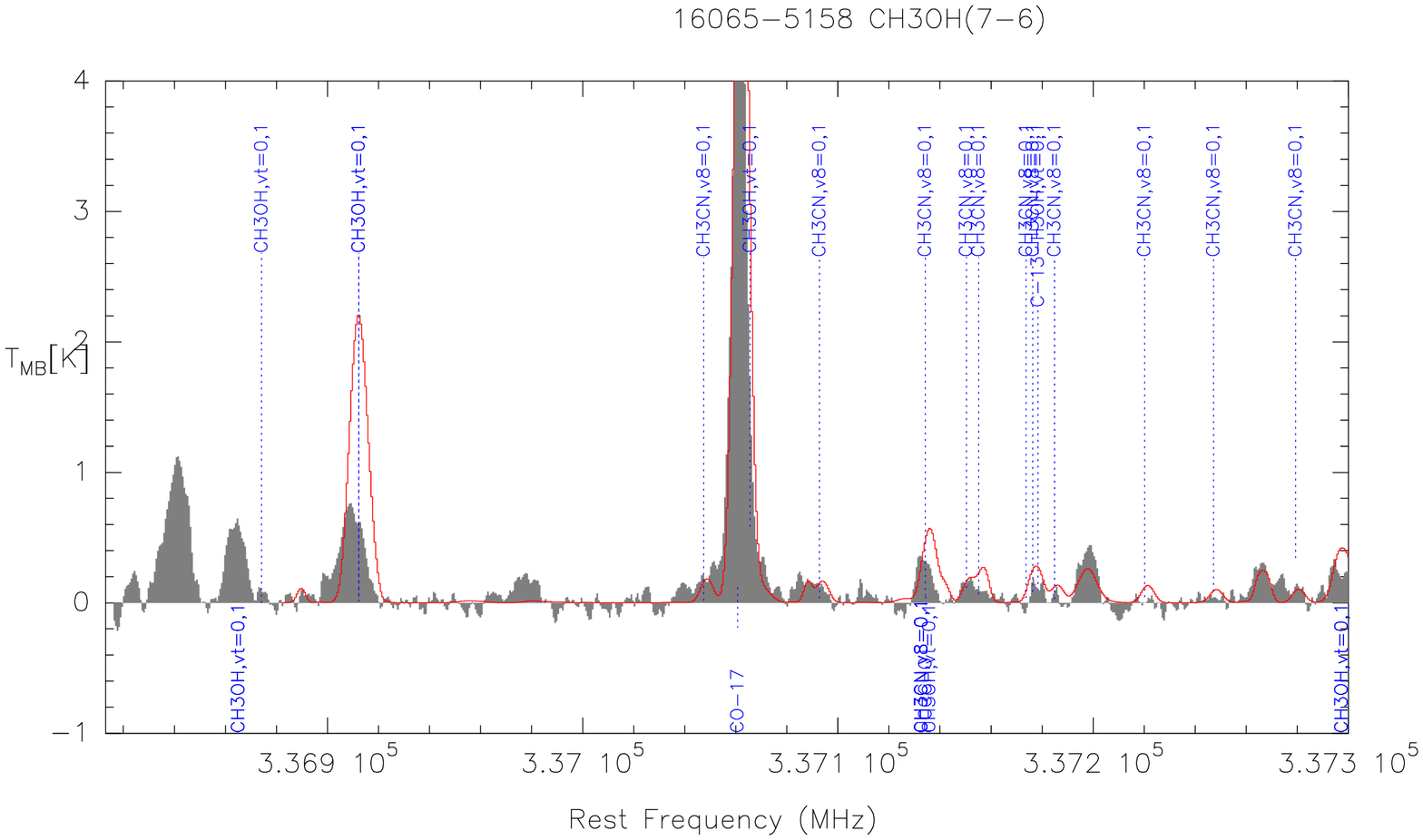}
\includegraphics[clip, trim=0 0 0 30,scale=0.37, angle=-90]{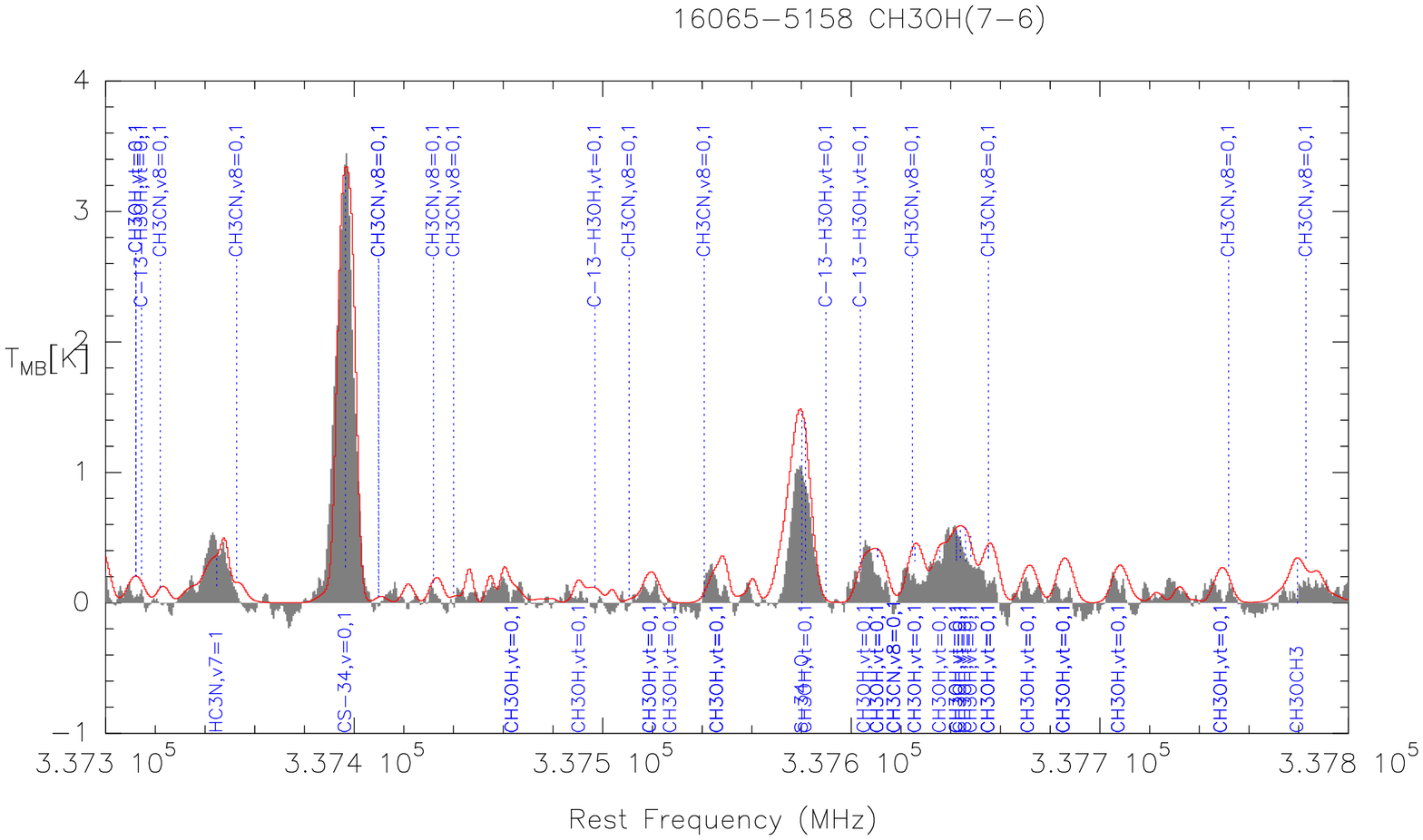}}
\centerline{
\includegraphics[clip, trim=0 0 0 30,scale=0.37, angle=-90]{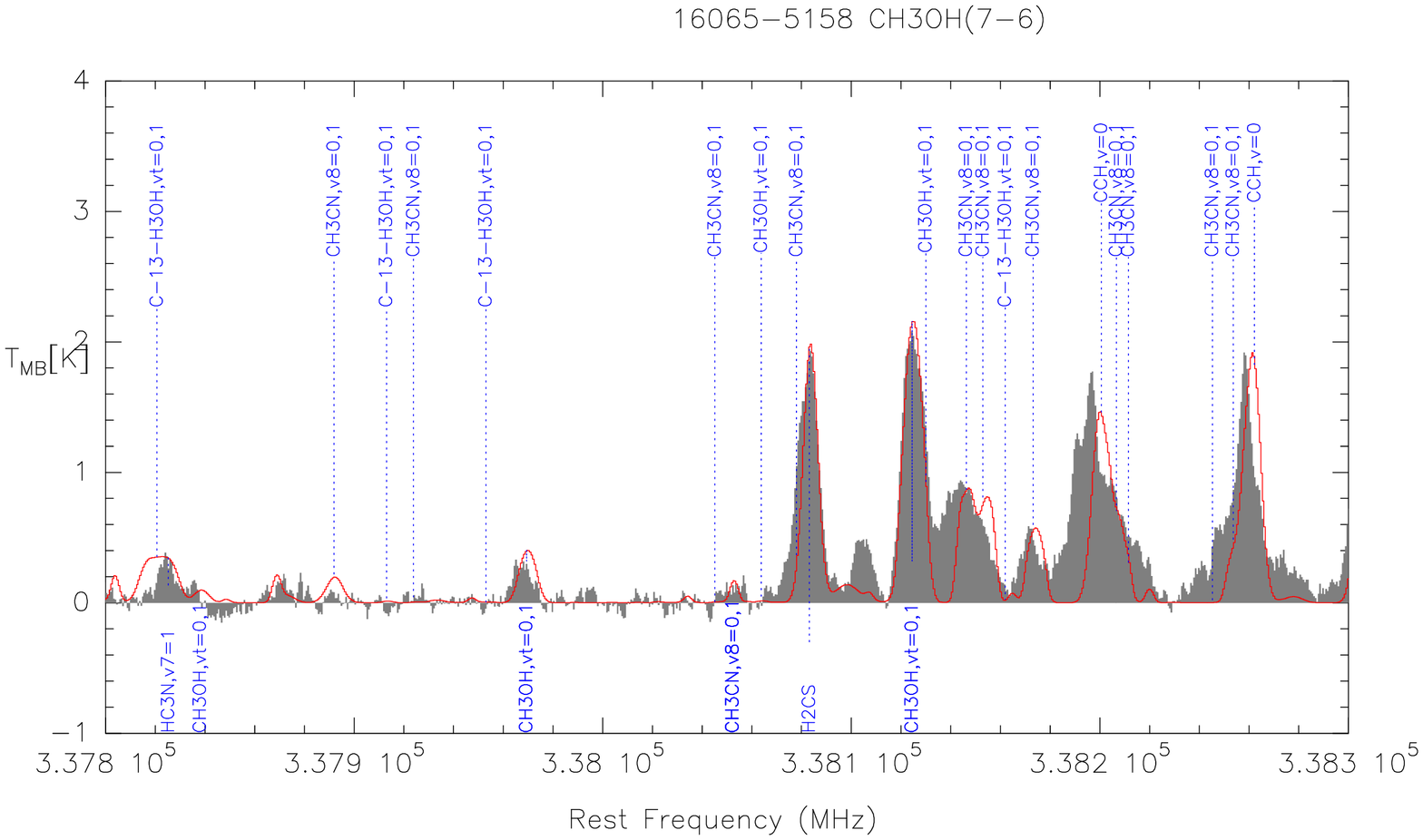}
\includegraphics[clip, trim=0 0 0 0,scale=0.37, angle=-90]{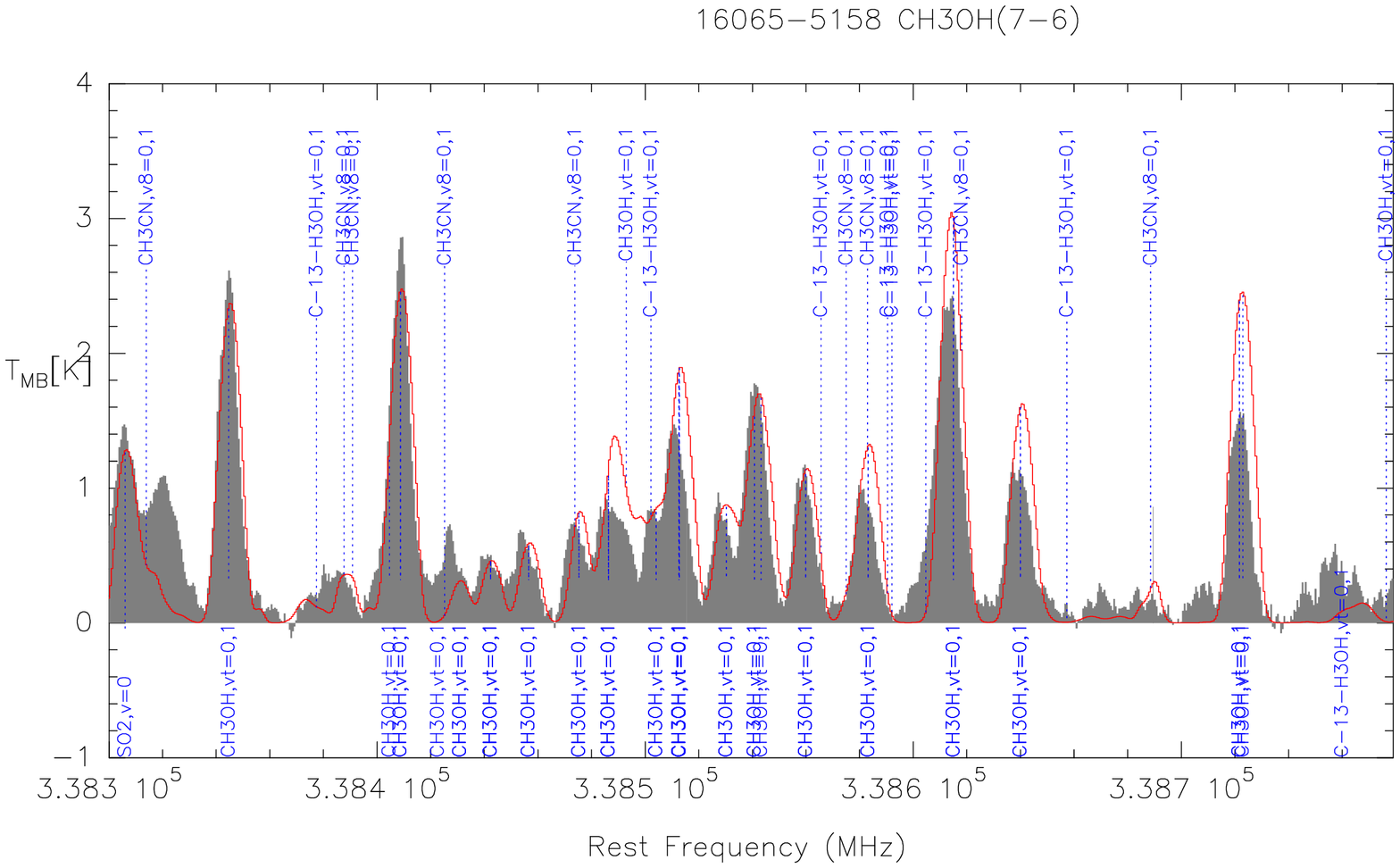}}
\caption{\label{fig:molem1b} Molecular emission of 16065-5158 around 338~GHz. The best-fit synthetic spectrum is overplotted (solid line).  }
\end{figure*} }

\onlfig{12}{
\begin{figure*}
\centerline{
\includegraphics[clip,trim=0 0 0 0, angle=-90,scale=0.37]{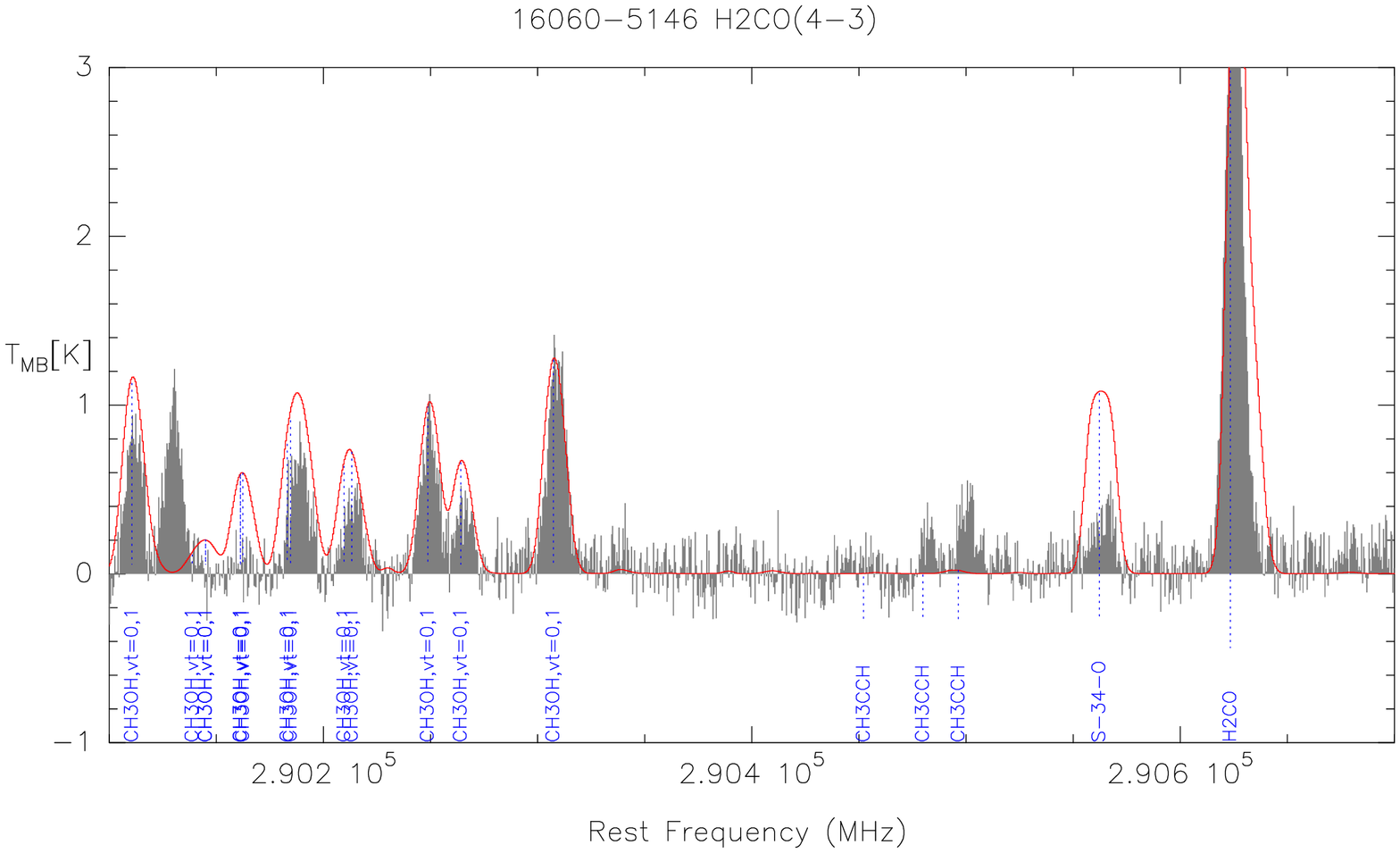}
\includegraphics[clip,trim=0 0 0 0, angle=-90,scale=0.37]{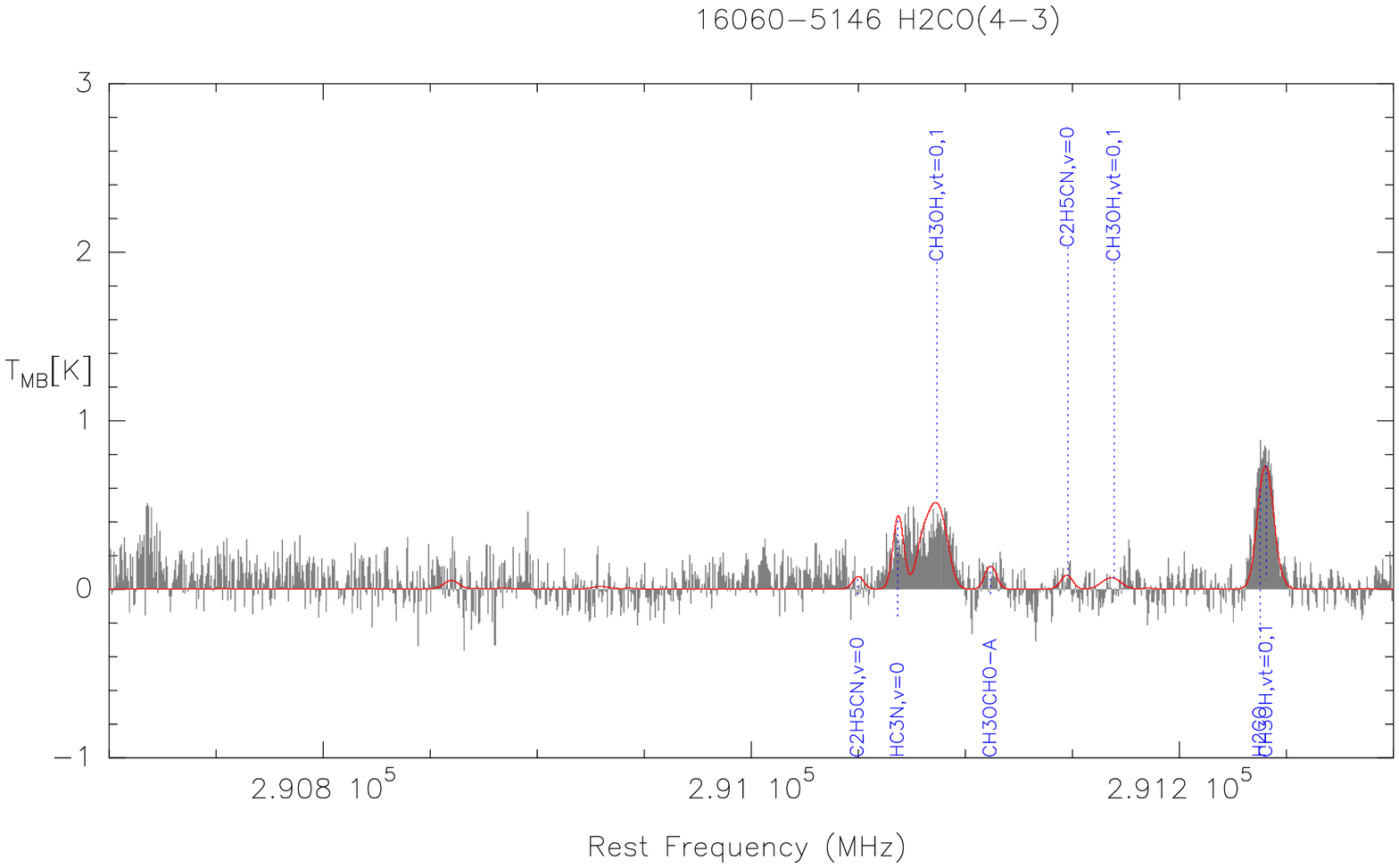}}
\centerline{
\includegraphics[clip,trim= 0 0 0 0, angle=-90,scale=0.37]{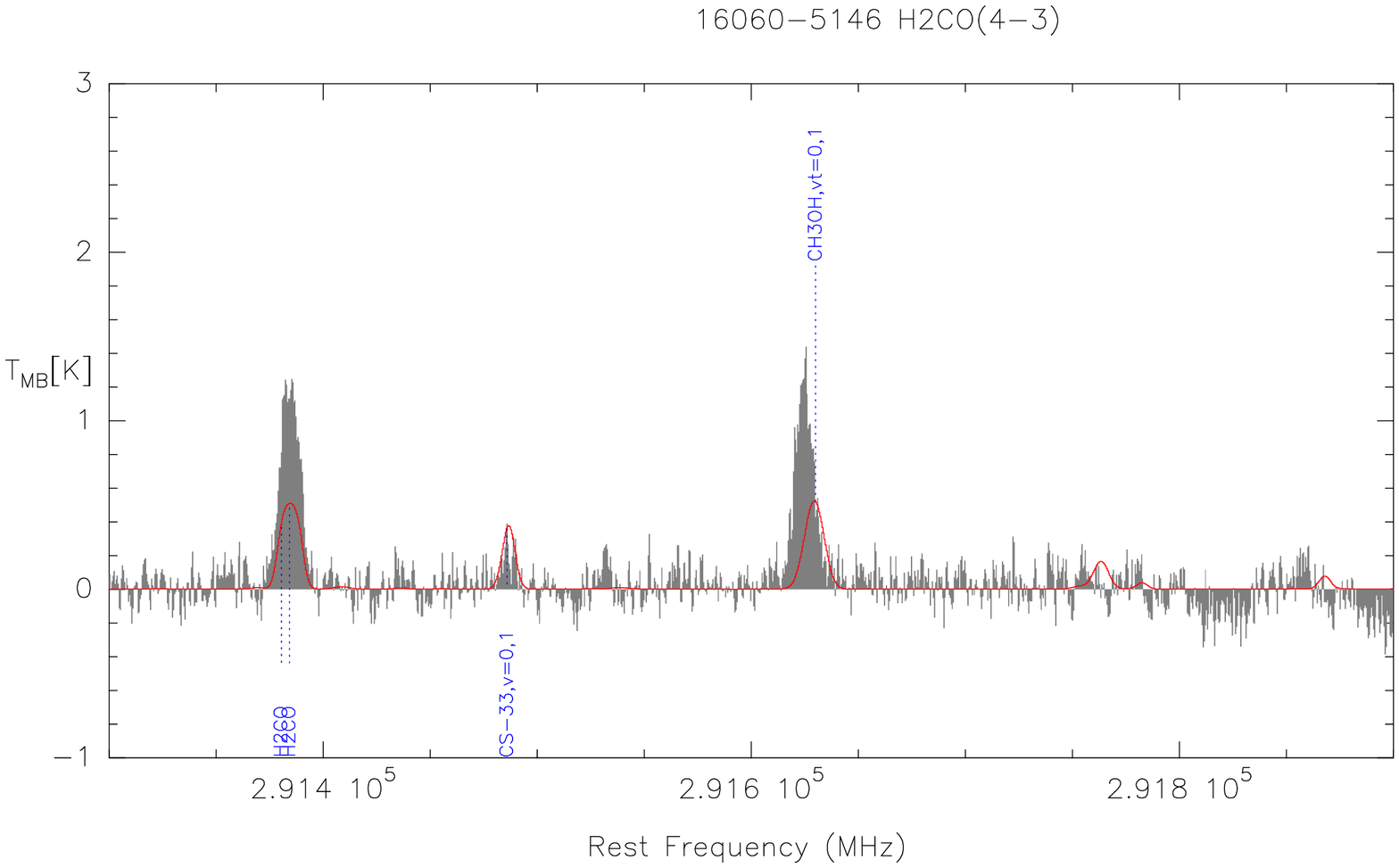}
\includegraphics[clip,trim= 0 0 0 30, angle=-90,scale=0.37]{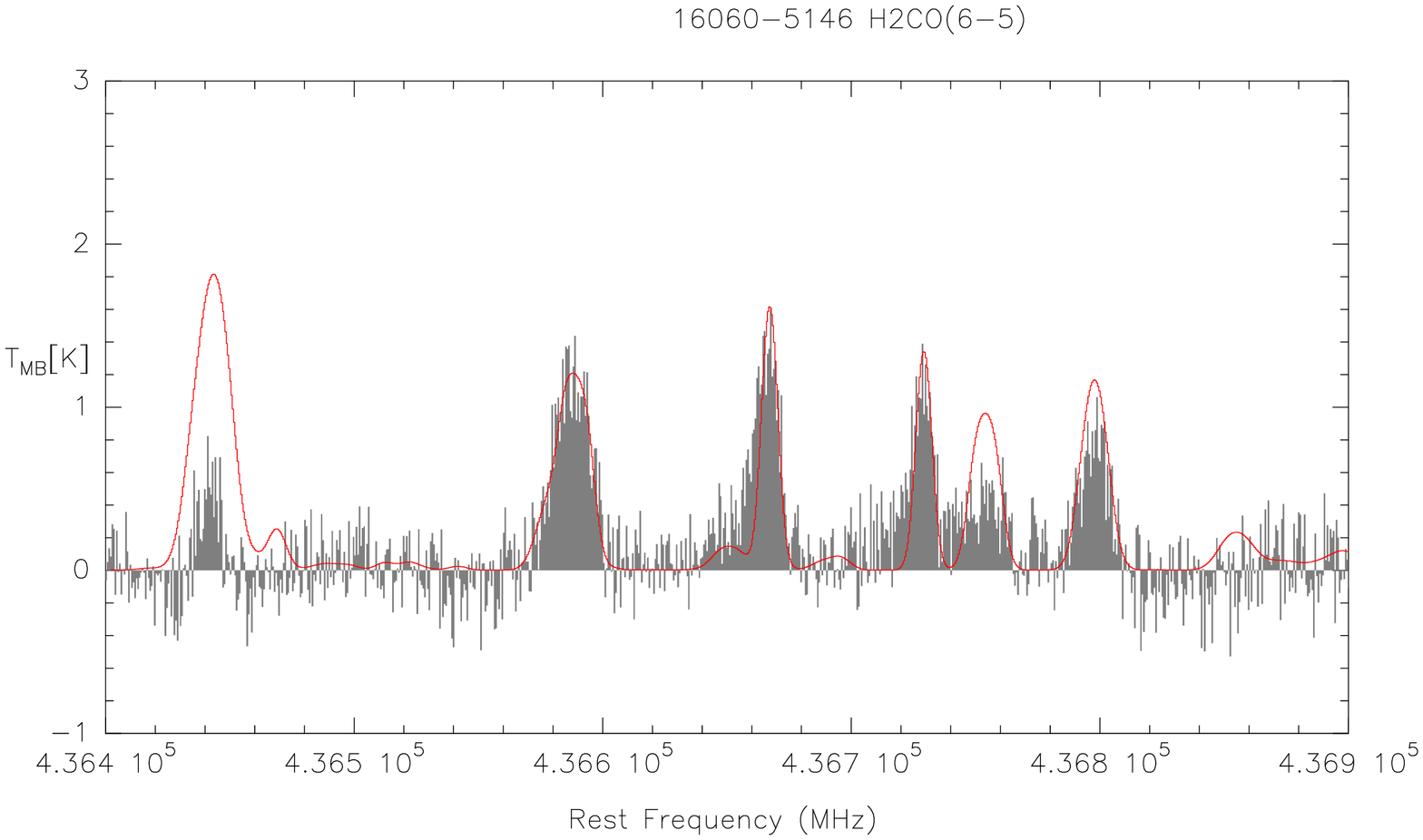}}
\centerline{
\includegraphics[clip,trim= 0 0 0 30,angle=-90,scale=0.37]{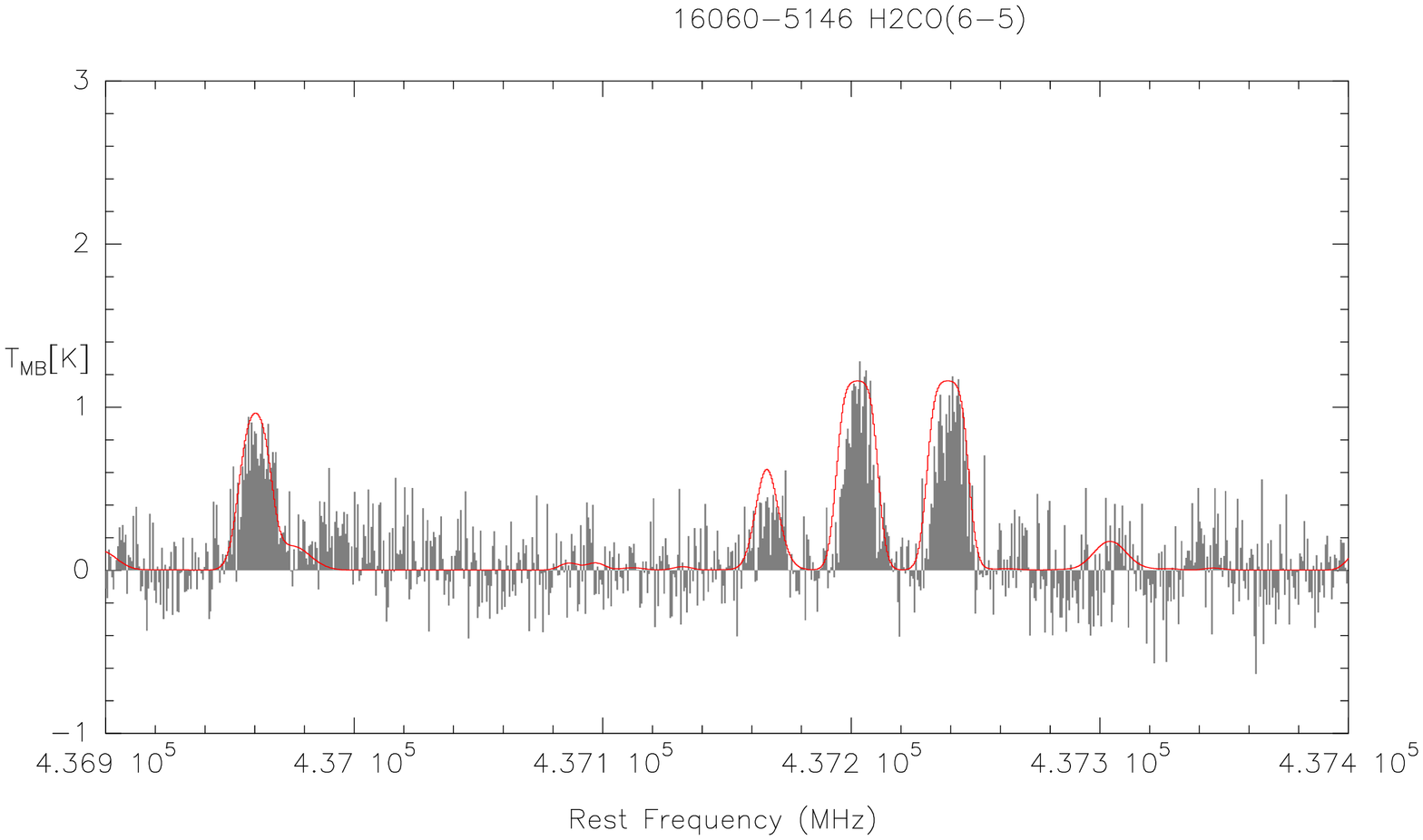}
}
\caption{\label{fig:molem2a} Molecular emission of 16060-5146 around 290~GHz
 and 436~GHz. The best-fit synthetic spectrum is overplotted (solid line). }
\end{figure*} }

\onlfig{13}{
\begin{figure*}
\centerline{
\includegraphics[clip,trim=0 0 0 30, angle=-90,scale=0.37]{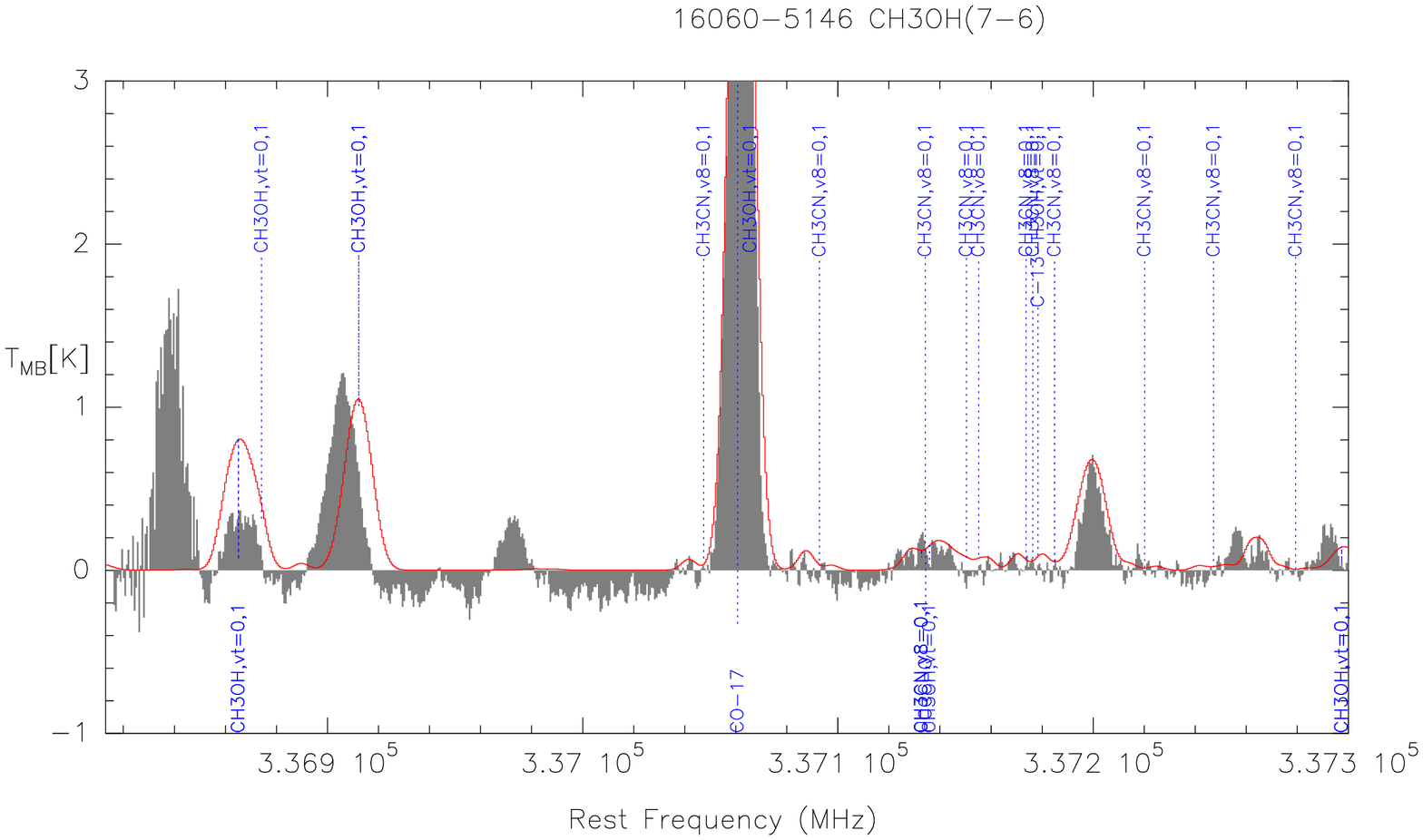}
\includegraphics[clip,trim=0 0 0 30,angle=-90,scale=0.37]{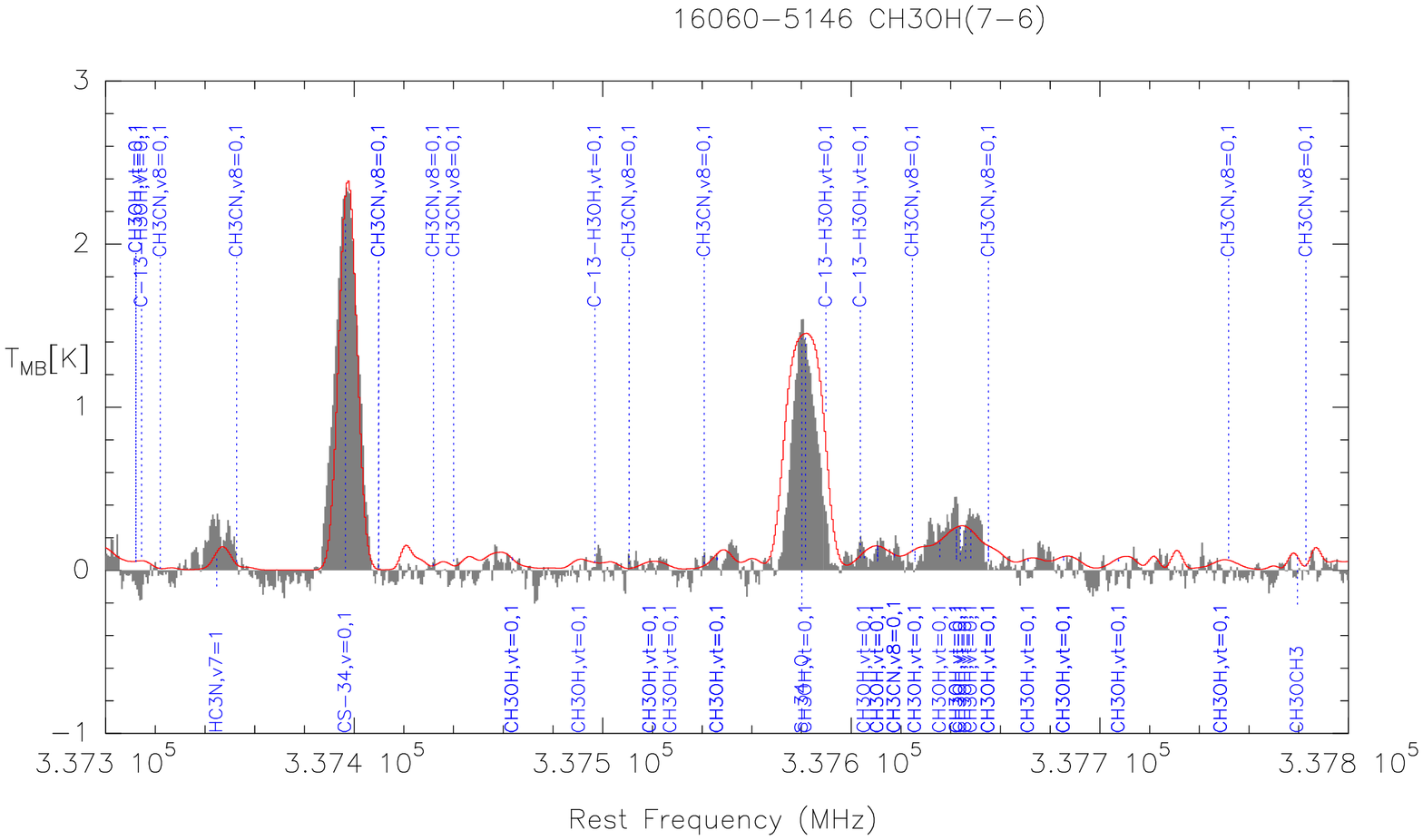}}
\centerline{
\includegraphics[clip,trim= 0 0 0 30, angle=-90,scale=0.37]{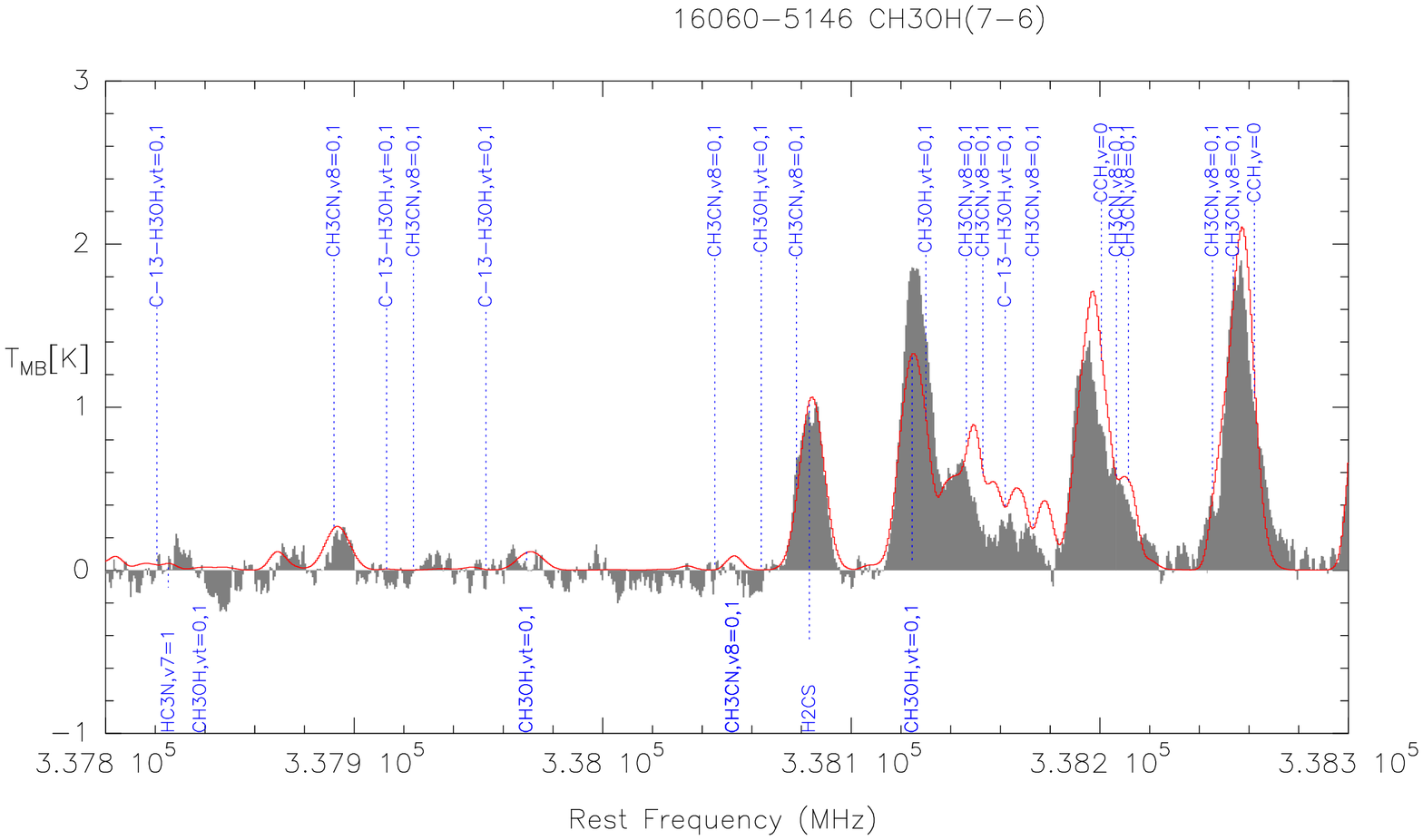}
\includegraphics[clip,trim= 0 0 0 0, angle=-90,scale=0.37]{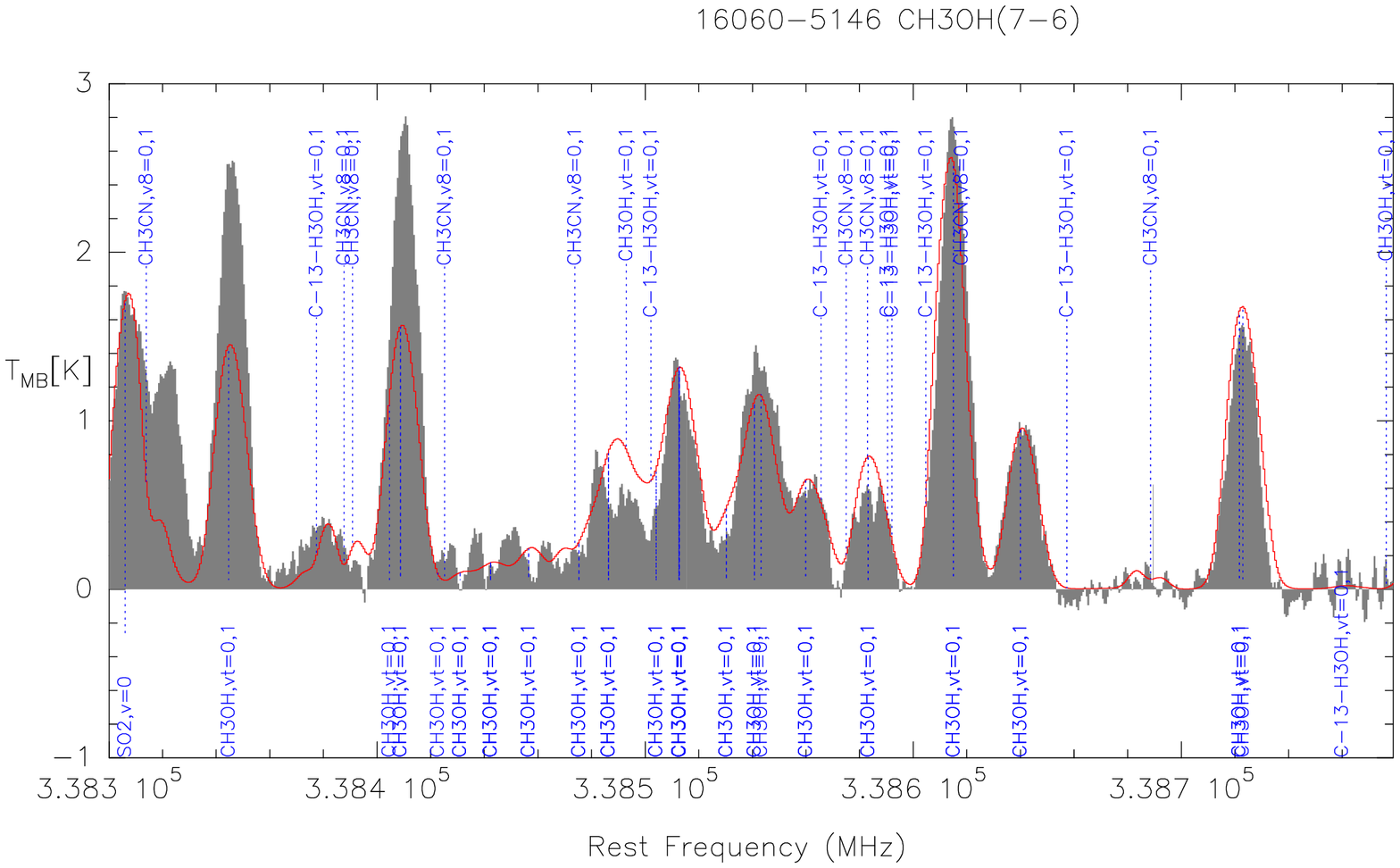}}
\caption{\label{fig:molem2b} Molecular emission of 16060-5146 around 338~GHz. The best-fit synthetic spectrum is overplotted (solid line). }
\end{figure*} }

\onlfig{14}{
\begin{figure*}
\centerline{
\includegraphics[angle=-90,scale=0.4]{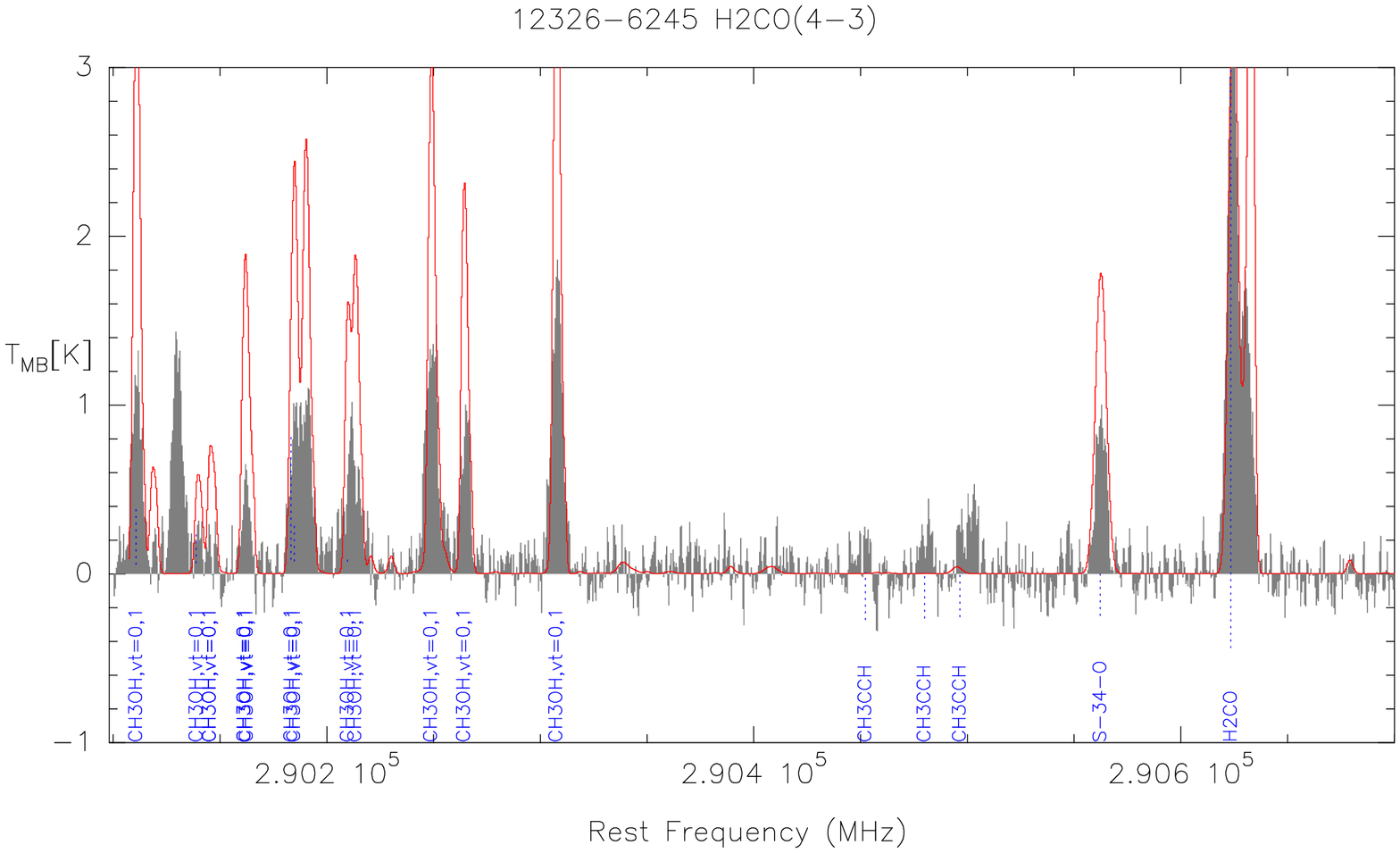} }
\centerline{
\includegraphics[angle=-90,scale=0.4]{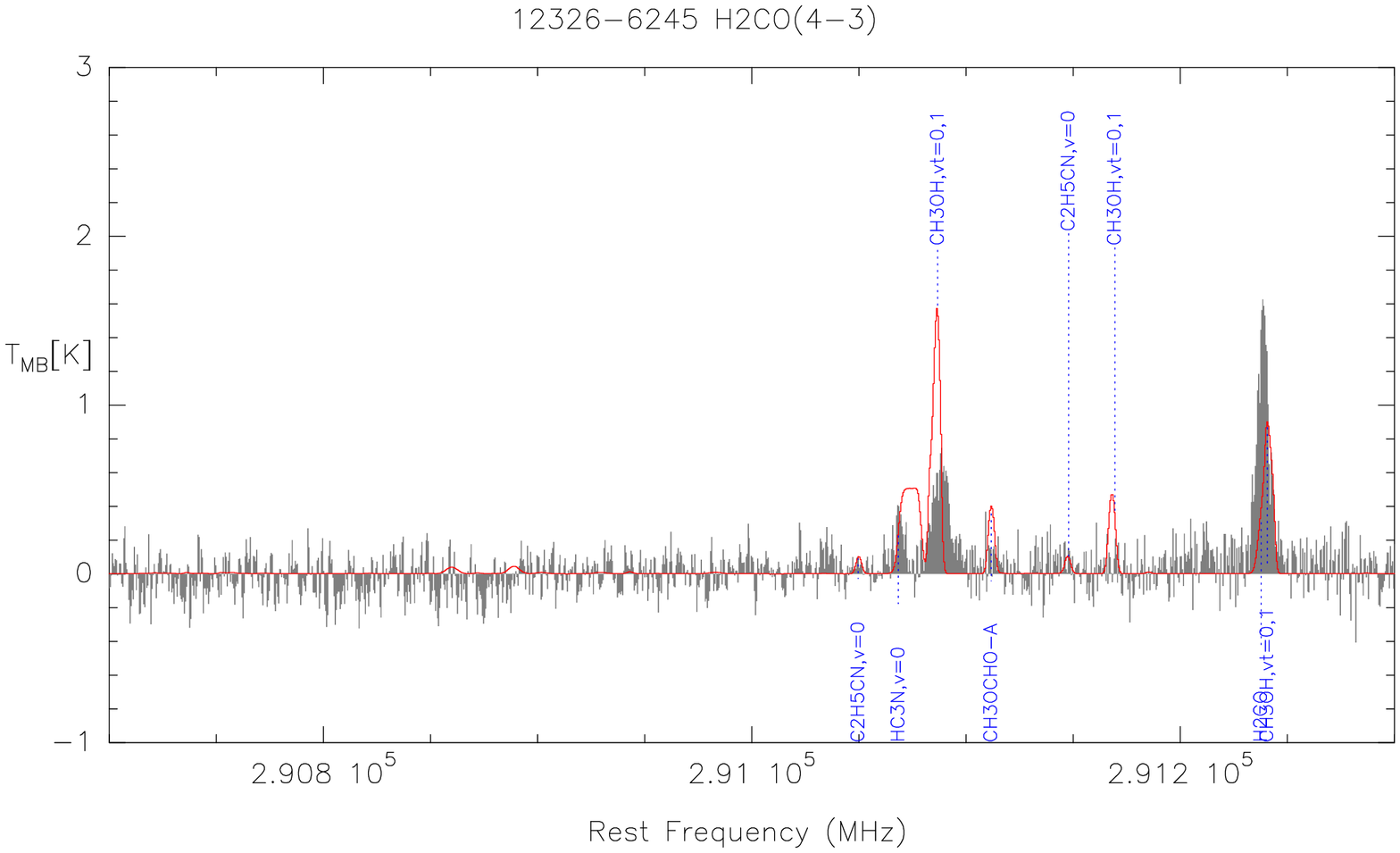} }
\centerline{
\includegraphics[angle=-90,scale=0.4]{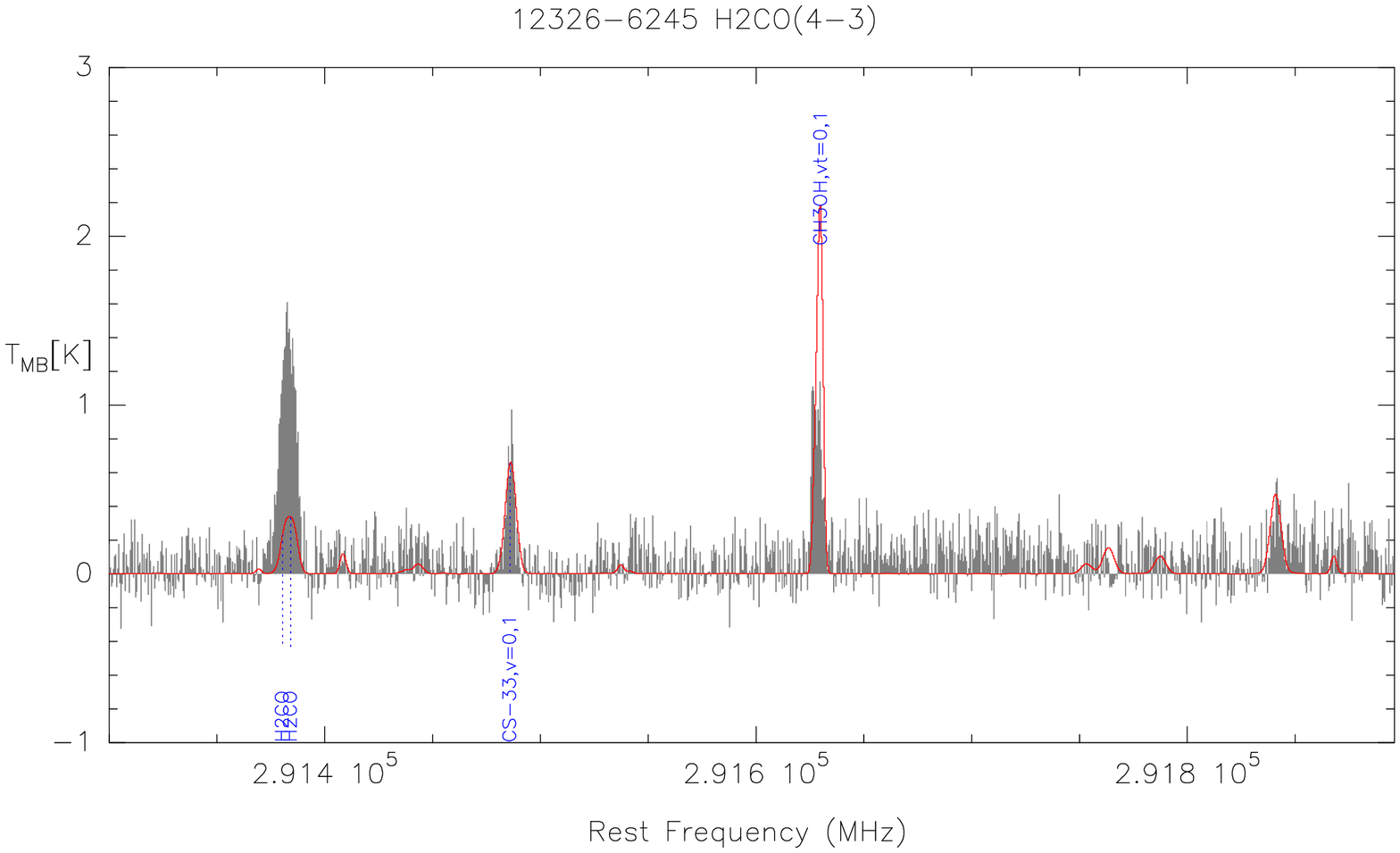} }
\caption{\label{fig:molem3a} Molecular emission of 12326-6245 around 290~GHz. The best-fit synthetic spectrum is overplotted (solid line). }
\end{figure*} }

\onlfig{15}{
\begin{figure*}
\centerline{
\includegraphics[angle=-90,scale=0.37]{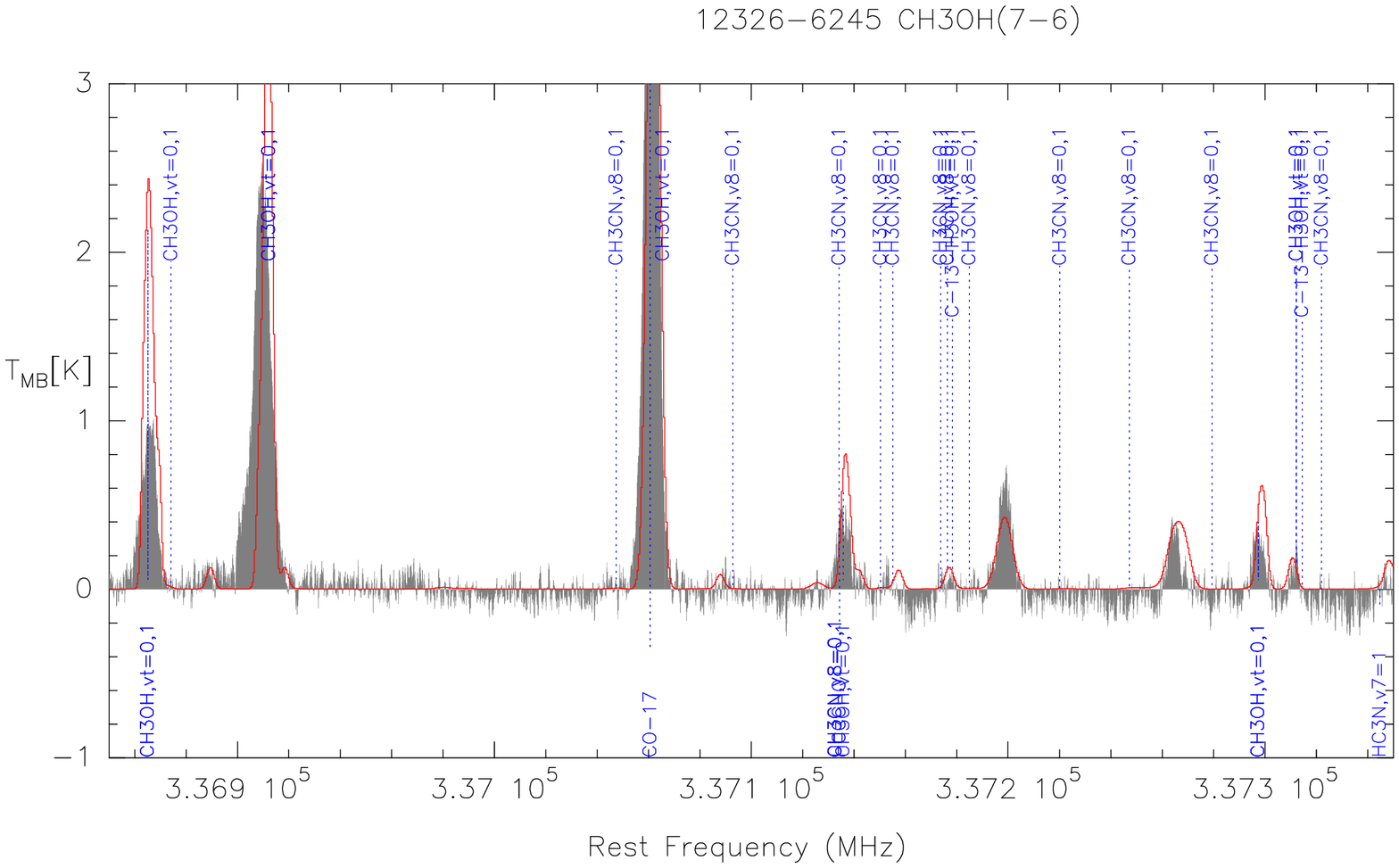}
\includegraphics[angle=-90,scale=0.37]{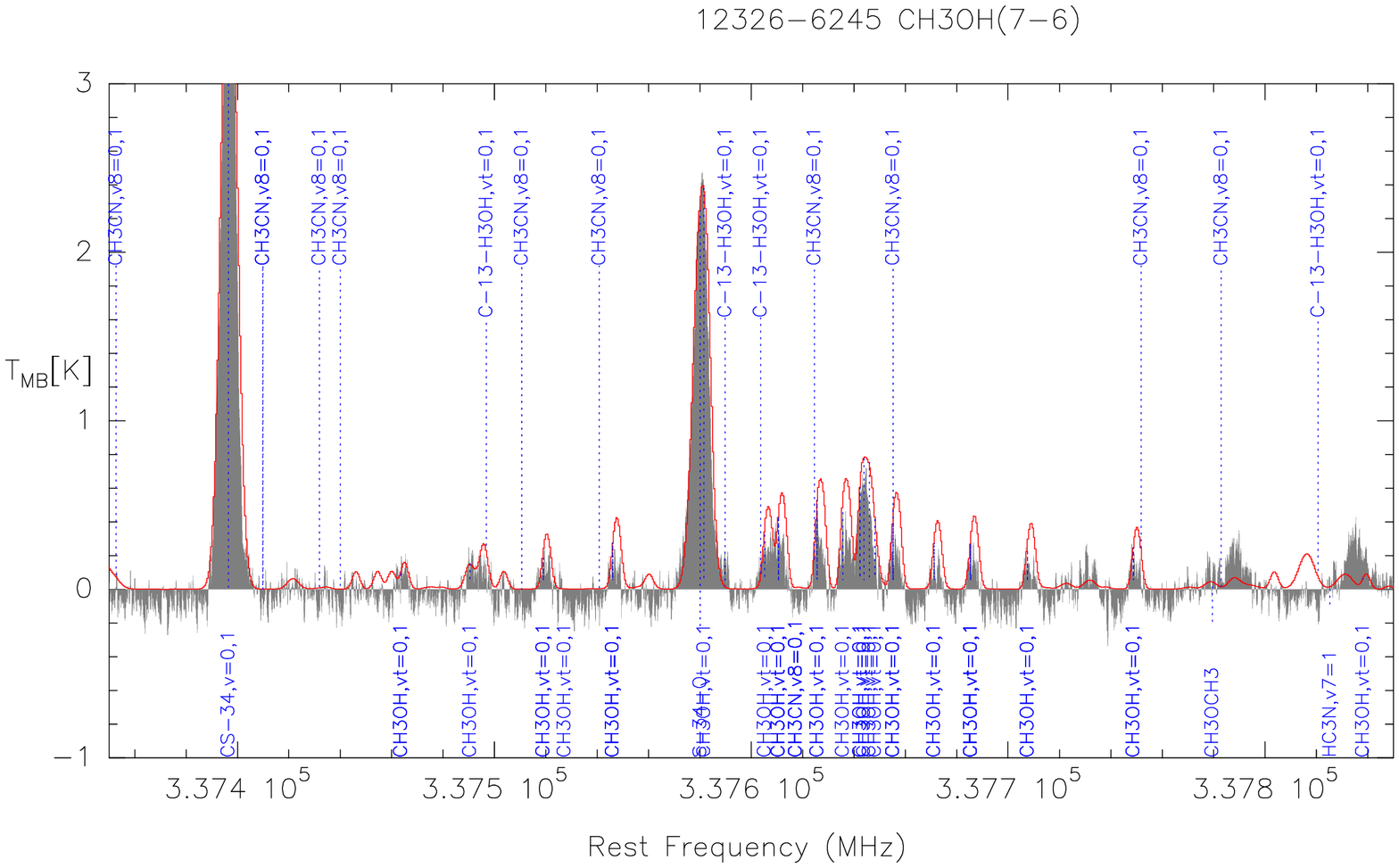}}
\centerline{
\includegraphics[angle=-90,scale=0.37]{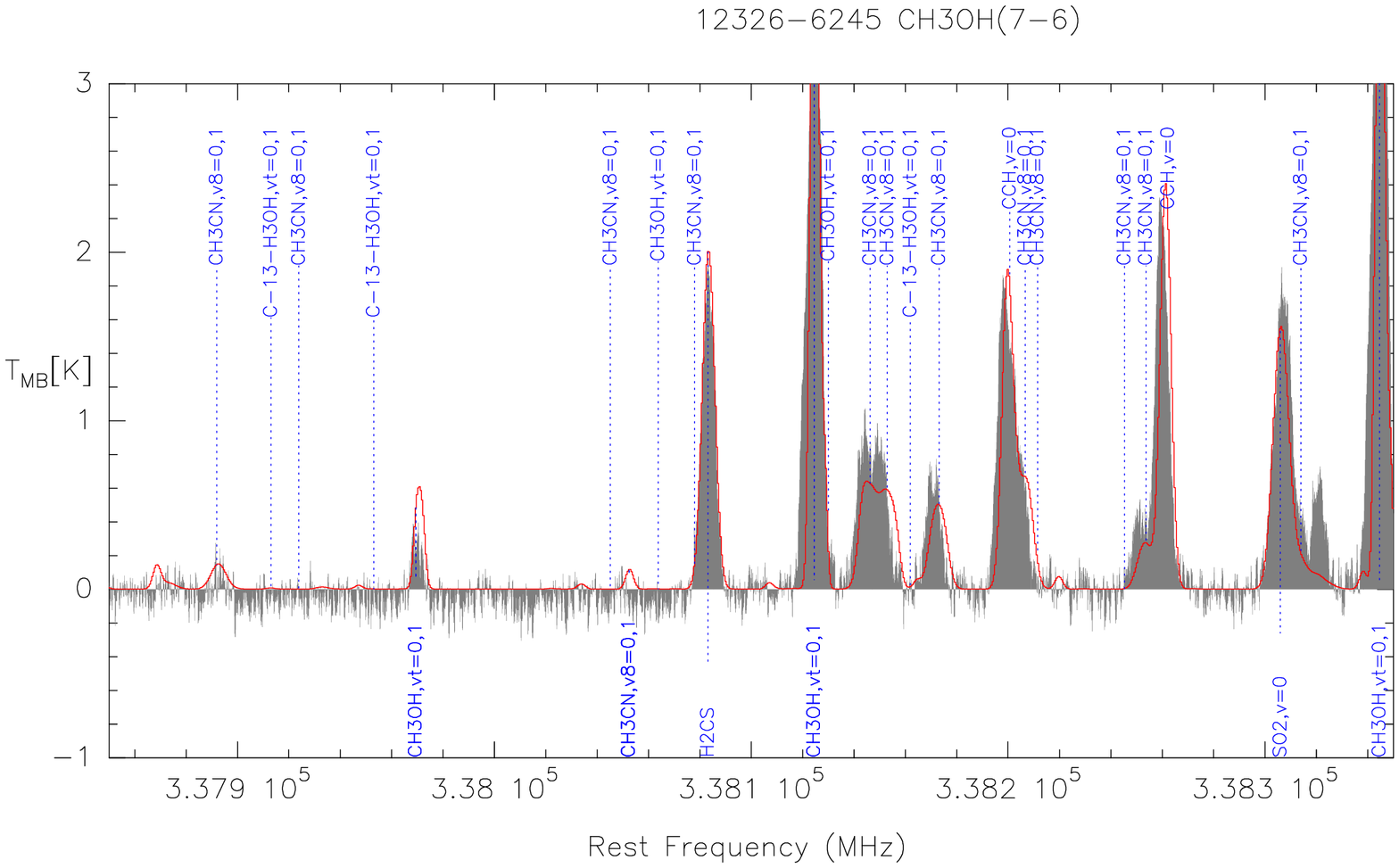}
\includegraphics[angle=-90,scale=0.37]{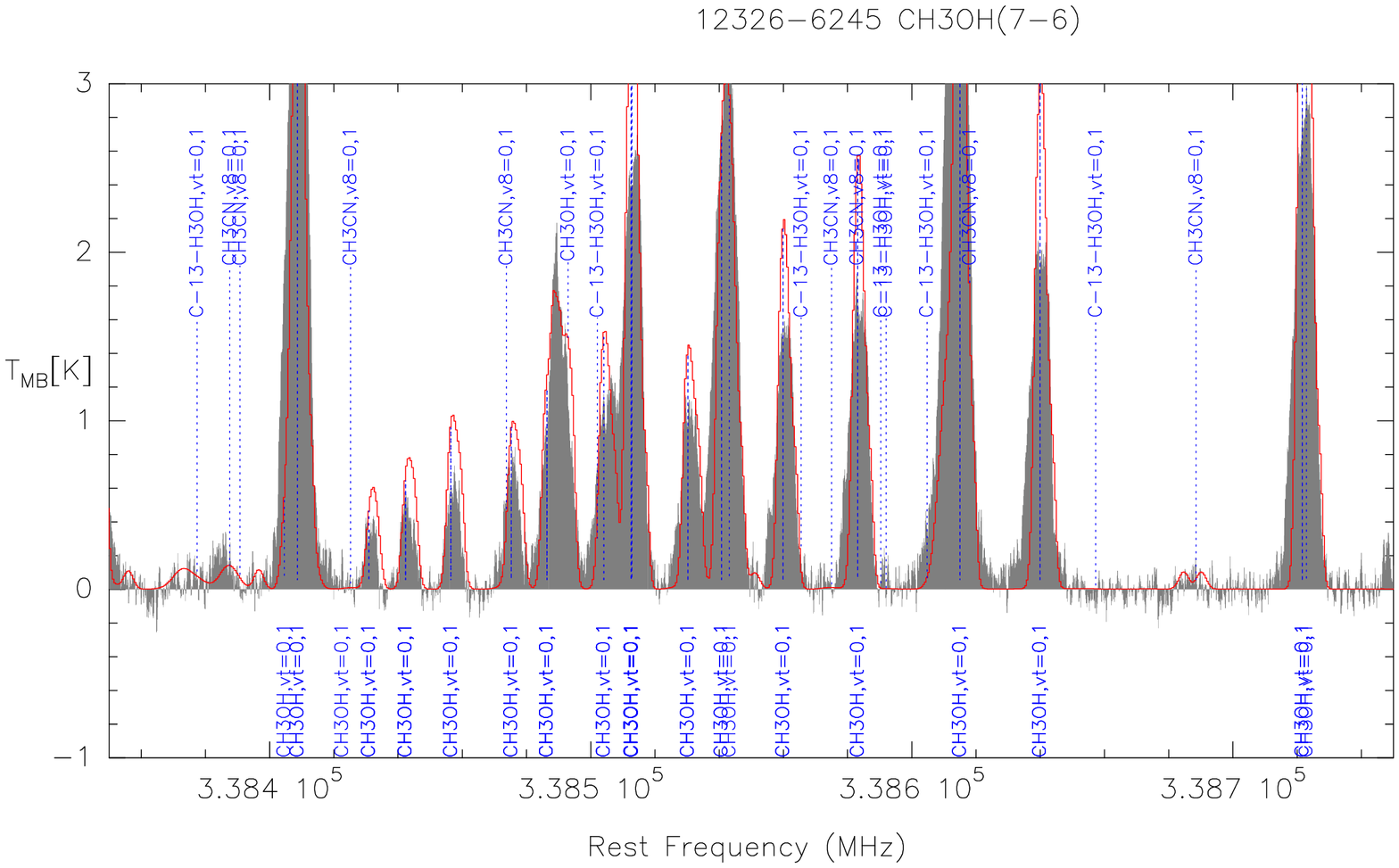}}
\caption{\label{fig:molem3b} Molecular emission of 12326-6245 around 338~GHz. The best-fit synthetic spectrum is overplotted (solid line). }
\end{figure*} }

\begin{table*}[htbp] 
\centering                                                                      
\caption{\label{tab_mol_16065} Results of the line modeling for 16065$-$5158 (-62.2 {\hbox{${\rm km\, s}^{-1}$}})}                          
\begin{tabular}{ccccccccc}                                                                           
\hline                                                                       
\hline               
Species  &  Size  &  $T_{\rm rot}$ & $N_{\rm b}$(mol) & $N_{\rm s}$(mol) & $\Delta$v & Offset &$N_{\rm b}$/$N_{\rm H_2}$  & \\ 
 &    (")  &  (K) & (cm$^{-2}$)   & (cm$^{-2}$)   & ({\hbox{${\rm km\, s}^{-1}$}}) & ({\hbox{${\rm km\, s}^{-1}$}}) & & \\                          
\hline                                                                                               
C$_2$H$_5$CN     &    \ldots $^a$            &  150$^a$          &  5.0(13)          &  \ldots                &  3.0              &  2.0              &  1.7(-10)         &  $^b$            \\  
CCH           &    \ldots $^a$            &  50$^a$           &  3.2(14)          &  \ldots                &  6.0              &  -1.0             &  1.1(-09)         &  $^c$            \\  
CH$_3$CCH     &    \ldots $^a$            &  50$^a$           &  3.0(15)          &  \ldots                &  7.0              &  0.0              &  1.0(-08)         &  $^c$            \\  
CH$_3$OCH$_3$    &    \ldots $^a$            &  200              &  1.3(15)          &  \ldots                &  2.5              &  3.0              &  4.5(-09)         &  $^b$            \\  
CH$_3$OCHO-a     &    \ldots $^a$            &  150$^a$          &  4.0(14)          &  \ldots                &  3.0              &  0.0              &  1.4(-09)         &  $^b$            \\  
CO               &    \ldots $^a$            &  50$^a$           &  4.0(19)          &  \ldots                &  5.5              &  -0.5             &  1.4(-04)         &  $^c,f$          \\  
CS               &  4.0              &  90               &  3.3(15)          &  7.0(16)          &  5.0              &  0.0              &  1.1(-08)         &  $^c,f$          \\  
H$_2$CS          &    \ldots $^a$            &  50$^a$           &  3.0(14)          &  \ldots                &  6.0              &  0.0              &  1.0(-09)         &  $^c$            \\  
HC$_3$N          &  0.8              &  240              &  7.9(13)          &  4.0(16)          &  7.0              &  1.0              &  2.7(-10)         &  $^c$            \\  
                &  10.0             &  40               &  4.7(14)          &  2.0(15)          &  7.0              &  1.0              &  1.6(-09)         &                  \\  
HCN              &  4.0              &  150$^a$          &  1.4(15)          &  3.0(16)          &  6.0              &  0.0              &  4.9(-09)         &  $^c,f$          \\  
HNCO             &    \ldots $^a$            &  150$^a$          &  1.0(14)          &  \ldots                &  6.0              &  1.0              &  3.5(-10)         &  $^c$            \\  
SO               &  8.0              &  30               &  8.2(15)          &  5.0(16)          &  7.0              &  1.0              &  2.9(-08)         &  $^e,f$          \\  
\hline                                                                                               
\end{tabular} 

\tablefoot{                                                                                     
$^a$Fixed Parameter; $^b$based on weak or partially blended lines only; $^c$based on one or a few lines only; $^d$includes vibrationally excited lines; $^e$non-LTE, see text; $^f$based on isotopologues, see text. The column labeled "offset" shows the offset in velocity relative to the systemic velocity of the source. $N_{\rm s}$ are the source-averaged column densities, $N_{\rm b}$ are the beam-averaged column densities and $N_{\rm b}$/$N_{\rm H_2}$ gives beam-averaged abundances.    }
\end{table*}    

\begin{table*}[htbp] 
\centering                                                                                       
\caption{\label{tab_mol_16060} Results of the line modeling for 16060$-$5146 (-91 {\hbox{${\rm km\, s}^{-1}$}})}                          
\begin{tabular}{ccccccccc}                                                                           
\hline                                                                       
\hline                
Species  &  Size  &  $T_{\rm rot}$ & $N_{\rm b}$(mol) & $N_{\rm s}$(mol) & $\Delta$v & Offset &$N_{\rm b}$/$N_{\rm H_2}$  & \\ 
 &    (")  &  (K) & (cm$^{-2}$)  & (cm$^{-2}$)  &  ({\hbox{${\rm km\, s}^{-1}$}}) & ({\hbox{${\rm km\, s}^{-1}$}}) & & \\                           
\hline                                                                                               
C$_2$H$_5$CN     &  10.0             &  150$^a$          &  2.4(13)          &  1.0(14)          &  3.0              &  0.0              &  4.2(-11)         &  $^b$            \\  
CCH           &   \ldots $^a$            &  50$^a$           &  5.0(14)          &  \ldots                &  10.0             &  -4.0             &  8.8(-10)         &  $^c$            \\  
CH$_3$CCH     &   \ldots $^a$            &  100              &  2.5(15)          &  \ldots                &  7.5              &  -1.50            &  4.4(-09)         &  $^c$            \\  
CH$_3$OCH$_3$    &   \ldots $^a$            &  150$^a$          &  1.0(14)          &  \ldots                &  2.0              &  2.0              &  1.8(-10)         &  $^b$            \\  
CH$_3$OCHO-a     &   \ldots $^a$            &  150$^a$          &  1.0(14)          &  \ldots                &  2.0              &  0.0              &  1.8(-10)         &  $^b$            \\  
CO               &   \ldots $^a$            &  50$^a$           &  6.3(19)          &  \ldots                &  9.0              &  -0.5             &  1.1(-04)         &  $^c,f$          \\  
CS               &  10.0             &  55               &  1.4(15)          &  6.0(15)          &  7.0              &  -0.5             &  2.5(-09)         &  $^c,f$          \\  
H$_2$CS          &   \ldots $^a$            &  50$^a$           &  2.6(14)          &  \ldots                &  10.0             &  -0.5             &  4.6(-10)         &  $^c$            \\  
HC$_3$N          &   \ldots $^a$            &  150$^a$          &  1.0(13)          &  \ldots                &  6.0              &  0.0              &  1.8(-11)         &  $^c$            \\  
HNCO             &  1.0$^a$          &  80               &  1.5(14)          &  5.0(16)          &  6.0              &  -2.0             &  2.6(-10)         &  $^c$            \\  
SO               &  4.0              &  40               &  1.9(16)          &  4.0(17)          &  6.0              &  -0.5             &  3.3(-08)         &  $^e,f$          \\  
\hline                                                                                               
\end{tabular} 
\tablefoot{                                                                                        
indices as in Table \ref{tab_mol_16065}   }                                                               
\end{table*}     

\begin{table*}[htbp]       
\centering                                                                                       
\caption{\label{tab_mol_12326} Results of the line modeling for 12326$-$6245 (-39.3 {\hbox{${\rm km\, s}^{-1}$}})}                       
\begin{tabular}{ccccccccc}                                                                           
\hline
\hline                                                                                               
Species  &  Size  &  $T_{\rm rot}$ & $N_{\rm b}$(mol) & $N_{\rm s}$(mol) & $\Delta$v & Offset &$N_{\rm b}$/$N_{\rm H_2}$  & \\ 
 &    (")  &  (K) & (cm$^{-2}$)   &(cm$^{-2}$)   &  ({\hbox{${\rm km\, s}^{-1}$}}) & ({\hbox{${\rm km\, s}^{-1}$}}) &  & \\                         
\hline                                                                                               
C$_2$H$_5$CN     &   \ldots $^a$            &  150$^a$          &  6.0(13)          &  \ldots                &  3.0              &  0.0              &  3.1(-10)         &  $^b$            \\  
CCH           &   \ldots $^a$            &  50$^a$           &  2.8(14)          &  \ldots                &  4.0              &  0.0              &  1.4(-09)         &  $^c$            \\  
CH$_3$CCH     &   \ldots $^a$            &  50$^a$           &  1.5(15)          &  \ldots                &  5.5              &  0.0              &  7.7(-09)         &  $^c$            \\  
CH$_3$OCH$_3$    &   \ldots $^a$            &  70               &  1.3(15)          &  \ldots                &  4.5              &  .0               &  6.6(-09)         &  $^b$            \\  
CH$_3$OCHO-a     &   \ldots $^a$            &  150$^a$          &  5.0(14)          &  \ldots                &  3.0              &  0.0              &  2.6(-09)         &  $^b$            \\  
CO               &   \ldots $^a$            &  50$^a$           &  2.8(19)          &  \ldots                &  4.5              &  -0.5             &  1.4(-04)         &  $^c,f$          \\  
CS               &   \ldots $^a$            &  50$^a$           &  2.6(15)          &  \ldots                &  6.0              &  0.0              &  1.3(-08)         &  $^c,f$          \\  
DCN              &   \ldots $^a$            &  150$^a$          &  3.0(13)          &  \ldots                &  6.0              &  0.0              &  1.5(-10)         &                  \\  
H$_2$CS          &   \ldots $^a$            &  50$^a$           &  2.5(14)          &  \ldots                &  5.0              &  0.0              &  1.3(-09)         &  $^c$            \\  
HC$_3$N          &  1.6              &  100              &  1.6(14)          &  2.0(16)          &  6.0              &  -5.0             &  8.2(-10)         &  $^c$            \\  
HNCO             &  1.6              &  75               &  2.4(14)          &  3.0(16)          &  6.0              &  0.0              &  1.2(-09)         &  $^c$            \\  
SO               &  10.0             &  50$^a$           &  7.1(15)          &  3.0(16)          &  6.0              &  0.0              &  3.6(-08)         &  $^e,f$          \\  
\hline                                                                                               
\end{tabular} 

\tablefoot{                                                                             
indices as in Table \ref{tab_mol_16065}    }                                                              
\end{table*}

\subsubsection{{\hbox{${\rm CH}_3{\rm CN}$}} \label{ch3cn_sec}}
The symmetric top molecule {\hbox{${\rm CH}_3{\rm CN}$}}\/ is an appropriate
temperature tracer \citep{2005ApJS..157..279A}.
The data and the synthetic spectra are displayed in Fig.
\ref{ch3cn_plots}.
Owing to the {\hbox{${\rm CH}_3{\rm CN}$}}\/ lines up to the $K$=7 transition in 16065$-$5158, we used a
compact hot component to model the higher excitation lines and additionally a more extended,
cooler component to model excess in the $K$=0 -- $K$=2 transitions. Beause this setup also
includes some {\hbox{${\rm CH}_3^{13}{\rm CN}$}} lines, we had both optically thick and thin
transitions to solve the $N$/$\theta$ degeneracy and obtain source sizes. 
While 16065$-$5158 and 12326$-$6245 can both be modeled with a two-component fit
displaying one hot, compact component and one extended, cooler one, it is also
possible to obtain reasonable results for 12326$-$6245 with only a hot compact
component. The two-component structure is attributed to a hot dense core
inside a colder, extended envelope. 16060$-$5146
had to be modeled with two hot, compact components with a velocity
difference of 9~{\hbox{${\rm km\, s}^{-1}$}}. This effect, which is not obvious in every species
studied here, suggests a complicated velocity
structure for parts of the molecular gas in this source.

We obtained physical sizes assuming near distances for 16060$-$5146 and 16065$-$5158
of 0.03~pc for the hot
component in 16065$-$5158, 12326$-$6245 and for the two components in
16060$-$5146. The extended components in 16065$-$5158 and 12326$-$6245 are 0.1~pc
each. Tables \ref{tab_16065}
to \ref{tab_12326} show the results for the {\hbox{${\rm CH}_3{\rm CN}$}}\/ modeling.

\begin{table*}          
\centering                                                                      
\caption{Properties of 16065$-$5158 (-62.2 {\hbox{${\rm km\, s}^{-1}$}}) as derived from LTE modeling of
 {\hbox{${\rm CH}_3{\rm CN}$}}, {\hbox{${\rm CH}_3{\rm OH}$}}\/, {\hbox{${\rm H}_2{\rm CO}$}}\/ and {\hbox{${\rm SO}_2$}}. }
\label{tab_16065} 

\begin{tabular}{lcccccccc}                                                                                            
\hline                                    
\hline                                                           
Species & Transition & Source Size & $T_{\rm rot}$  &  $N_{\rm s}$ & $N_{\rm b}$ & $\Delta$v & Offset & $N_{\rm b}$/$N_{\rm H_2}$ \\     
& & (\farcs) & (K)  & (cm$^{-2}$) & (cm$^{-2}$) & ({\hbox{${\rm km\, s}^{-1}$}}) & ({\hbox{${\rm km\, s}^{-1}$}}) & \\
\hline                            
  {\hbox{${\rm CH}_3{\rm CN}$}}   &      19--18   &   0.7   &   250   &      3.0(15)   &    4.5(12)   &   4.5   &  -1.0   & 1.6(-11)  \\ 
  {\hbox{${\rm CH}_3{\rm CN}$}}   &      19--18   &     5.0   &    30   &    1.0(17)   &    7.2(15)   &   4.5   &  -1.0   & 2.5(-08)  \\ 
  {\hbox{${\rm CH}_3{\rm CN}$}}   &      16--15   &   1.3   &   220   &    1.4(15)   &    7.3(12)   &   4.5   &     1.0   & 2.5(-11)  \\ 
  {\hbox{${\rm CH}_3{\rm CN}$}}   &      16--15   &     5.0   &    30   &    9.4(15)   &    6.7(14)   &   4.5   &     1.0   & 2.3(-09)  \\ 
    {\hbox{${\rm CH}_3{\rm OH}$}}   &        6--5   &   1.8   &   140   &      2.0(18)   &    2.0(16)   &     6.0   &     0.0   & 6.9(-08)  \\ 
    {\hbox{${\rm CH}_3{\rm OH}$}}   &        6--5   &    13.0   &    30   &      5.0(15)   &    1.7(15)   &     6.0   &     0.0   & 6.0(-09)  \\ 
    {\hbox{${\rm CH}_3{\rm OH}$}}   &        7--6   &   1.8   &   140   &      7.0(17)   &    6.9(15)   &     6.0   &   1.5   & 2.4(-08)  \\ 
    {\hbox{${\rm CH}_3{\rm OH}$}}   &        7--6   &    13.0   &    30   &      5.0(15)   &    1.7(15)   &     6.0   &     1.0   & 6.0(-09)  \\ 
    {\hbox{${\rm H}_2{\rm CO}$}}   &        4--3   &   2.1   &   160   &      1.0(16)   &    1.3(14)   &     6.0   &  -0.5   & 4.7(-10)  \\ 
    {\hbox{${\rm H}_2{\rm CO}$}}   &        4--3   &     \ldots   &    50   &          \ldots   &      4.0(14)   &     6.0   &  -0.5   & 1.4(-09)  \\ 
    {\hbox{${\rm H}_2{\rm CO}$}}   &        6--5   &   2.1   &   160   &    4.8(16)   &    6.4(14)   &     6.0   &  -0.5   & 2.2(-09)  \\ 
    {\hbox{${\rm H}_2{\rm CO}$}}   &        6--5   &     \ldots   &    50   &          \ldots   &      5.0(13)   &     6.0   &  -0.5   & 1.7(-10)  \\ 
    {\hbox{${\rm SO}_2$}}   &     338~GHz   &     \ldots   &    30   &          \ldots   &      3.0(16)   &     8.0   &     0.0   & 1.0(-07)  \\ 
    {\hbox{${\rm SO}_2$}}   &     338~GHz   &     \ldots   &   250   &          \ldots   &      1.0(15)   &     8.0   &     0.0   & 3.5(-09)  \\ 
    {\hbox{${\rm SO}_2$}}   &     430~GHz   &     \ldots   &    30   &          \ldots   &      3.0(16)   &     8.0   &     0.0   & 1.0(-07)  \\ 
    {\hbox{${\rm SO}_2$}}   &     430~GHz   &     \ldots   &   250   &          \ldots   &      1.0(15)   &     8.0   &     0.0   & 3.5(-09)  \\ 
\hline   
\end{tabular}
\tablefoot{
We display in
 the table the rotational temperature $T_{\rm rot}$ , source-averaged column density $N_{\rm s}$, beam-averaged column density $N_{\rm b}$, velocity width
 $\Delta$v, velocity offset from systemic velocity, and abundances. If two entries per transition are given, they
refer to the cold extended envelope and the hot compact core.}

\end{table*}

\begin{table*}  

\caption{Same as in Table \ref{tab_16065}, but for 16060$-$5146 (-91 {\hbox{${\rm km\, s}^{-1}$}}). }
\label{tab_16060} 

\centering                                                                          
\begin{tabular}{lcccccccc}                                                                                            
\hline                
\hline                                                                               
Species &Transition & Source Size & $T_{\rm rot}$  &  $N_{\rm s}$ &$N_{\rm b}$ & $\Delta$v &
Offset & $N_{\rm b}$/$N_{\rm H_2}$ \\     
& & (\farcs) & (K)  & (cm$^{-2}$) & (cm$^{-2}$) & ({\hbox{${\rm km\, s}^{-1}$}}) & ({\hbox{${\rm km\, s}^{-1}$}}) & \\
\hline                            
  {\hbox{${\rm CH}_3{\rm CN}$}}   &      19--18   &   1.1   &   170   &      1.0(16)   &    3.7(13)   &   5.5   &     2.0   & 6.5(-11)  \\ 
  {\hbox{${\rm CH}_3{\rm CN}$}}   &      19--18   &     1.0   &   180   &      2.0(16)   &    6.2(13)   &   5.5   &    -6.0   & 1.1(-10)  \\ 
  {\hbox{${\rm CH}_3{\rm CN}$}}   &      16--15   &   1.1   &   170   &    1.5(16)   &    5.6(13)   &   5.5   &     4.0   & 9.8(-11)  \\ 
  {\hbox{${\rm CH}_3{\rm CN}$}}   &      16--15   &     1.0   &   180   &      3.0(16)   &    9.2(13)   &   5.5   &    -5.0   & 1.6(-10)  \\ 
    {\hbox{${\rm CH}_3{\rm OH}$}}   &        7--6   &     1.0   &   170   &      5.0(17)   &    1.5(15)   &     3.0   &     5.0   & 2.7(-09)  \\ 
    {\hbox{${\rm CH}_3{\rm OH}$}}   &        7--6   &     1.0   &   170   &      5.0(17)   &    1.5(15)   &     3.0   &    -5.0   & 2.7(-09)  \\ 
    {\hbox{${\rm CH}_3{\rm OH}$}}   &        7--6   &     \ldots   &    30   &          \ldots   &      4.0(15)   &     7.0   &    -2.0   & 7.0(-09)  \\ 
    {\hbox{${\rm CH}_3{\rm OH}$}}   &        6--5   &     1.0   &   170   &      1.0(17)   &    3.1(14)   &     3.0   &     5.0   & 5.4(-10)  \\ 
    {\hbox{${\rm CH}_3{\rm OH}$}}   &        6--5   &     1.0   &   170   &      1.0(17)   &    3.1(14)   &     3.0   &    -5.0   & 5.4(-10)  \\ 
    {\hbox{${\rm CH}_3{\rm OH}$}}   &        6--5   &     \ldots   &    30   &          \ldots   &      2(15)   &     7.0   &    -2.0   & 3.5(-09)  \\ 
    {\hbox{${\rm H}_2{\rm CO}$}}   &        4--3   &   1.3   &   170   &      5.0(16)   &    2.6(14)   &     7.0   &     0.0   & 4.6(-10)  \\ 
    {\hbox{${\rm H}_2{\rm CO}$}}   &        4--3   &     \ldots   &    70   &          \ldots   &      3.0(14)   &    10.0   &     0.0   & 5.3(-10)  \\ 
    {\hbox{${\rm H}_2{\rm CO}$}}   &        6--5   &   1.3   &   170   &      5.0(16)   &    2.6(14)   &     7.0   &     0.0   & 4.6(-10)  \\ 
    {\hbox{${\rm H}_2{\rm CO}$}}   &        6--5   &     \ldots   &    70   &          \ldots   &      6.0(13)   &    10.0   &     0.0   & 1.1(-10)  \\ 
    {\hbox{${\rm SO}_2$}}   &     338~GHz   &     \ldots   &    30   &          \ldots   &      5.0(16)   &    10.0   &     0.0   & 8.8(-08)  \\ 
    {\hbox{${\rm SO}_2$}}   &     430~GHz   &     \ldots   &   100   &          \ldots   &      1.0(15)   &     8.0   &     0.0   & 1.8(-09)  \\ 
\hline                                                                                               
\end{tabular} 
\tablefoot{
Because of its line profiles, 16060$-$5146 was modeled with two
hot compact components in some molecules instead of the core/envelope model used for the other
two sources. }

\end{table*}

\begin{table*}    
\centering                                                                        
\caption{Same as in Table \ref{tab_16065}, but for 12326$-$6245 (-39.3 {\hbox{${\rm km\, s}^{-1}$}}).}
\label{tab_12326} 

\begin{tabular}{lcccccccc}                                                                                            
\hline                                                                       
\hline  

Species & Transition & Source Size & $T_{\rm rot}$  &  $N_{\rm s}$ & $N_{\rm b}$ & $\Delta$v & Offset & $N_{\rm b}$/$N_{\rm H_2}$ \\     
& & (\farcs) & (K)  & (cm$^{-2}$) &(cm$^{-2}$) & ({\hbox{${\rm km\, s}^{-1}$}}) & ({\hbox{${\rm km\, s}^{-1}$}}) & \\
\hline                         
  {\hbox{${\rm CH}_3{\rm CN}$}}   &      19--18/16--15   &   1.6   &   100   &      8(15)   &    6.3(13)   &   5.4   &     0.0   & 3.2(-10)  \\ 
  {\hbox{${\rm CH}_3{\rm CN}$}}   &      19--18/16--15   &   7.0   &    20   &      1(15)   &    1.3(14)   &   5.4   &     0.0  & 6.7(-10)  \\ 
    {\hbox{${\rm CH}_3{\rm OH}$}}   &        6--5   &    1.0   &   240   &      4(18)   &    1.2(16)   &     4   &   0.5   & 6.3(-08)  \\ 
    {\hbox{${\rm CH}_3{\rm OH}$}}   &        6--5   &    12.0   &    50   &    1.5(16)   &    4.6(15)   &     4   &     0.0   & 2.4(-08)  \\ 
    {\hbox{${\rm CH}_3{\rm OH}$}}   &        7--6   &   1.3   &   200   &      1(18)   &    5.2(15)   &     3   &    -1.0   & 2.7(-08)  \\ 
    {\hbox{${\rm CH}_3{\rm OH}$}}   &        7--6   &   12.0   &    50   &      1(16)   &    3.1(15)   &     4   &     0.0   & 1.6(-08)  \\ 
    {\hbox{${\rm H}_2{\rm CO}$}}   &        4--3   &     \ldots   &    75   &          \ldots   &      4(14)   &     5   &     0.0   & 2.0(-09)  \\ 
    {\hbox{${\rm SO}_2$}}   &     290~GHz   &     \ldots   &    35   &          \ldots   &    2.5(16)   &     6   &  -0.5   & 1.3(-07)  \\ 
    {\hbox{${\rm SO}_2$}}   &     338~GHz   &     \ldots   &    35   &          \ldots   &    1.8(16)   &     6   &   0.5   & 9.2(-08)  \\ 
\hline                                                                                               
\end{tabular} 

\end{table*}

\subsubsection{{\hbox{${\rm CH}_3{\rm OH}$}} }
For all three sources, data for the {\hbox{${\rm CH}_3{\rm OH}$}}\/(7--6) band were obtained in a setup at 338~GHz,
including also the torsionally excited $v_t$=1 {\hbox{${\rm CH}_3{\rm OH}$}}\/ lines at 337.6~GHz. The
{\hbox{${\rm CH}_3{\rm OH}$}}(6--5) lines at 290~GHz were observed in all three sources, while for 12326$-$6245, the torsionally excited
lines at 289~GHz were covered in a different frequency setup.
Both the {$J$}=6--5 and the {$J$}=7--6 series were modeled with XCLASS under the
assumption of LTE.\\
\indent To obtain the temperature for the hot compact component in the {$J$}=7--6 band, the
optically thin torsionally excited lines and the higher excitation lines in
the $v_t$=0 bands were used. In combination with
optically thick lines in the band, the $N$/$\theta$ degeneracy could be resolved.
The
torsionally excited lines at 289 and 337~GHz also contain a line with
higher optical depth, from which we could obtain an estimate of the
source size for this component.\\
\indent For details of the model in 12326$-$6245, see
Figs. \ref{ch3oh1_plots} and \ref{ch3oh2_plots},
while plots of the remaining two sources can be found in the appendix. The synthetic model spectrum in Fig. \ref{ch3oh1_plots} overestimates the
 torsionally exited lines, which is because of the added complexity of modeling
 the torsionally exited $v_t=1$ lines simultaneously with the hot component of
the $v_t=0$ lines. \\
\indent In 16060$-$5146, the line profiles are much broader
owing to a blend of two hot compact components and one extended component,
and the spectra suffer from a worse baseline than the spectra for the other two
sources. We could obtain a good fit for two hot components, but because of the complexity of the data, only a beam-averaged column density could be obtained for the cold component. The results of the {\hbox{${\rm CH}_3{\rm OH}$}}\/ modeling can be
found in Tables \ref{tab_16065}--\ref{tab_12326}.

\subsubsection{H$_2$CO}
Because it is a slightly asymmetric rotor, {\hbox{${\rm H}_2{\rm CO}$}}\/ is a good tracer of kinetic
temperatures \citep{1993ApJS...89..123M} and it is quite ubiquitous in regions of star
formation \citep{1990ApJ...348..542M,1993ApJS...89..123M}.\\
\indent In this survey, we observed {\hbox{${\rm H}_2{\rm CO}$}}\/ in all three sources in the {\hbox{${\rm H}_2{\rm CO}$}}(4--3)
transitions at 291~GHz, 
and in 16065$-$5158 and 16060$-$5146
also in {\hbox{${\rm H}_2{\rm CO}$}}(6--5) lines at 437~GHz.
One can obtain a reasonable fit to the 
two transitions with a hot
compact core and a cold extended envelope component. The results of the {\hbox{${\rm H}_2{\rm CO}$}}\/ modeling can be
found in Tables \ref{tab_16065}--\ref{tab_12326}, while the spectra and
synthetic models can be
found in Figs. \ref{fig:molem1a}, \ref{fig:molem2a}, and \ref{fig:molem3a} in the
appendix. For 12326$-$6245, only the {\hbox{${\rm H}_2{\rm CO}$}}(4--3) transition was observed. In
16060$-$5146, we could not model the hot {\hbox{${\rm H}_2{\rm CO}$}}\/ emission with two velocity components, as we did for {\hbox{${\rm CH}_3{\rm OH}$}}\/ and {\hbox{${\rm CH}_3{\rm CN}$}}.

\subsubsection{S-bearing species}
We observed the lines from the S-bearing species {\hbox{\rm CS}}\/, {\hbox{${\rm H}_2{\rm CS}$}}, {\hbox{\rm SO}}\/, and {\hbox{${\rm SO}_2$}}\/ in our
frequency setups. For the {\hbox{${\rm H}_2{\rm CS}$}}(10--9) line, only the beam-averaged column
density for a fixed temperature of 50~K could be determined, while the other
species were observed in several lines.  \\
\indent We observed the isotopologic {\hbox{${\rm C}^{33}{\rm
     S}$}}(9--8), {\hbox{${\rm C}^{33}{\rm S}$}}(6--5), and {\hbox{${\rm C}^{34}{\rm S}$}}(7--6) transitions in {\hbox{\rm CS}}\/, while no {\hbox{${\rm C}^{32}{\rm S}$}} line was included in any of the setups. Assuming isotopic ratios of $^{32}$S/$^{34}$S=23 and
$^{32}$S/$^{33}$S=127 \citep{1994ARA&A..32..191W,1996A&A...305..960C}, we could model {\hbox{\rm CS}}\/ and {\hbox{\rm SO}}\/ consistently
over the frequency bands. The same was not possible for {\hbox{${\rm SO}_2$}}.
Apart from the {\hbox{${\rm SO}_2$}}(5--4) line at 351.3~GHz, which has a lower level energy
of 19~K above ground state, the other {\hbox{${\rm SO}_2$}}\/ lines cover higher excitation
conditions, which are between 76~K and 770~K above ground state energy. 
The results of the {\hbox{${\rm SO}_2$}}\/ modeling can be
found in Tables \ref{tab_16065}--\ref{tab_12326}, while the results for the
remaining species are given in Tables
\ref{tab_mol_16065}--\ref{tab_mol_12326}.

\subsubsection{Uncertainty estimates}

To estimate the errors of the modeled parameters, we performed an $\chi^2$ analysis for the  {\hbox{${\rm CH}_3{\rm CN}$}}(16--15) band  in 12326$-$6245 with a fixed source size, varying temperature $T$,
and column densities $N$. This indicated uncertainties of up to 40$\%$ in $T$ and up to
23$\%$ in $N$ within the 3$\sigma$
confidence limit, which we consider as typical for only one frequency setup and no line blends from either sideband.
It is evident from Tables  \ref{tab_16065}--\ref{tab_12326} that in some cases the simultaneous modeling of the hot or cold component of a species could not be done consistently in two different frequency setups. This discrepancy is smaller than a factor of 5 (apart from the cold component of {\hbox{${\rm CH}_3{\rm CN}$}}(19--18) in 16065--5158) though, and could have several reasons: \\
\indent First, there can be uncertainties in the flux calibration and pointing, because the data were taken at different times. In addition, in 16060--5158 and 16060--5146 the two setups were observed with a positional offset of 8\farcs to each other. While at these positions the emission is still covered in the beam, influences of the source geometry might already play a role.   \\

Based on the modeling results of the {\hbox{${\rm CH}_3{\rm CN}$}}(16--15) band alone, a common line width for the hot and cold component
was assumed for the {\hbox{${\rm CH}_3{\rm CN}$}}(19--18) transitions. This could be another potential cause for discrepancies, but the heavy line blending of the {\hbox{${\rm CH}_3{\rm CN}$}}(19--18) band does not allow us to derive a reliable line width for
the hot component.  The
line widths from the XCLASS modeling listed in Tables \ref{tab_16065} to \ref{tab_12326} are intrinsic
line widths, and most of the lines are influenced by a high optical depth (see also observed average line widths in the results section), which complicates the determination of the line widths.

\subsection{Spectral energy distributions}

\begin{figure} 
\includegraphics[angle=-90,width=7.5cm]{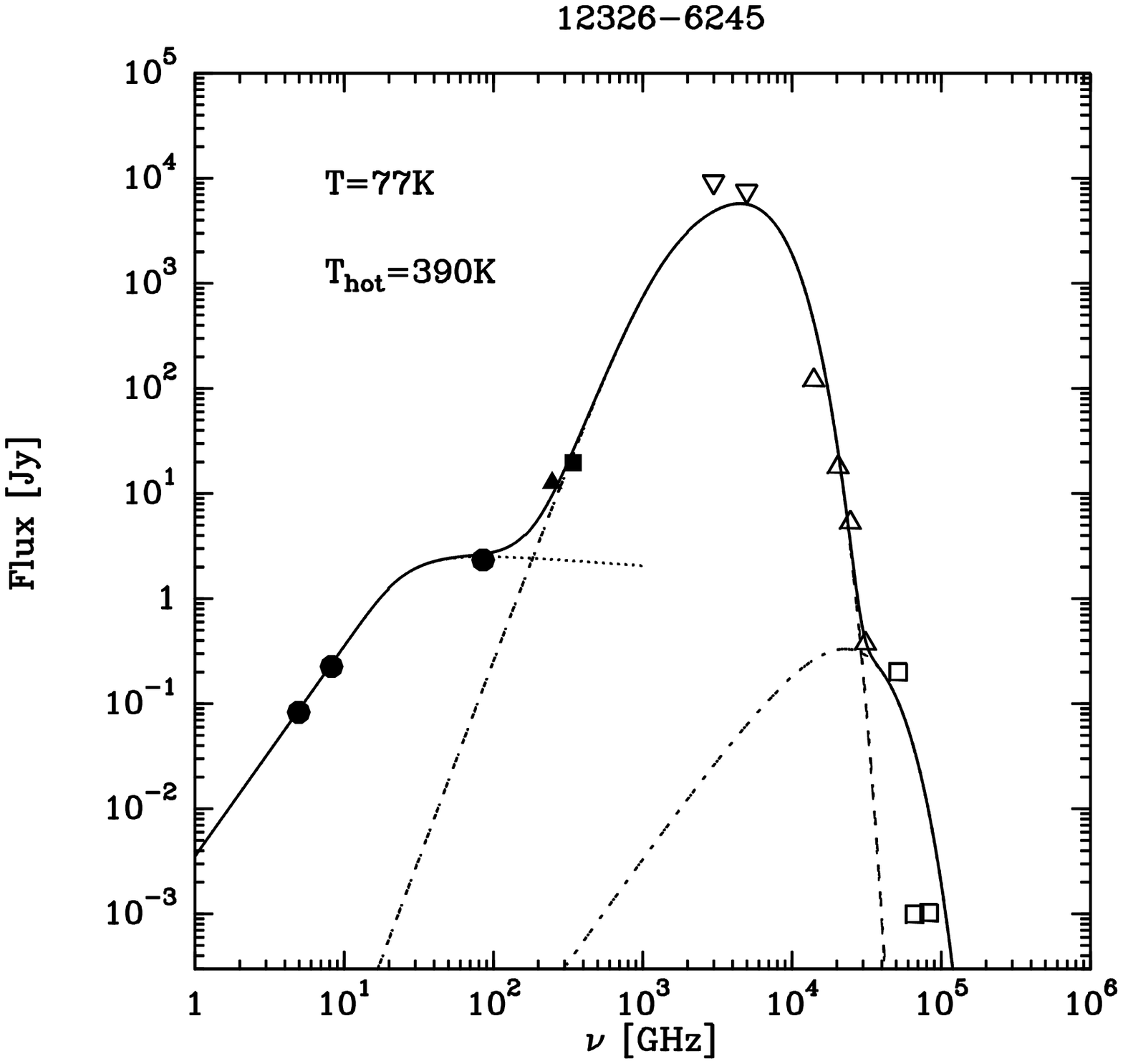}
\includegraphics[angle=-90,width=7.5cm]{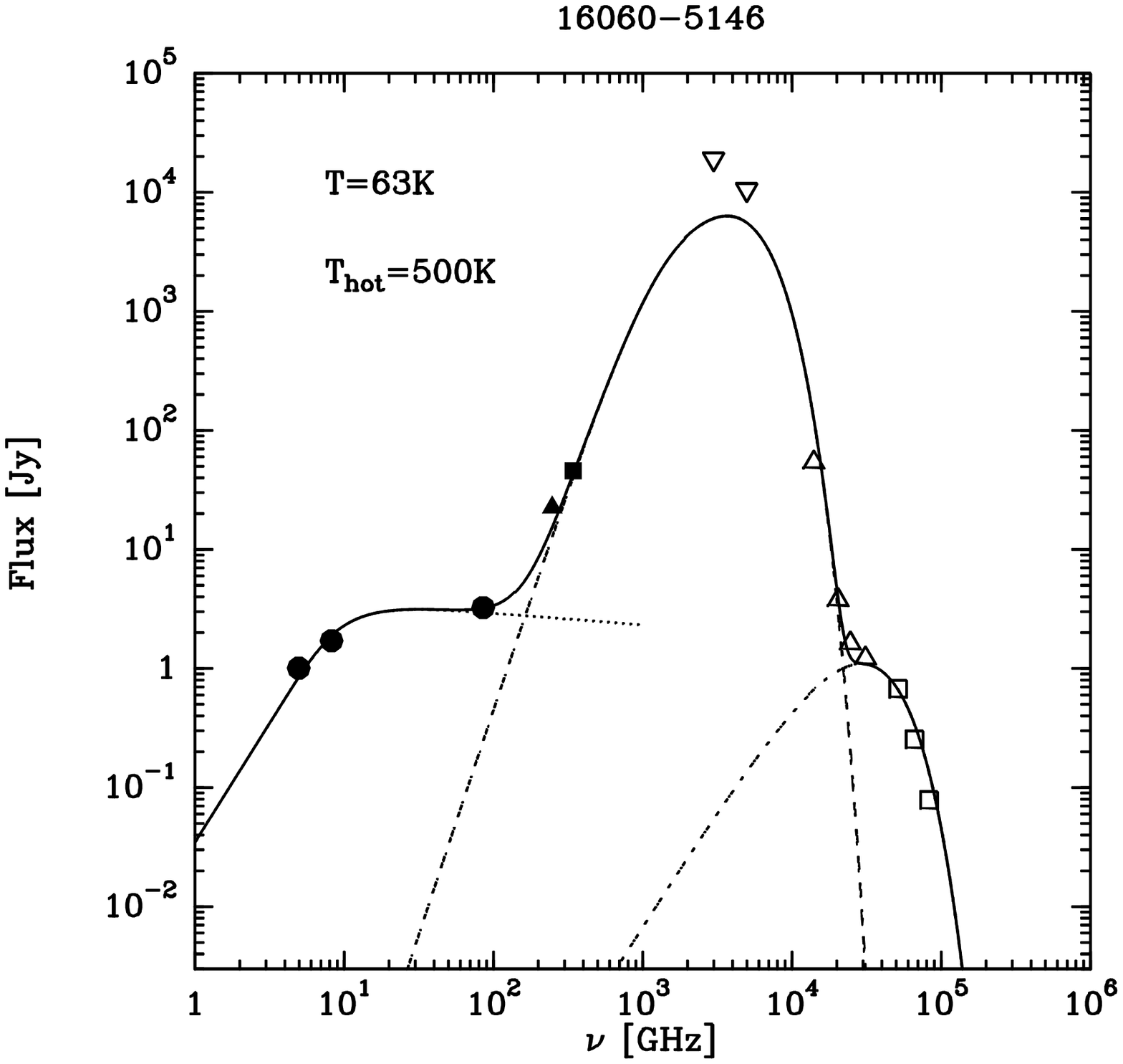}
\includegraphics[angle=-90,width=7.5cm]{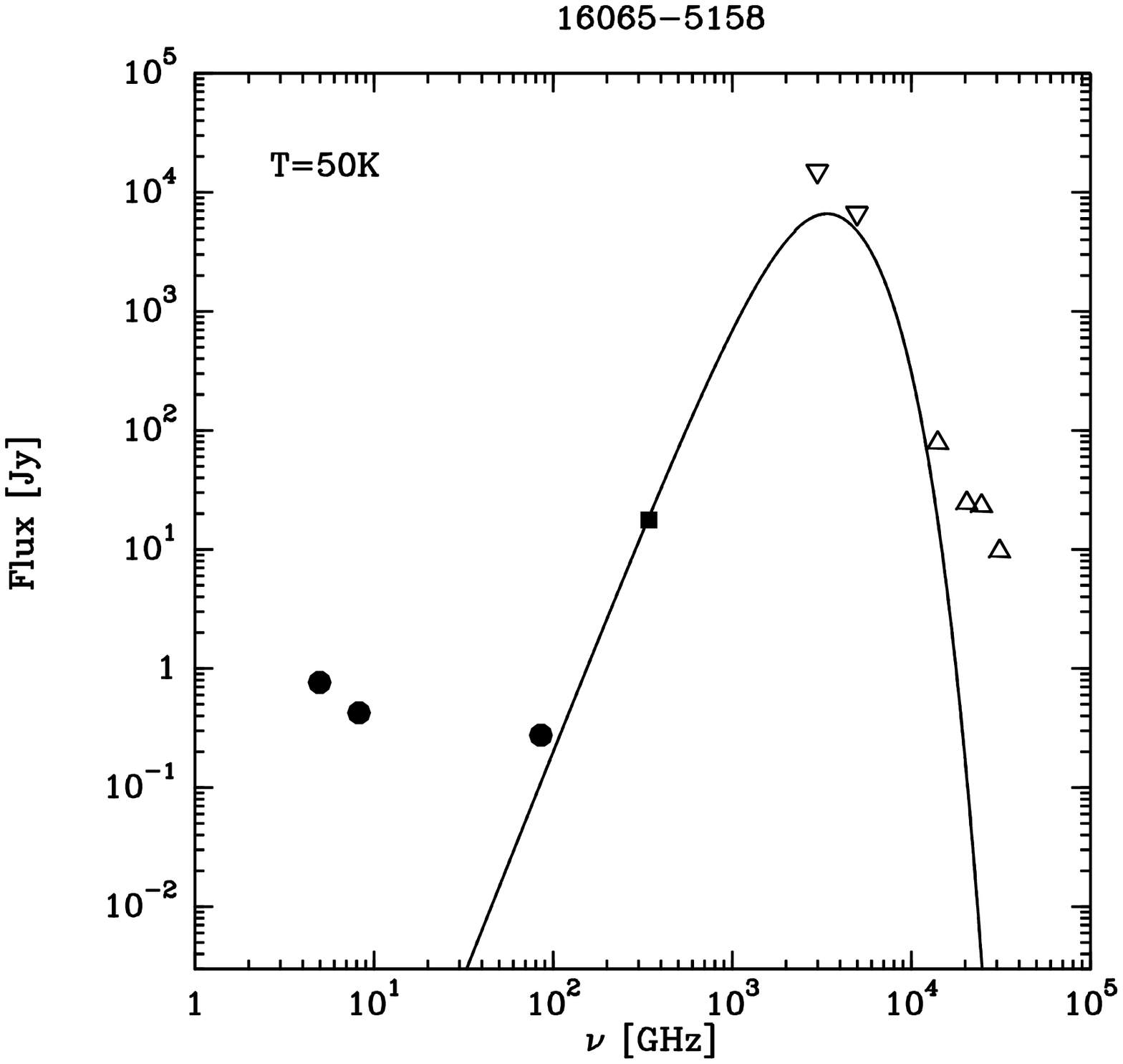}
\caption{\label{fig:sed} Spectral energy distributions for the three
 sources. The IRAS fluxes (downward pointing triangles) are taken from the
 high-resolution IRAS Galaxy Atlas (IGA) data. The empty squares and triangles show the MSX and {\it Spitzer} GLIMPSE
 fluxes respectively, the filled circles the ATCA fluxes and the filled
 triangle and square the bolometer fluxes from SIMBA and LABOCA.}
\end{figure}

To obtain the dust temperatures and luminosities for the three sources,
the spectral energy distribution (SED; i.e., flux density vs. frequency) was analyzed for each source.
Estimates of the dust temperatures were obtained by fitting graybody plus
free-free spectra to the SEDs (see Fig. \ref{fig:sed}). For all three sources
a power-law dependence of the dust opacity with $\beta=1.8$ was used (see
Sec. \ref{lab:obsres}), based on the dust properties used to derive the masses
and column densities of the three sources \citep{1994A&A...291..943O}. We
also estimated total luminosities by integrating the fluxes under the SEDs. Note that the GLIMPSE fluxes are not corrected for extinction.

12326--6245 and 16060--5146 were modeled using three different components, while one was used for 16065$-$5159. The data at 3.6 --
12~{$\mu\hbox{m}$}\/ trace the hot, compact radiation source that is already well
developed in the infrared, while the data points between 24~{$\mu\hbox{m}$}\/ and 3~mm
represent the contribution from the colder,  more extended dust envelope. The IRAS points
at 60 and 100~{$\mu\hbox{m}$}\/ are taken from the high-resolution IRAS Galaxy Atlas
(IGA) data. They represent upper limits to the IRAS flux. The cm continuum data, which represent the contribution of
free-free emission from an associated  {\hbox{UCH{\sc ii}}}\/ region, is modeled as third
component. In 16065$-$5158, only the cold, extended envelope was modeled, because the radio continuum and infrared data are offset from the dust peak.\\ 

For 12326$-$6245, a dust temperature of 77~K was derived with the graybody fit for the extended
component. For the hot
component traced by the {\it Spitzer} IRAC points, the blackbody fit resulted in a
temperature of 390~K. The luminosity of the 77~K extended dust component is
1.0$\times 10^{5}$~\Lsun\/ and the
one derived from the 390~K compact component 74~\Lsun .\\
16060$-$5146 has a dust temperature of 63~K, a temperature of the compact component
of 500~K and luminosities of 2.4$\times 10^{5}$~\Lsun\/ and 490~\Lsun\/ for
the extended and compact components, respectively.\\
For 16065$-$5158, the dust temperature was derived to be 50~K, but in this source,
it is hard to constrain the parameters, because only the ATCA 3~mm emission and
the LABOCA 870~{$\mu\hbox{m}$}\/ emission are associated with it, because IRAS and MSX values
give but upper limits. The luminosity of the extended component is 0.6$\times
10^{5}$~\Lsun\/ .\\
The spectral energy distributions show that the 3~mm sources in 12326$-$6245 and 16060$-$5146 are dominated by free-free
emission and by dust radiation for 16065$-$5158.\\

\subsection{Spectral types}
To obtain the spectral type of the mm-sources, we estimated the Lyman continuum flux  from the cm observations of
\citet{2007A&A...461...11U}, which trace the free-free emission of the  {\hbox{UCH{\sc ii}}}\/
region.\\
Because that emission is optically thick from both regions, we obtained their parameters by
fitting the spectral energy distribution with the theoretical spectrum of a
homogeneous constant density plasma \citep{2006ApJ...651..914G}. 
For 12326$-$6245, an emission measure
of 2.6$\times10^{9}$ pc~cm$^{-6}$ and a Lyman continuum flux of 8.2$\times10^{48}$
s$^{-1}$ were obtained, leading to a spectral type between O6.5 and a stellar
luminosity of $L=1.5\times 10^5$~\Lsun ,
according to \citet{1973AJ.....78..929P}. \\
For 16060$-$5146, the emission measure is 4.0$\times10^{8}$ pc~cm$^{-6}$, the
Lyman continuum flux 1.5$\times10^{49}$s$^{-1}$ and the spectral type was
found to be between O6 and O5.5 with $L=2.5 - 4.0 \times 10^5$~\Lsun.\\
This kind of analysis works with the assumption, that the Lyman continuum luminosity of
the source is dominated by the emission of one object.

\subsection{Mass estimates}
Table \ref{dustm} lists the dust masses derived from the 870~{$\mu\hbox{m}$}\/ LABOCA
observations as well as the virial masses derived from the line velocity
dispersion of
the {\hbox{\rm C$^{17}$}{\rm O}}(3--2) line. $M_{\rm vir}$ in \Msun\/ was derived according to
\citet{1988ApJ...333..821M} and \citet{2006A&A...460..721M}

\begin{equation}\label{eq:virmass}
M_{\rm vir}=0.5 \Delta v^2 d \Theta_{10} \kappa  ~~[\Msun].
\end{equation}

The beam-deconvolved source sizes at the 10$\%$ flux contour were taken as the angular source sizes
$\Theta_{10}$ in \farcs. $\kappa\sim 1.3$ is a correction factor to correct
for the average radial density dependence.  
We estimated the dust masses in two ways. $M_{\rm warm}$ gives a measure of the
masses of the warmer component associated with the IRAS source, as traced by the
graybody fit of the SED, while $M_{\rm ext}$
traces the masses of the extended cooler envelopes.   
The dust masses were derived using Eq. \ref{eq:mass}, following \citet{2006A&A...460..721M}, with a dust absorption
coefficient, $\kappa_{\rm d}$, of 0.176~{\rm m}$^2$/kg \citep{1994A&A...291..943O}
(model V)
and a dust-to-gas ratio $R_{\rm d}=\frac{1}{100}$. $B_{\nu}$(T$_{\rm d}$) is the Planck
function of a blackbody at dust temperature {\hbox{$T_{\rm d}$}} , $S_{\nu}$ the integrated
flux in Jy and $d$ the kinematic distance in kpc.

For the warm
component, {\hbox{$T_{\rm d}$}} \/ as derived from the SED and the peak fluxes were used, while the mass of the
extended dust emission was derived with the integrated intensity within the
10\% contour and assuming a dust temperature of 20~K for
12326$-$6245 and 30~K for 16065$-$5158 and 16060$-$5146, respectively. They were assumed to be
equal to the gas kinetic temperatures derived in the {\hbox{${\rm CH}_3{\rm CN}$}}\/ modeling,
following \citet{2006A&A...460..721M}. For 16060$-$5146, where the
{\hbox{${\rm CH}_3{\rm CN}$}}\/ was modeled without a cold component, the temperature derived from
{\hbox{${\rm CH}_3{\rm OH}$}}\ was taken instead.  

\begin{equation}
M_{\rm gas}=\frac{S_{\nu}d^2}{B_{\nu}({\hbox{$T_{\rm d}$}} ) \kappa_{\rm d}  R_{\rm d}} ~~[\Msun]
\label{eq:mass}
\end{equation}

\begin{table*}[htbp] 

\caption{\label{dustm} Dust masses derived from LABOCA 870~{$\mu\hbox{m}$}\/ emission. }  

\begin{center}                                                                                       
\begin{tabular}{lccccccc}                                                                     
\hline                                                                                               
\hline 

Source & $M_{\rm warm}$ & $M_{\rm ext}$    & $M_{\rm vir}$ & $R_{\rm warm}$ &
$R_{\rm ext}$&$n_{\rm warm}$(H$_2$)&$n_{\rm ext}$(H$_2$)\\
& (10$^3$\Msun) & (10$^3$\Msun) & (10$^3$\Msun) & (pc) & (pc) & (10$^8$cm$^{-3}$) & (10$^4$cm$^{-3}$)\\
\hline
12326-6245 & 0.42  & 1.6    & 3.0    & 0.03& 0.6 & 6.7& 3.6  \\
16060-5146 & 1.9/5.5 & 4.1/12 & 13/23 & 0.04/0.07&0.6/1.1 & 12/6.9 & 6.5/3.7 \\
16065-5158 & 0.51/3.7 & 2.0/14 &  5.9/17 & 0.03/0.08& 0.6/2.1 & 8.2/3.1& 1.7/0.63  \\
\hline 

\end{tabular}                                                                   
\end{center}                                                                                         
\tablefoot{The
 first entry gives the mass of the hot component traced by the SED, the second of the whole
 dust envelope
 traced at 870~{$\mu\hbox{m}$}\/ and the third column lists the virial masses derived
 from {\hbox{\rm C$^{17}$}{\rm O}}(3--2) emission. In the last four columns, the radii for the hot compact component (taken from the XCLASS line
 modeling) and the extended dust emission as well
 as the H$_2$ number densities for both components are listed. If two values
 are listed, they indicate near/far distances used for the calculations.}
\end{table*}

\section{Discussion}

\subsection{Molecular line data}
The considerable band width of the FFT spectrometers at APEX and the huge number of molecular species that we detected in each source enabled us to study their chemical composition. 
To facilitate a comparison, the molecular emission was divided into two
classes, the hot and the cold molecules following the analysis described in Sect. 4.1, with rotational temperatures above
100~K and below 100~K respectively, which corresponds to the ice evaporation
temperature for complex organic molecules \citep{2007A&A...465..913B}.\\

\begin{figure} 
\includegraphics[angle=0,width=8.0cm]{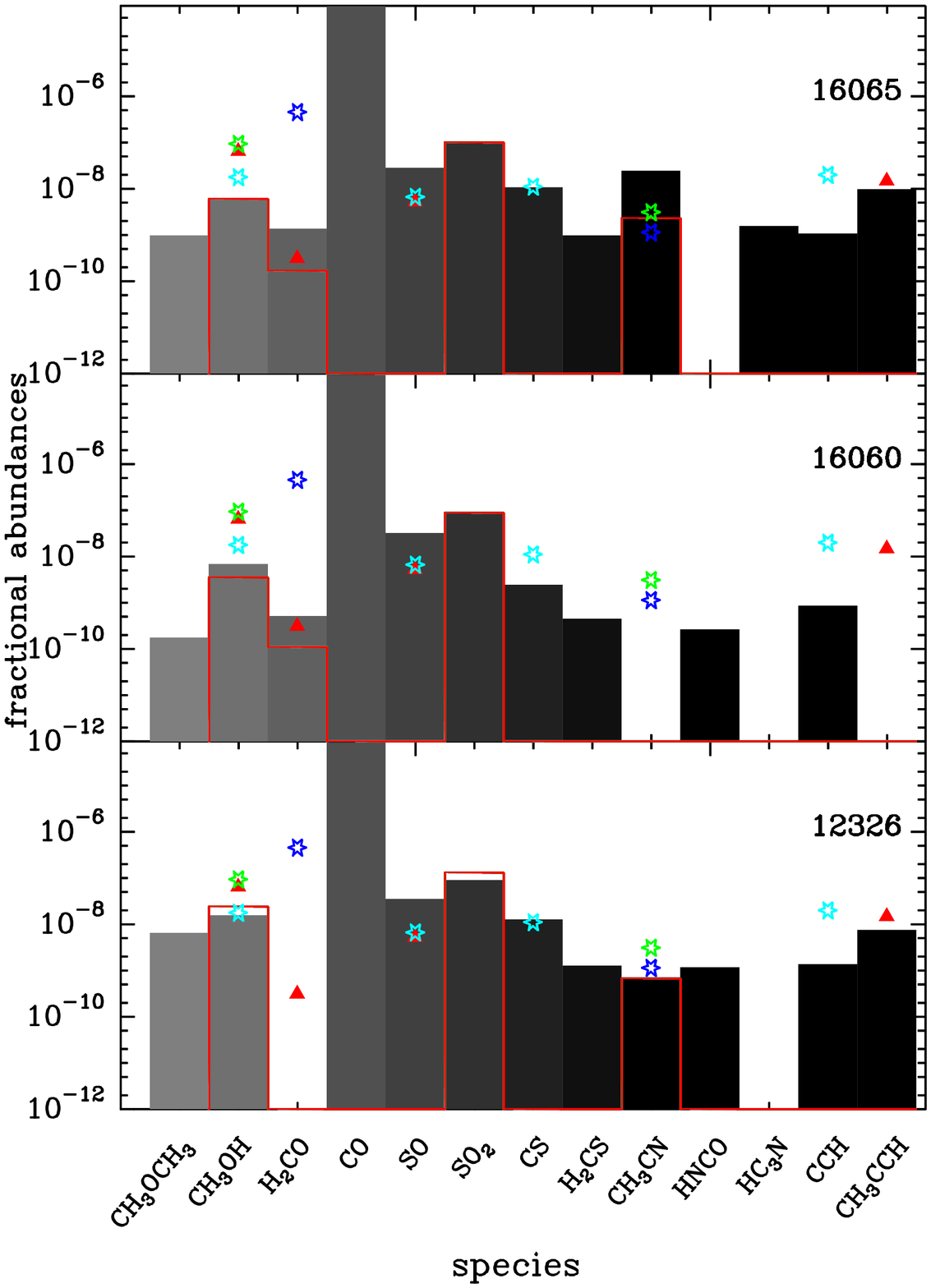}
\includegraphics[angle=0,width=8.0cm]{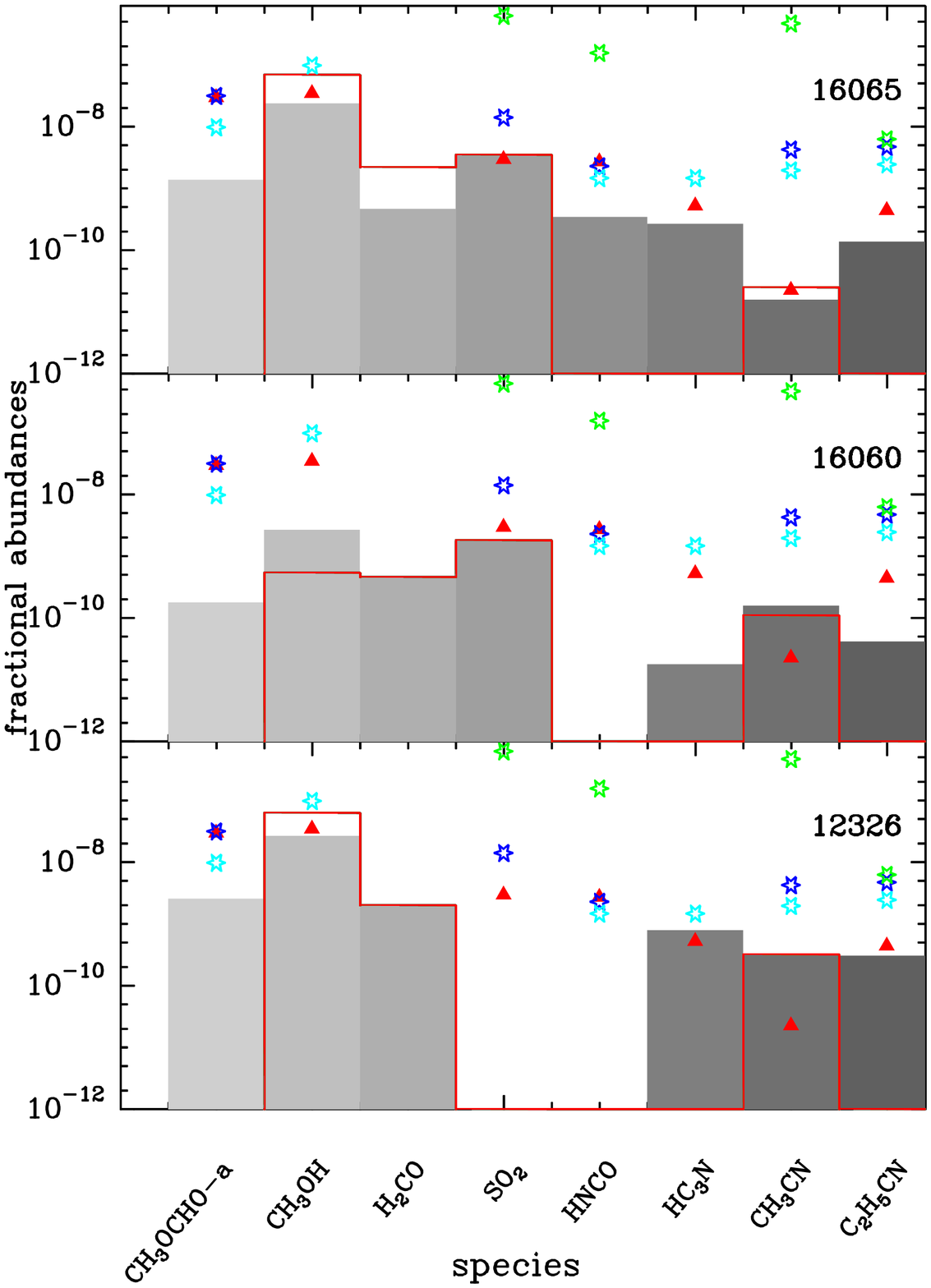}
\caption{\label{abun_comp} Fractional abundances in the three sources for T $<$ 100~K (cold
 component, top plot) and T $>$ 100~K (hot component, bottom plot).  Red triangles show abundances for G34.3+0.15 (Macdonald et al. 1996),
 green stars for NGC6334I, blue stars for G327.3-0.6 (both Schilke et
 al. 2006) and turquoise stars for 17233-3606 (Leurini et al. 2007). The red lines indicate those molecules where the modeling of
different transitions resulted in different column densities (see Tables
\ref{tab_12326} to \ref{tab_16060}).} 
\end{figure}

Several molecules, {\hbox{${\rm CH}_3{\rm OH}$}}, {\hbox{${\rm CH}_3{\rm CN}$}}, {\hbox{${\rm H}_2{\rm CO}$}}\/, and {\hbox{${\rm SO}_2$}}\/ were observed in
more than one frequency setup. In these cases it was not always possible to
model the species with one consistent set of temperatures and column
densities. The reason for this might be either calibration uncertainties
between the bands or deviations from local thermal equilibrium. \\  
In Fig. \ref{abun_comp} the abundances in the hot and cold component are
shown. In 16060--5146, only {\hbox{${\rm CH}_3{\rm OH}$}} and {\hbox{${\rm CH}_3{\rm CN}$}} were modeled with two hot components. In the remaining species, it was impossible to resolve two hot components, even though large line widths often point in this direction. In all three sources, one can see a trend of observing higher
abundances in the O-bearing species than in the N-bearing species in the hot component.

Chemical differentiation between N- and O-bearing species has been
observed in several high resolution studies of hot molecular cores, among them W3(OH)
\citep{1999ApJ...514L..43W} and Orion-KL \citep{1995ApJS...97..455S}. In the
latter, the ``hot core'' is abundant in complex, saturated N-bearing species, while the
``compact ridge'' shows high abundances of complex O-bearing
molecules. With interferometric observations, \citet{2005IAUS..231..217L}
resolve a similar differentiation in the sources E and F in G9.62+0.19, a high-mass star-forming region with similar distance and luminosity to the three
sources presented here. Because the chemical differentiation is only resolved
with high spatial resolution observations, one would not expect to find the
difference in a single dish beam. This leaves the question why one sees a
higher abundance of O-bearing species in the three sources of our sample.  
One possibility is that the complex O-bearing molecules stem from a more
extended region and their emission is therefore less beam-diluted.

\citet{2001ApJ...546..324R} argue that a higher abundance of
O-bearing species might suggest an earlier evolutionary stage where the hot
core has developed into a rich O-bearing chemistry in the first 10$^4$ yr
before evolving into a nitrogen rich state, but placing our sources at this young state is inconsistent with the high
abundance of {\hbox{${\rm CH}_3{\rm CN}$}}\/ found here.  The total fraction of N observed is on the order of $3\times 10^{-8}$, $3\times
10^{-9}$ and $5\times 10^{-10}$ for 16065$-$5158, 12326$-$6245
and 16060$-$5146. This fraction was made up of {\hbox{${\rm CH}_3{\rm CN}$}}, {\hbox{\rm HNCO}}, {\hbox{${\rm HC}_3{\rm N}$}}, and {\hbox{${\rm C}_2{\rm H}_5{\rm CN}$}}.  Most of
the nitrogen in hot cores is however locked up in the N-bearing species
{\hbox{${\rm N}_2{\rm H}^{+}$}} and {\hbox{${\rm NH}_3$}}, which were not covered by any of our receiver setups.

On the other hand, it is conceivable that the complex N-bearing species, which
according to the above-mentioned interferometric studies are expected to be
found closest to the hot core, have already been destroyed by the radiation of
the young star, placing the sources at a slightly later stage of the hot core
phase. For 16060$-$5146 and 12326$-$6245, this would be consistent with the discovery of
cm continuum emission at the position of the 870~{$\mu\hbox{m}$}\/ peak. Another origin
of the high abundance of O-bearing species could be a
different composition of the initial grain mantles as discussed by
\citet{1992ApJ...399L..71C} and modeled by \citet{1993ApJ...408..548C}
with thermal effects.\\

In Fig. \ref{abun_comp}, the beam averaged molecular abundances with respect
to {\hbox{\rm H$_2$}} are compared with those derived for
G327.3--0.6 (blue), NGC6334I \citep[green, ][]{2006A&A...454L..41S} and
17233--3606 \citep[cyan,][]{2008A&A...485..167L}), which were also obtained with APEX and analyzed in
the same way as our data. 
To compare with a well-studied northern hemisphere
hot core, the abundances of G34.3+0.15
\citep{1996A&AS..119..333M} were also included.
Overall, one can see good agreement between our data and the other four hot
cores for the cold envelope molecules, apart from the slight overabundance
of {\hbox{\rm SO}}\/ and {\hbox{${\rm SO}_2$}}\/ in our sources. \\
Comparing the hot species, it is obvious that while they are line rich,
our sources are weaker in their line emission than the other
sources. The deficiency of nitrogen-bearing species in the hot component is
pronounced when comparing with the other sources. \citet{2009AJ....137..406B}
do not detect any N-bearing species toward two young sources in their high-resolution study of hot cores performed with the SMA. They explain the lack of
N-bearing species by saying that these need more time to evolve
or warmer gas. Because we are detecting lines from gas with $T~>~200$~K, it is not
likely that the gas is not warm enough for nitrogen chemistry. The
association of 16060$-$5146 and 12326$-$6245 with  {\hbox{UCH{\sc ii}}}\/ regions also rules out
the hypothesis that they are still at an early evolutionary stage.\\

A prominent feature of the three sources is the high abundance of sulfur
species. According to \citet{2004MNRAS.354.1141V}, the sulfur species are expected to show a
time-dependent signature. In their models, the authors predict the
{\hbox{\rm SO}}/{\hbox{${\rm H}_2{\rm CS}$}} ratio to be around one at earlier times and then lowering, which is not observed in any
of our sources. 

 \citet{2003A&A...412..133VA} model the sulfur chemistry in the
envelopes of young massive stars. The abundances they derive for {\hbox{{\rm
     CS}}}\/ and {\hbox{${\rm H}_2{\rm CS}$}} agree with those found in our work. {\hbox{${\rm SO}_2$}}\/ and {\hbox{\rm SO}}\/, however, are
observed here at much higher abundances. The derived {\hbox{${\rm SO}_2$}}\/ abundances
compare to those of the {\hbox{${\rm SO}_2$}}\/ "jump" model of the inner envelopes by \citet{2003A&A...412..133V}. \citet{2009AJ....137..406B} discuss the use of {\hbox{{\rm CS}}}, {\hbox{\rm SO}}\/ and
{\hbox{${\rm SO}_2$}}\/ as chemical clocks, with {\hbox{{\rm CS}}}\/ as a suitable tracer for the earlier
stages, while {\hbox{\rm SO}}\/ and {\hbox{${\rm SO}_2$}}\/ better trace the more evolved stages. This
picture is consistent with our data. \\

\subsection{Continuum data}
Modeling of the spectral energy distributions of the three sources revealed
that the 3~mm continuum in 16060$-$5146 and 12326$-$6245 is dominated by free-free
emission, pointing toward a more evolved stage of the sources. This is
consistent with the infrared data and the radio continuum data observed by \citet{2007A&A...461...11U}, which also point toward 16065$-$5158 as the most embedded and least developed object of the three. \\
The luminosities derived from the ionized regions in 12326$-$6245 and 16060$-$5146
differ by about a factor of 2 from the FIR luminosities derived from the
IRAS fluxes. As \citet{2006ApJ...651..914G} discuss, these differences might result because part of the stellar luminosity is absorbed by the dust or there might be stars in the
region contributing to the FIR luminosity, but which are not hot enough to
ionize the gas. The derived luminosity also indicates that the three
sources are most likely placed at their near distance,  because they would
otherwise have to harbor an extremely massive star more luminous than O4. \\   
Comparing the dust mass estimates for the three sources, the virial and
continuum masses agree well for the near distance, indicating that the sources
are likely bound. To
get an estimate of the upper limit for the mass that might be concentrated at the innermost hot
component, we determined the mass using the temperature derived from the SED
model and the peak flux of the 870~{$\mu\hbox{m}$}\/ continuum. These masses are a factor
of about four lower than the envelope masses for 12326$-$6245 and 16065$-$5158,
while they are about a factor of 2 lower for 16060$-$5146, indicating that this
source is very centrally peaked. \\

All three sources have been observed by \citet{2004A&A...426...97F}, while 12326$-$6245 and 16060$-$5146 have also been observed by
\citet{2006A&A...460..721M} and \citet{2005MNRAS.363..405H}. Taking into
account the different ways the integrated fluxes used, our masses agree well
within factors of about 2 with the estimates by these authors.\\ They are,
however, more massive than the average sources in the samples of
\citet{2004A&A...417..115W}, \citet{2006A&A...447..221B} and
\citet{2002ApJS..143..469M}, whose samples contain generally less massive and
luminous sources. \\

\section{Conclusions}
We observed the three sources 12326$-$6245, 16060$-$5146 and 16065$-$5158 in several
molecular lines and in the mm and sub-mm continuum as a pilot study for a
forthcoming line and continuum survey of a large sample of 47 high-mass star-forming regions.  
The goal of the pilot study was, first, to define and test useful frequency setups to be used
with the APEX telescope for the two surveys and second, to study the three
sources in more detail, because they all show a hot-core-type molecular
spectrum, which is found in high-mass star-forming regions.\\

We chose a setup with {\hbox{${\rm HCO}^+$}}(4--3) and {\hbox{{\rm CO}}}(3--2) as our tools to pinpoint the location of the dense gas peak as a result of this pilot study.
These same tools will also provide information about the kinematics of the regions from the line shapes of the two
tracers. The sources were then observed in a setup targeted at the
{\hbox{${\rm CH}_3{\rm OH}$}}(7--6) series of lines to
obtain abundances for the envelope component and a potential hot core
component. \\ Once LABOCA was operational, the sources were imaged at 870~{$\mu\hbox{m}$}\/
to study their dust content. \\ 
The sources of the pilot study were observed in five frequency setups centered
at {\hbox{${\rm CH}_3{\rm OH}$}}(6--5),
{\hbox{${\rm CH}_3{\rm OH}$}}(7--6), {\hbox{${\rm H}_2{\rm CO}$}}(4--3), {\hbox{${\rm H}_2{\rm CO}$}}(6--5), and {\hbox{${\rm CH}_3{\rm CN}$}}(16--15) bands. The line surveys
resulted in the detection of lines from 19 different species.  Following
\citet{1998A&AS..133...29H}, the sources are all three classified as line-rich
sources because of the detection of high-excitation {\hbox{${\rm CH}_3{\rm OH}$}}\/ transitions. Because we
also detected complex organic species, the sources were classified as hot
cores. We modeled lines from an extended cooler envelope component as well as those from a hot compact component. \\
All three sources show an overabundance of oxygen-bearing species compared to
nitrogen-bearing species and a high abundance of sulfur
species. 

16060$-$5146 and 16065$-$5158 both show secondary clumps in their
surroundings. While they are all three located in regions of active ongoing
star formation, the 870~{$\mu\hbox{m}$}\/ peak of 16065$-$5158 is not associated with any
FIR source. Spectral
energy distributions including the ATCA 3~mm data were modeled for all three
sources to derive their dust properties.\\ 
The three sources imaged in 870~{$\mu\hbox{m}$}\/ and 3~mm continuum data are massive,
luminous hot cores. While 16065$-$5158 seems to be a very young, deeply embedded
object at the center of a young association, 12326$-$6245 and 16060$-$5146 seem more evolved and have already developed
 {\hbox{UCH{\sc ii}}}\/ regions. The star-forming activity in the region of 16060$-$5146 might
have been triggered by the large bubble at whose edge it is located.

\bibliography{prom}
\bibliographystyle{aa}

\begin{acknowledgements}

The authors thank an anonymous referee for useful comments.
Paola Caselli and Malcolm Walmsley provided additional discussion and comments, which helped to improve the paper. 
We thank the staff at the APEX and ATCA telescopes and the
people who maintain the CDMS and JPL molecular spectroscopy
databases. C. Dedes acknowledges financial support from the Studienstiftung des Deutschen
Volkes and the International Max Planck Research School for Astronomy and
Astrophysics during her thesis work. S. Thorwirth acknowledges support by the Deutsche Forschungsgemeinschaft (TH 1301/3-1).
This research has made use of NASA's Astrophysics Data System
and the NASA/ IPAC Infrared Science Archive, which is operated by the Jet Propulsion Laboratory, California Institute of Technology, under contract with the National Aeronautics and Space Administration.

\end{acknowledgements}

\end{document}